\documentclass[preprint,times]{elsarticle}	

\usepackage{color, colortbl}
\usepackage[T1]{fontenc}
\usepackage{bm}
\usepackage{hyperref}
\usepackage{placeins}
\usepackage{siunitx}
\usepackage{subfig}
\usepackage{booktabs}
\usepackage[margin=2cm]{geometry}
\usepackage{main-defs}

\biboptions{sort&compress}

\graphicspath{{figures}}

\begin{document}

\title{CaloShowerGAN, a Generative Adversarial Networks model for fast calorimeter shower simulation}

\author[1]{Michele Faucci Giannelli\corref{cor1}\href{https://orcid.org/0000-0003-3731-820X}{\includegraphics[scale=0.5]{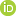}}}
\ead{michele.faucci.giannelli@cern.ch}
\author[2]{Rui Zhang\corref{cor1}\href{https://orcid.org/0000-0002-8265-474X}{\includegraphics[scale=0.5]{orcid_16x16.png}}}
\ead{rui.zhang@cern.ch}

\address[1]{Istituto Nazionale di Fisica Nucleare (INFN), Sezione di Roma Tor Vergata, Roma, 00133, Italy}
\address[2]{Department of Physics, University of Wisconsin, Madison, Wisconsin 53706, USA}

\cortext[cor1]{Equal contribution}

\begin{abstract}
In particle physics, the demand for rapid and precise simulations is rising.
The shift from traditional methods to machine learning-based approaches has led to significant advancements in simulating complex detector responses.
CaloShowerGAN is a new approach for fast calorimeter simulation based on Generative Adversarial Network (GAN).
We use Dataset 1 of the Fast Calorimeter Simulation Challenge 2022 to demonstrate the efficacy of the model to simulate calorimeter showers produced by photons and pions.
The dataset is originated from the ATLAS experiment, and we anticipate that this approach can be seamlessly integrated into the ATLAS system.
This development brings a significant improvement compared to the deployed GANs by ATLAS and could offer great enhancement to the current ATLAS fast simulations.
\end{abstract}

\flushbottom
\maketitle

\tableofcontents
\clearpage

\section{Introduction}

Modern particle and nuclear physics programs require extensive, high-precision Monte Carlo (MC) simulations for modelling the response to particles that travel through the detector materials.
This task is traditionally accomplished via a comprehensive detector simulation, utilising the \texttt{Geant4} toolkit~\cite{Agostinelli:2002hh}. 
The most time-consuming part of the simulation process is within the calorimeters, a sub-detector that measures energy deposits of particles.
When the initial particle interacts with the dense material in the calorimeters, it generates secondary particles.
This is a cascading process that can produce thousands of particles and form what is commonly known as a calorimeter shower.
The large number of particles to be simulated is the origin of the time and resource-intensive nature of the \GEANT simulation.
Overall, this aspect dominates the simulation time in collider experiments.
As an example, in a typical event of a top and anti-top quark pair production simulated in the ATLAS experiment~\cite{PERF-2007-01} at the Large Hadron Collider (LHC), the calorimeter shower simulation takes about \SI{80}{\%} of the total simulation time~\cite{SOFT-2010-01}.
In the upcoming High Luminosity LHC program, the increased data volumes are expected to surpass the available computing capabilities for producing the necessary amount of MC events used in physics analyses~\cite{ATLAS:2020pnm,Software:2815292}.
To uphold the consistent MC-to-data ratio, it becomes essential to substitute the calorimeter simulation, with a quicker alternative.
This requirement has encouraged the creation of fast and high-fidelity calorimeter simulation techniques.

Numerous efforts have been undertaken to speedup the simulation of calorimeter response while maintaining acceptable physics accuracy.
The FastCaloSim method~\cite{ATL-PHYS-PUB-2010-013,ATL-SOFT-PUB-2014-001}, developed within the ATLAS Collaboration, is an example of such attempts.
It involves the formulation of parameterised responses for the calorimeter, tailored to specific types of incoming particles.
By employing this parametrisation, it accelerates the speed of simulating an event by approximately a factor of ten, effectively bypassing the intricate shower development process carried out by \texttt{Geant4}.
In similar researches, including Refs.~\cite{de_Oliveira_2017,PhysRevLett.120.042003,CALOGAN,deOliveira:2017rwa,erdmann2018generating,Erdmann_2019,3dimGAN,Belayneh:2019vyx,Krause:2021ilc,Krause:2021wez,Krause:2022jna, Mikuni:2022xry,ATL-SOFT-PUB-2018-001,ATL-SOFT-PUB-2020-006,Buhmann:2020pmy, buhmann2021fast, Buhmann:2021lxj, Buhmann:2021caf, amram2023calodiffusion}, machine learning approaches using cutting-edge generative techniques are proposed for generating the calorimeter response.
In the recently developed AtlFast3~\cite{SIMU-2018-04}, the new generation of high-accuracy fast simulation in ATLAS, a combination of parametric and machine learning approaches (FastCaloSimV2~\cite{SIMU-2018-04} and FastCaloGAN~\cite{ATL-SOFT-PUB-2018-001}, respectively) is adopted to achieve optimal performance in terms of both speed and simulation accuracy across the detector's full phase space.

In this context, the research community organised the Fast Calorimeter Simulation Challenge 2022~\cite{challenge}, hereinafter referred to as CaloChallenge.
It is a newly introduced community challenge that aims at motivating the development of generative algorithms to address the calorimeter simulation challenge.
Standardised datasets and tools to facilitate training and validating processes are provided too.

In this paper, we present a fast calorimeter simulation model using Generative Adversarial Networks (GANs) technique, named \texttt{CaloShowerGAN}, with a focus on utilising the first dataset provided by the CaloChallenge.
Other models that participated in the CaloChallenge using this dataset, at the time of writing this paper, are documented in Refs.~\cite{Krause:2022jna, amram2023calodiffusion, 2023arXiv230308046H, 2023arXiv230803847M}.

The paper is organised as follows.
\Sect{\ref{sec:dataset}} briefly describes the datasets used in this study provided by the challenge.
The \CSG model description, hyper-parameter optimisation, and training procedure are detailed in \Sect{\ref{sec:model}}.
Further optimisations are described in \Sect{\ref{sec:furtherOpt}}.
The results and performances of \CSG are presented in \Sect{\ref{sec:performance}}.
Future research directions are presented in \Sect{\ref{sec:future}} followed by conclusions in  \Sect{\ref{sec:conclude}}.

\section{Input datasets}
\label{sec:dataset}

\begin{figure}[tbh]
\centering
    \includegraphics[width=1\textwidth]{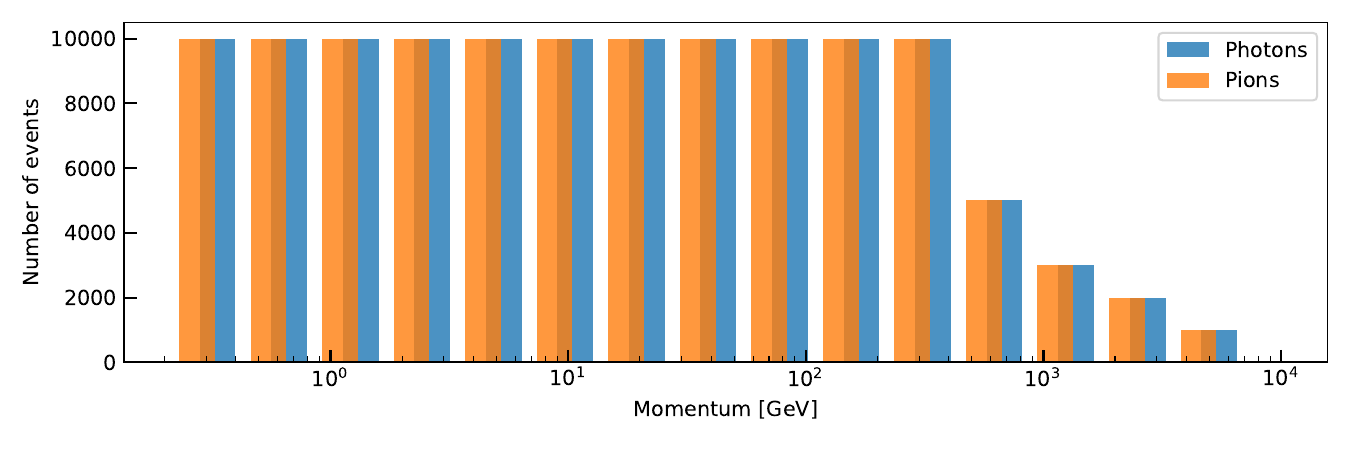}
    \caption{Statistics of Dataset 1 in the CaloChallenge under each incident momentum.}
    \label{fig:statistics}
\end{figure}

The first dataset provided by the challenge is part of the ATLAS open dataset~\cite{opendata} used in AtlFast3~\cite{SIMU-2018-04}.
It comprises two distinct subsets representing different particle types, the photon and the charged pion sets.
These data are generated by ATLAS using \GEANT with the official ATLAS detector geometry, ensuring that they accurately represent genuine electromagnetic and hadronic showers.
Each subset consists of 15 samples with different incident momenta generated at the calorimeter surface, followed by noise-free simulation to facilitate the training of accurate showers.
The incident momentum of the samples ranges from \SI{256}{\MeV} to \SI{4}{\TeV}, increasing in powers of two.
In the range from \SI{256}{\MeV} to \SI{256}{\GeV}, 10000 events are simulated at each energy value, while for higher momenta, the statistical count decreases.
An overview of these statistics is depicted in \Fig{\ref{fig:statistics}}.

All events are generated within the $|\eta|$ range of $[0.20, 0.25]$, aligning with the ATLAS' chosen strategy for parameterising the complete detector response.
Further insights into the reasoning behind this strategy and comprehensive sample details can be found in Ref.~\cite{SIMU-2018-04}.
In each event, the spatially distributed energy deposits simulated by \GEANT are referred to as ``hits''.
These hits are initially defined in Cartesian coordinates and subsequently transformed into cylindrical coordinates ($r$, $\alpha$, layer) along the particle's flying direction.
Here, $r$ is the distance of the hit from the intersection point between the extrapolation of the generated particle and the layer, while $\alpha$ denotes the polar angle in cylindrical coordinates.
The coordinate ``layer'' corresponds to the physical instrumented layer within the ATLAS calorimeter, indicating the extent of particle propagation from the origin of the detector.
The first four layers, numbered 0--3 according to the ATLAS naming scheme, are electromagnetic calorimeters dedicated to the measurement of electromagnetic showers while the following layers, denoted as 12, 13, and 14, are part of hadronic calorimeters used to measure hadronic showers.
Subsequently, the hits within each layer are aggregated into volumes referred to as ``voxels'', with the energy within a voxel being the cumulative sum of the energies contributed by all the associated hits.
The number of voxels in each layer for the two subsets is summarised in \Tab{\ref{tab:voxel}}.
The energy deposits in voxels are used as the input for the \CSG and are the information that the generative model aims to reproduce.

\begin{table}[tbh]
    \centering
    \caption{
        Configuration of voxels in the training dataset in each layer.
        Note that photons do not penetrate significantly in the hadronic calorimeter; therefore, they comprise fewer layers.
        $N_r$ and $N_\alpha$ are the number of bins in the $r$ and $\alpha$ directions, respectively.
        $N_{\text{voxel}}$ is the number of voxels.
    }
    \label{tab:voxel}
    \begin{tabular}{crrrr}
        \toprule
        Particle    & Layer    & $N_r$ & $N_\alpha$    & $N_{\text{voxel}}$\\
        \midrule
        \photon     & $0$       & $8$   & $1$           & $8$ \\
                    & $1$       & $16$  & $10$          & $160$ \\
                    & $2$       & $19$  & $10$          & $190$ \\
                    & $3$       & $5$   & $1$           & $5$ \\
                    & $12$      & $5$   & $1$           & $5$ \\
        Total       &           &       &               & $368$ \\
        \midrule
        \pion     & $0$       & $8$   & $1$           & $8$ \\
                    & $1$       & $10$  & $10$          & $100$ \\
                    & $2$       & $10$  & $10$          & $100$ \\
                    & $3$       & $5$   & $1$           & $5$ \\
                    & $12$      & $15$  & $10$          & $150$ \\
                    & $13$      & $16$  & $10$          & $160$ \\
                    & $14$      & $10$  & $1$           & $10$ \\
        Total       &           &       &               & $533$ \\
        \bottomrule
    \end{tabular}
\end{table}

\section{Model and hyperparameters}
\label{sec:model}

\CSG is designed to have a similar structure to FastCaloGAN available in Ref.~\cite{FastCaloGAN_code} so that it can be easily integrated by the ATLAS collaboration.
At the same time, the tool diverges from FastCaloGAN, leading to better performance.
This is achieved by optimisation in training data pre-processing and adjustments in model architecture and hyperparameters.
Several of these enhancements are owning to insights in understanding the details of how showers interact within the calorimeter.

\subsection{\CSG}
\begin{figure}[htb]
    \centering
        \includegraphics[width=0.3\textwidth]{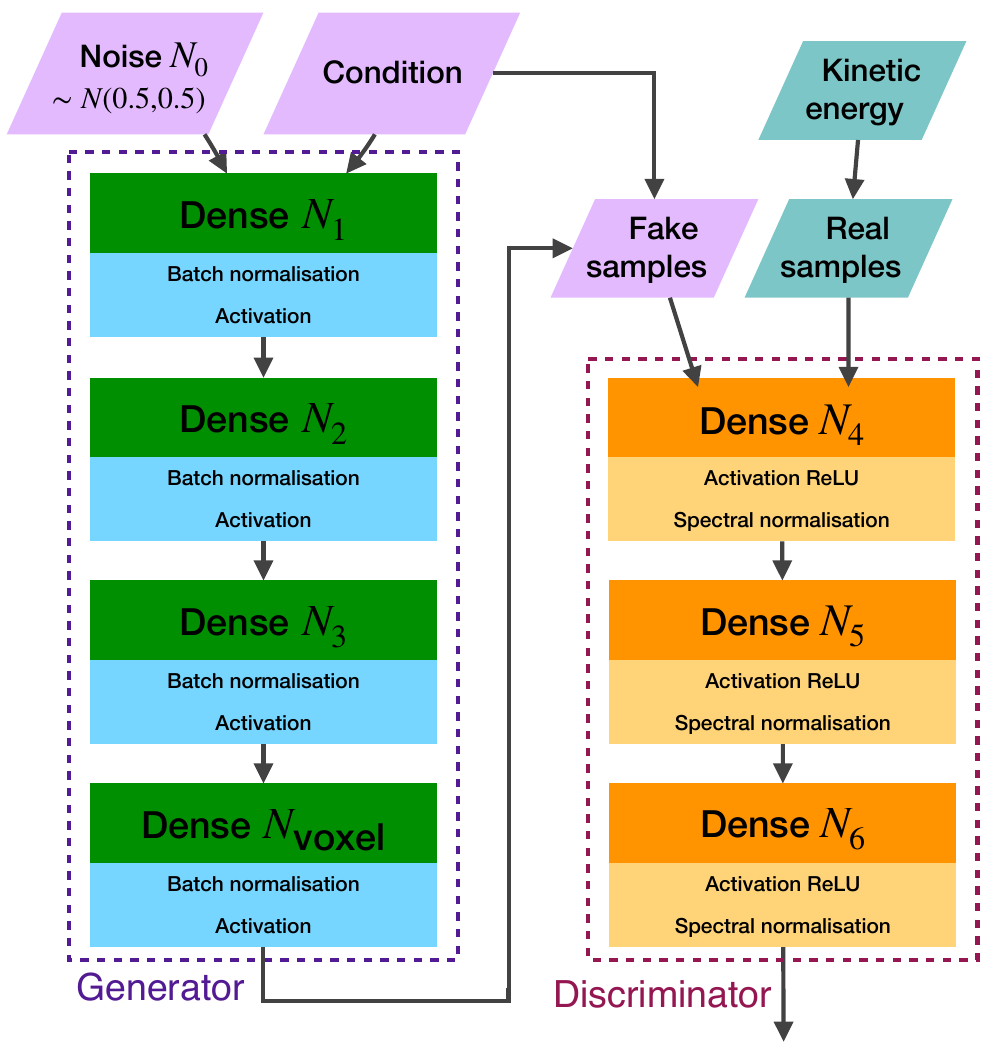}
        \caption{
            Schematic view of the \CSG architecture.
            The generator comprises three hidden layers with node counts of $N_1$, $N_2$, and $N_3$, respectively, along with an output layer sized to match the number of voxels.
            Each layer consists of a dense layer, followed by a batch normalisation operation and an activation operation.
            A 1-dimensional condition representing kinetic energy is used, alongside a 50-dimensional latent vector randomly sampled from a normal distribution with both mean and standard deviation of 0.5.
            The generator's output, concatenated with the condition label, is input to the discriminator.
            The discriminator comprises three dense layers with a ReLU activation function. 
            Spectral normalisation operation is used within the discriminator to stabilise the training.
            The generator's activation function and both networks' sizes vary between photon and pion GANs.
        }
    \label{fig:gan}
\end{figure}

\CSG is constructed on the foundation of the Conditional Wasserstein GAN algorithm~\cite{arjovsky2017towards,arjovsky2017wasserstein}, which has been established for delivering good performance and training stability.
To effectively simulate calorimeter showers across a broad range of incident momenta spanning multiple orders of magnitude, \CSG is conditioned on the true kinetic energy\footnote{
The relativistic relationship between kinetic energy ($E_\text{kin}$) and four-momentum ($\vect{p}$) is represented by $E_\text{kin} = \sqrt{\vect{p}^2 + m^2} - m$, where $m$ is the mass of the particle.
} of the incoming particle.
This variable is preferred because the characteristics of the shower are directly proportional to the logarithm of the kinetic energy of the particle rather than the momentum.
The architecture of \CSG employed in this study is depicted in \Fig{\ref{fig:gan}}.
The generator comprises three hidden layers and one output layer.
Each layer incorporates a dense layer, followed by batch normalisation~\cite{ioffe2015batch} and an activation operation.
The generator receives a noise vector randomly sampled from a high-dimensional normal distribution, where each dimension has a mean and standard deviation of $0.5$.
The condition label of the generated event is simultaneously fed as input.

The output of the generator aligns with the number of voxels and is subsequently input into the discriminator as ``fake'' events, along with the concatenated condition label.
The ``real'' events are taken from the dataset outlined in \Sect{\ref{sec:dataset}}, joined with their actual condition labels.
The discriminator encompasses three dense layers with a ReLU activation function.
No batch normalisation operation is employed, as it is found not to enhance performance.
Instead, spectral normalisation~\cite{miyato2018spectral} is adopted to stabilise the training.

\subsection{Data preprocessing}
\label{sec:preprocessing}

The input data comprises energy deposits in voxels, measured in megaelectronvolts (MeV) and structured in an $n \times m$ matrix.
Here $n$ represents the number of voxels, and $m$ denotes the number of events.
Furthermore, the energy of each voxel is  normalised based on the kinetic energy of the particle.
This is similar to what is done in FastCaloGAN.
This normalisation procedure allows to standardise all values within the input vector to a similar order of magnitude for all input momenta, eliminating the significant difference between the momenta of the samples.
In this way, the GAN can focus on reproducing the shape of the showers rather than its absolute value.
The condition label is transformed to a normalised range of $[0, 1]$ using the following equation:
\begin{equation}
    \hat{E} = \frac{\log{\frac{E_\text{kin}}{E_\text{min}}}}{\log{\frac{E_\text{max}}{E_\text{min}}}}.
\end{equation}
Here $E_\text{min}$ ($E_\text{max}$) is the minimum (maximum) kinetic energy of the incoming particle in the training data. 
It has been observed that this scheme yields a large improvement over an alternative method employed in FastCaloGAN.

\subsection{Training}

The training process for \CSG involves independent training on distinct particle samples to maximise performance.
Each training employs a batch size of $1024$ and runs for a total of $10^{6}$ iterations.
Model checkpoints are created at intervals of $10^3$ iterations.
Due to the adversarial nature of GAN training, the final iteration does not necessarily yield the best outcome.
To address this, the approach of saving multiple iterations is adopted.
This approach has two-fold benefits: it enables quick training without evaluation during the process, and it offers flexibility in retrieving the optimal iteration using different strategies without the need for GAN re-training.

Inspired by the methodology used in FastCaloGAN, the total energy distributions associated with each of the 15 incident momentum points are picked as the assessment metric.
The $\chi^2$ value for each GAN model is computed between the binned distributions of the \GEANT sample and generated sample by the model and then normalised by the number of degrees of freedom used in each distribution (\chisq). 
The model that gives the lowest \chisq among the saved iterations is considered the best.
This selection process has proven to be a reliable metric for the overall quality of a shower.
It is found that the shape of the generated showers consistently improves in models with lower \chisq values.

The GAN architecture comprises various hyperparameters that require optimisations.
Beginning with the values employed in FastCaloGAN, the optimisation process is applied to several parameters, including the learning rate and momentum for both the generator and discriminator optimisers, the batch size, the discriminator-to-generator (D/G) ratio, the $\lambda$ that controls the penalty contribution in the Wasserstein GAN, and the choice of activation functions.
The D/G ratio, quantifying the number of times the discriminator is trained relative to a single training pass of the generator in each iteration, is found to play an important role.
Generator and discriminator sizes, ranging from one-quarter of the determined size to as much as four times the size, are explored.
Optimisation algorithms are also investigated beyond the commonly used Adam~\cite{kingma2014adam} optimiser including RAdam~\cite{liu2019variance} coupled with LookAhead~\cite{zhang2019lookahead}, as well as AdamW~\cite{loshchilov2017decoupled}.
It was observed that these alternative optimisers yielded sub-optimal performance when compared to Adam.

Eventually, the Adam optimiser is selected for both the generator and discriminator, utilising a learning rate of $10^{-4}$ and a momentum value of $\beta_1 = 0.5$.
These values are used for both pion and photon GANs.

\textbf{\textit{Hyper-parameters for Photon \CSG.}}
The utilisation of the Swish~\cite{swish} activation function in the photon \CSG yields better performance compared to more common choices like ReLU~\cite{agarap2018deep}.
While Swish activation can potentially introduce training instability, this drawback appears to be mitigated when used with the Glorot Normal~\cite{glorot2010understanding} initialisation method for the generator neuron weights.
Conversely, the ReLU activation, employed in conjunction with the He Uniform~\cite{he2015delving} initialisation, serves better in the discriminator.
Notably, a higher D/G ratio proves advantageous in strengthening the discriminator's potency against the generator, particularly when using the Swish activation function.

The size of the networks is chosen as follows.
The latent dimension is set to $100$, allowing for intricate data representation.
The width of the generator layers is increasing in the three hidden layers, from $100$, $200$, to $400$.
This scale is twice that of the generators seen in the FastCaloGANs, offering a substantial leap in capacity and potential.
The discriminator size of all three hidden layers is fixed to the number of voxels, i.e.\ $368$.
The value of $\lambda$ is chosen to be $3$.

\textbf{\textit{Hyper-parameters for Pion \CSG.}}
The pion \CSG employs the ReLU activation due to its better performance compared to Swish.
Both the latent dimension and the size of the generator layers have been carefully optimised.
The latent dimension is chosen to be $200$.
The width of the generator layers is determined to be $200$, $400$, and $800$, respectively.
Note the dimensions are larger than the photon \CSG to account for the larger voxel count employed for pions, as well as the higher complexity and variety of hadronic showers.
Larger networks are tested but fail to yield large gain and extend the training time significantly, hence they are not considered.
The optimal discriminator size mirrors that of the generator, although it assumes a distinct configuration from the one employed for photons.
The discriminator sizes sequentially progress from $533$ voxels in the input layer to $800$, $400$, and $200$ in subsequent layers.
Maintaining a low D/G ratio and a relatively higher value of $\lambda$ are found to be optimal.
The introduction of batch normalisation does not provide an enhancement the performance, but it is applied to be consistent with the photon \CSG.

An overview of the hyperparameters used in the photon and pion \CSG is shown in \Tab{\ref{tab:hp}}.

\begin{table}[htb]
    \centering
    \caption{
        Determined hyperparameter values for the photon and pion GANs.
    }
    \label{tab:hp}
    \begin{tabular}{l cc}
        \toprule
        Hyperparameter                                  & Photon                & Pion \\
        \midrule
        Latent space size                         & $100$ & $200$ \\
        Generator size ($N_1, N_2, N_3$)  & $100$, $200$, $400$  & $200$, $400$, $800$ \\
        Discriminator size ($N_4$, $N_5$, $N_6$)  & $368$, $368$, $368$  & $800$, $400$, $200$ \\
        Generator optimiser                       & Adam  & Adam \\
        \hspace{1em} Learning rate                & \num{1e-4}      & \num{1e-4} \\
        \hspace{1em} $\beta_1$                    & \num{0.5}       & \num{0.5} \\
        \hspace{1em} $\beta_2$                    & \num{0.999}     & \num{0.999} \\
        Discriminator optimiser                   & Adam  & Adam \\
        \hspace{1em} Learning rate                & \num{1e-4}      & \num{1e-4} \\
        \hspace{1em} $\beta_1$                    & \num{0.5}       & \num{0.5} \\
        \hspace{1em} $\beta_2$                    & \num{0.999}     & \num{0.999} \\
        Batch size   & $1024$ & $1024$ \\
        D/G ratio   & $8$ & $5$ \\
        $\lambda$   & $3$ & $20$ \\
        Activation (generator) & Swish & ReLU \\
        Activation (discriminator) & ReLU & ReLU \\
        Neuron weight initialisation (generator)        & Glorot Normal  & He Uniform \\
        Neuron weight initialisation (discriminator)    & He Uniform     & He Uniform \\
        Trainable parameters (generator, discriminator) & $261$k, $408$k & $871$k, $829$k \\
        \bottomrule
    \end{tabular}
\end{table}

\subsection{Intermediate results after hyperparameter optimisation}
\label{sec:result}

\def\bestPhotonSingleGAN{BNswish_hpo4-M1}
\def\bestPhotonSingleGANckpt{733}
\def\bestPionSingleGAN{BNReLU_hpo27-M1}
\def\bestPionSingleGANckpt{653}

\def\bestPhotonLow{BNLeakyReLU_hpo31-M-P-L-Sle12.3}
\def\bestPhotonLowckpt{719}
\def\bestPhotonMid{BNswish_hpo101-M-P-L-Sge12le18.2}
\def\bestPhotonMidckpt{922}
\def\bestPhotonHigh{BNswish_hpo101-M-P-L-Sge18}
\def\bestPhotonHighckpt{337}

\def\bestPhotonLayerNormLow{BNswishCustActiv_hpo113-M-Pnormlayer2-L-Sle12.1}
\def\bestPhotonLayerNormLowckpt{452}
\def\bestPhotonLayerNormMid{BNswishCustActiv_hpo113-M-Pnormlayer2-L-ge12le18.1}
\def\bestPhotonLayerNormMidckpt{797}
\def\bestPhotonLayerNormHigh{BNswishCustActiv_hpo113-M-Pnormlayer2-L-ge18.1}
\def\bestPhotonLayerNormHighckpt{951}

\def\bestPionSingleGANLayerNorm{BNReLUCustActiv_hpo27-M-Pnormlayer2.5}
\def\bestPionSingleGANLayerNormckpt{1965}

The performance of the selected \CSG is presented in \Figrange{\ref{fig:totalEphotonSingleGAN}}{\ref{fig:totalEpionSingleGAN}} where the distribution of the total energy for generated events and the input \GEANT sample are compared for all momentum points for photons and pions respectively.
Each pad contains the $\chi^2_i/\text{NDF}_i$ for the specific momentum $i$ while the total \chisq is displayed at bottom right, calculated as $\Sigma_i {\chi^2} / \Sigma_i \text{NDF}$.

\def\captionEtotalStart{The calorimeter response for the \GEANT simulation (solid black line) compared to \CSG (dashed red line) }

\def\captionEtotalEnd{The $\chi^2$ values in each sub-panel are calculated from the distributions in that incident energy and the final $\chi^2$ is calculated from the concatenation of the histograms in all energies.
}
\begin{figure}[htb]
\centering
    \includegraphics[width=0.7\textwidth]{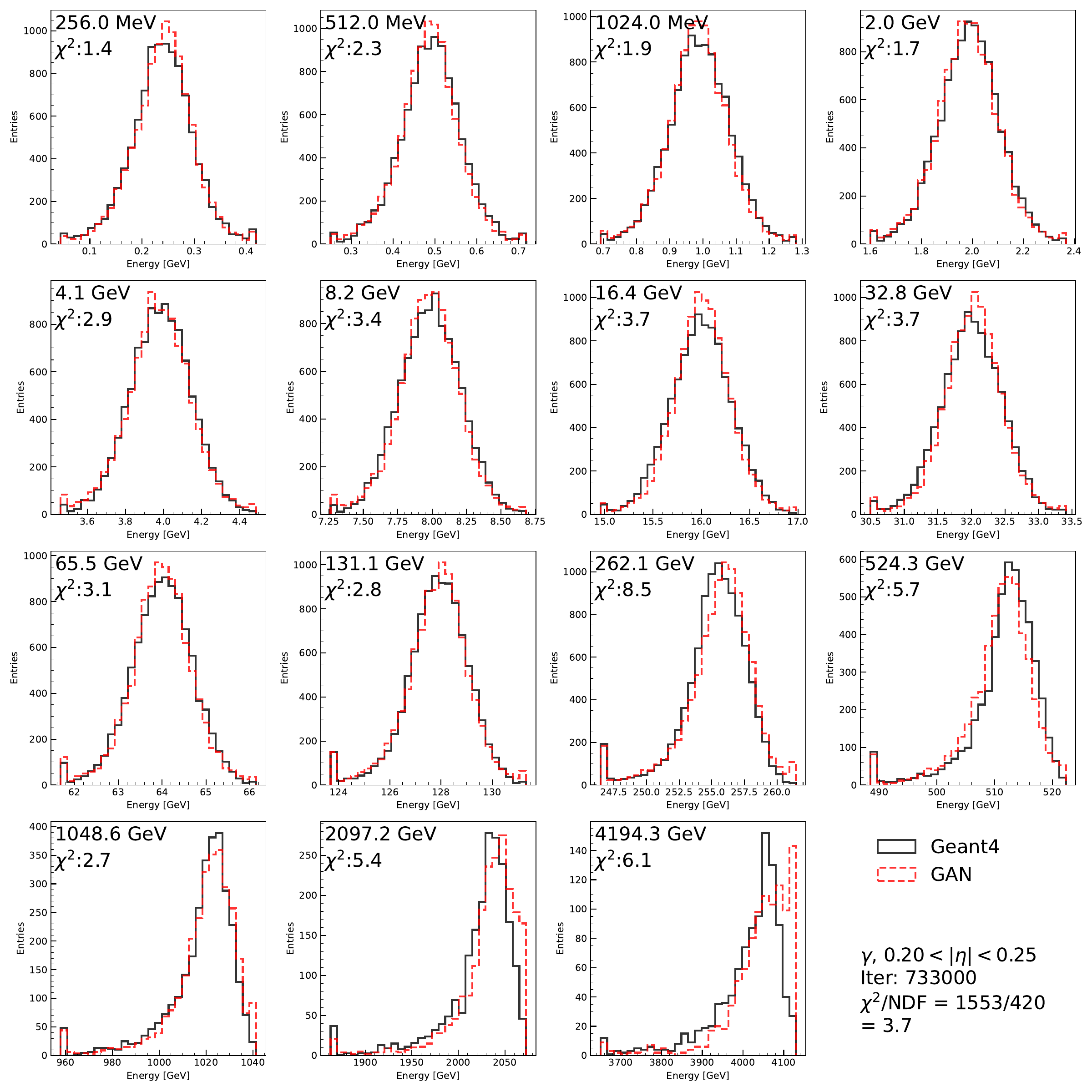}
    \caption{
        \textbf{Photon \CSG.}
        \captionEtotalStart for photons in the full momentum range. 
        \captionEtotalEnd
        }
    \label{fig:totalEphotonSingleGAN}
\end{figure}

\begin{figure}[htb]
\centering
    \includegraphics[width=0.7\textwidth]{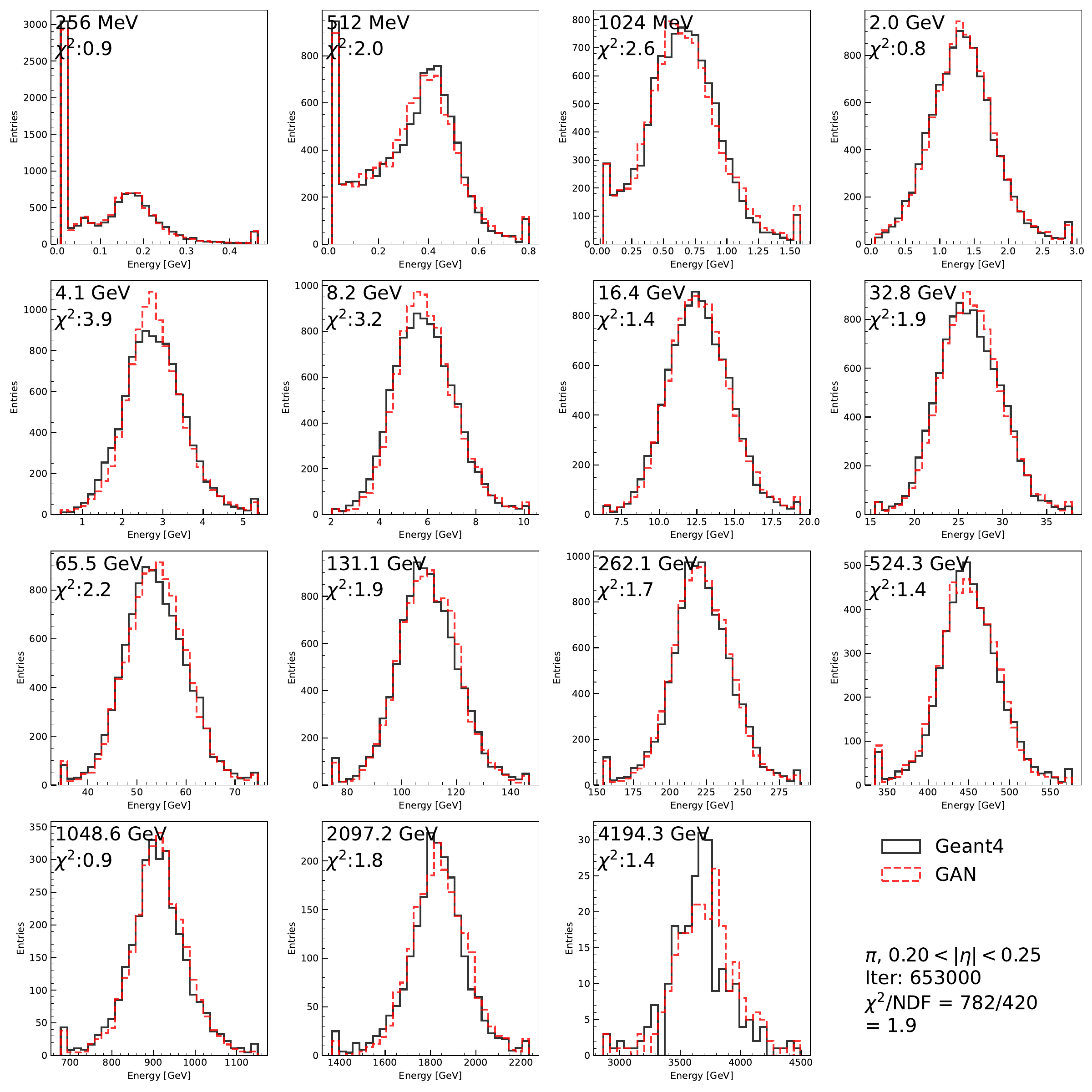}
    \caption{
        \textbf{Pion \CSG.}
        \captionEtotalStart for pions in the full momentum range. 
        \captionEtotalEnd
        }
    \label{fig:totalEpionSingleGAN}
\end{figure}

\CSG yields large improvement in comparison to the similar distributions presented in FastCaloGAN~\cite{SIMU-2018-04}.
The results obtaioned for pions show agreement across all momentum points, yielding a total \chisq value of 1.9\footnote{Our studies have revealed that the \chisq value can exhibit an error of approximately 0.1 as a result of variations stemming from random seed choices in generation.}.
Only a few distributions show small deviations from the \GEANT distributions, and there is no pronounced distinction in modelling either high or low momenta.
The level of agreement achieved for photons GAN is a little worse, giving a total \chisq value of 3.7, nearly twice that achieved for pions.
There is a clear trend in the values of the individual \chisq which are worse for the higher momenta.
The agreement worsened above 262~\GeV with a visible shift in the generated distributions.
These are likely caused by the difficulty in reproducing the asymmetric distribution of the total energy.
These drawbacks motivate the further development below.

\CSG represent a significant improvement not only in terms of physics performance but also in training speed when compared to FastCaloGAN.
The optimised \CSG requires approximately 6 hours while FastCaloGAN required over 16 hours on a GPU card with a similar computational setup.

\FloatBarrier

\section{Further optimisation of \CSG}
\label{sec:furtherOpt}

\subsection{Momentum split for photon GAN}
As depicted in \Fig{\ref{fig:totalEphotonSingleGAN}}, the performance of the photon \CSG reveals a dependence on the photon momentum, revealing three distinct momentum regions.
These regions are distinct by specific features of the electromagnetic showers within the ATLAS calorimeter, which can pose challenges to the training process of \CSG:

\begin{enumerate}
   \item In the low-momentum region, i.e.\ for momenta up to 4~\GeV, particles deposit almost all their energy in the initial two layers of the calorimeter.
   In the remaining layers, the voxels have minimal or negligible energy deposits, accompanied by significant event-to-event fluctuations.
   These fluctuations, absent in higher-momentum regions where all voxels are populated, can potentially confuse \CSG during the learning process. 
   \item In the medium-momentum range between 8~\GeV and 262~\GeV, the energy is deposited predominantly in layer 2 as this is the layer with the largest amount of material. This presents a contrasting scenario compared to the lower momentum range.
   Here, the first two layers, while containing some energy, contribute insignificantly to the overall energy deposit.
   \item Lastly, the samples with a momentum above 262~\GeV are characterised by an asymmetric response in the total energy. 
   This asymmetry is attributable to the shower extending beyond the boundary of the voxel volume and is affected by non-linearities in the calorimeter's response, which are more significant at higher energy samples.
\end{enumerate}

The mixture of events from these three distinct momentum groups during training introduces complexity to the learning process.
Thus, the generation of photon showers in \CSG is separated in three GANs, initialised with the previously outlined parameters, within the energy intervals of [256~MeV, 4~GeV], [4~GeV, 262~GeV], and [262~GeV, 4~TeV].
These GANs share a common momentum point to allow seamless interpolation across all momentum points in generation.

An additional limited HP scan is conducted to fine-tune the three GANs.
The outcomes confirms that the majority of the employed hyperparameters are indeed optimal for all three GANs, requiring only minor adjustments to achieve better performance.
In the low-momentum range, the ReLU activation function takes over Swish, while the He Uniform initializer replaces Gloroth.
For the two GANs trained in the higher momentum ranges, superior results are achieved using a downscaled generator network along with a latent space size reduced by half compared to the single GAN configuration.

The adoption of this approach yields a clear enhancement, evidenced by the lowered \chisq values for each individual GAN as compared to the single GAN's value of 3.7.
In the low-momentum GAN, \chisq values across all momentum points are improved, with the exception of the 4~\GeV sample where a visible distribution shift contributes to a raised \chisq.
In the medium-momentum range, all momentum points give a smaller \chisq compared to the corresponding values obtained using the single GAN.
However, the most substantial improvement is observed within the high-momentum range, where the previously observed disparity between generated events and input samples is effectively mitigated.
The \chisq results are summarised in \Tab{\ref{tab:energysplit}} where a further result is derived using two GANs.
While this alternative offers decreased precision compared to the three-GAN approach, it might still be valuable to consider by an experiment as it demands shorter training times and less memory during detector simulation when conducting inference.

\definecolor{Gray}{gray}{0.9}

\begin{table}[htbp]
    \centering
    \caption{
        Performance of the photon \CSG when splitting in different momentum ranges. The ``sum''-med $\chi^2$ and NDF are the sum of these values for the GANs considered. 
    }
    \begin{tabular}{cc
        S[round-mode=places, round-precision=0]
        S[round-mode=places, round-precision=0]
        S[round-mode=places, round-precision=1]
    }
        \toprule
        Name        & {momentum range}              & {$\chi^2$} & {NDF}  & {$\chi^2/\text{NDF}$} \\
        \toprule
        \rowcolor{Gray}
        Single GAN  & [\SI{256}{\MeV}, \SI{4}{\TeV}] & 1553             & 420    & 3.7 \\
        \midrule
                    & [\SI{256}{\MeV}, \SI{4}{\GeV}] & 360              & 140	  & 2.6 \\ 
                    & [\SI{4}{\GeV}, \SI{4}{\TeV}]   & 1042             & 308    & 3.4 \\ 
        \rowcolor{Gray}
        Two GANs    &     Sum          & 1402             & 448    & 3.1 \\
        \midrule
                    & [\SI{256}{\MeV}, \SI{4}{\GeV}] & 360              & 140	  & 2.6 \\ 
                    & [\SI{4}{\GeV}, \SI{262}{\GeV}] & 528              & 196    & 2.7 \\ 
                    & [\SI{262}{\GeV}, \SI{4}{\TeV}] & 299              & 140    & 2.1 \\ 
        \rowcolor{Gray}
        Three GANs  &    Sum           & 1187             & 476    & 2.5 \\
        \bottomrule
    \end{tabular}
\label{tab:energysplit}
\end{table}

This approach is not applicable to pions, primarily due to the inherent characteristics of hadronic showers.
Even at lower energies, hadronic showers tend to distribute some energy across all layers.
Splitting pion training into multiple GANs results in a similar overall performance and is therefore not recommended.

\FloatBarrier

\subsection{Layer-energy normalisation}
Despite the enhanced results stemming from the 3-GAN approach in the photon GAN, the overall performance still falls short of that demonstrated by CaloFlow~\cite{Krause:2022jna}, which currently stands as the leading model for photons.
At the time of writing, CaloFlow attains a remarkable \chisq value of 1.17 for photons and 1.32 for pions.
In pursuit of further enhancements, various strategies were explored.
The most significant improvement is achieved by adopting a different normalisation strategy used in Ref.~\cite{SIMU-2020-04} for the input data used during \CSG training.
Additional studies were performed and some of them are detailed in \Sect{\ref{sec:future}}.

The normalisation detailed in \Sect{\ref{sec:preprocessing}} lacks constraints on physics quantities, such as layer-specific energy and total energy.
While this information is implicitly present in the data, not explicitly providing it as part of the inputs to the GANs makes the learning process more challenging.
This information can be encoded using the following normalisation procedure:
Firstly, the voxel energy is normalised with respect to the total energy in the corresponding layer.
Subsequently, the energy in each layer is normalised to the total energy in the shower, resulting in 5 (7) new input dimensions for the photon (pion) GANs.
Another input is incorporated, representing the ratio between the total shower energy and the kinetic energy of the incident particle.

Adopting this normalisation strategy requires modifications to the final generator layer.
The output nodes are grouped based on the voxel count in each layer, and a SoftMax activation function is applied to each group to enforce layer-wise normalisation.
Similarly, 5 (7) nodes, corresponding to the total deposits of photons (pions) in each layer, are grouped, employing another SoftMax activation.
The last node, linked to the normalised total deposits in the entire calorimeter, uses a ReLU activation function.
In the discriminator, only the input layer size is altered to accommodate the additional values.
In the case of pions, it is worth noting that longer training times can be particularly beneficial; actually, pions with lower momenta continue to show significant learning improvements even after 1 million iterations.
Therefore, for pions, the number of iterations used for the training is extended to 2 million and so doubles the training time to 12 hours.
As a result, the total training time for \CSG is approximately 30 hours, which is only slightly less than the 32 hours used for FastCaloGAN.

\FloatBarrier

\section{Performance of \CSG}
\label{sec:performance}

\subsection{Total energy}

\begin{figure}[htb]
\centering
    \begin{minipage}[b]{.49\linewidth}
        \subfloat[low-momentum, 256~\MeV--4~\GeV\label{fig:totalEphotonLayerNormLow}]{
            \includegraphics[width=\textwidth]{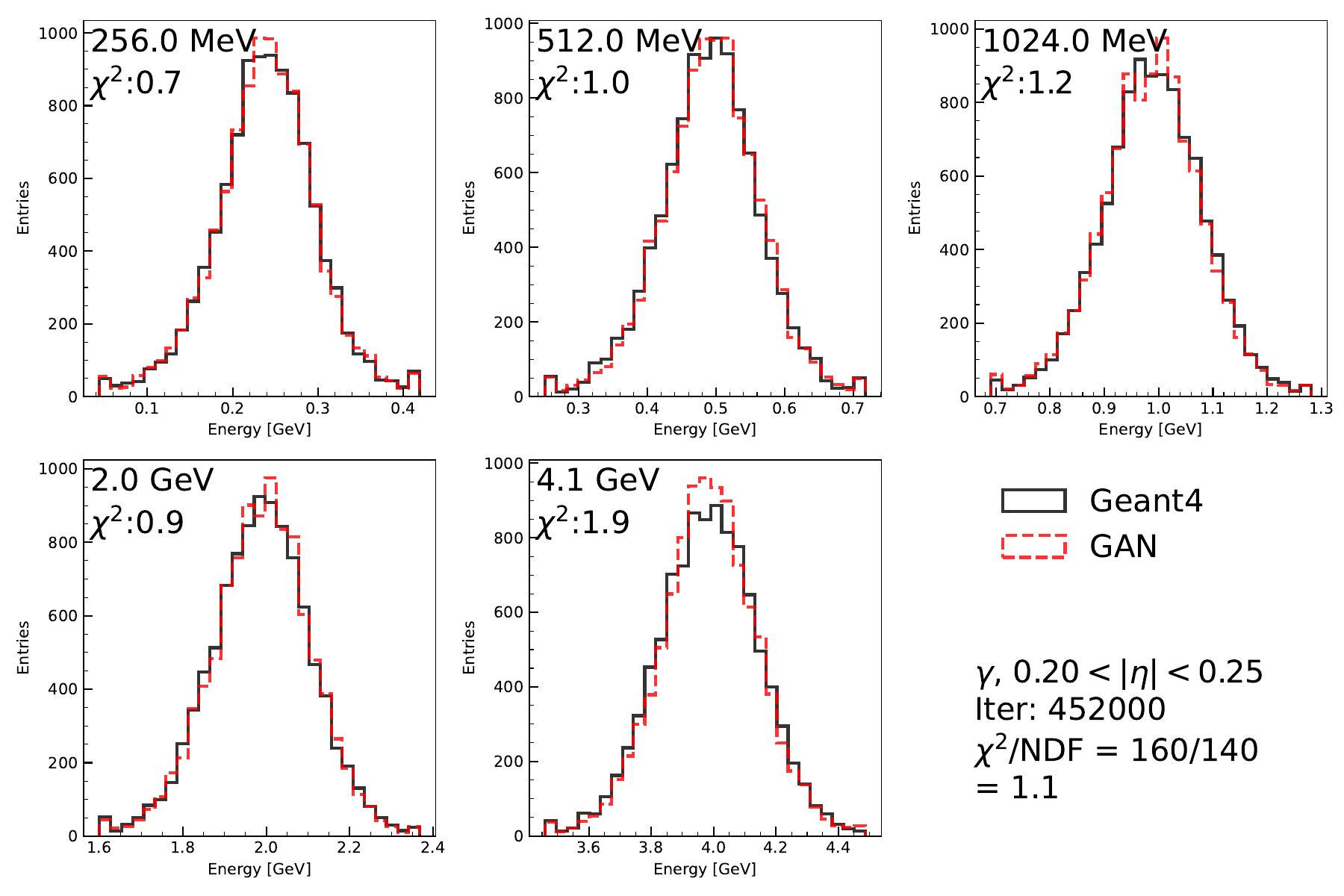}
        }
        \\
        \subfloat[high-momentum, 262~\GeV--4~\TeV\label{fig:totalEphotonLayerNormHigh}]{
            \includegraphics[width=\textwidth]{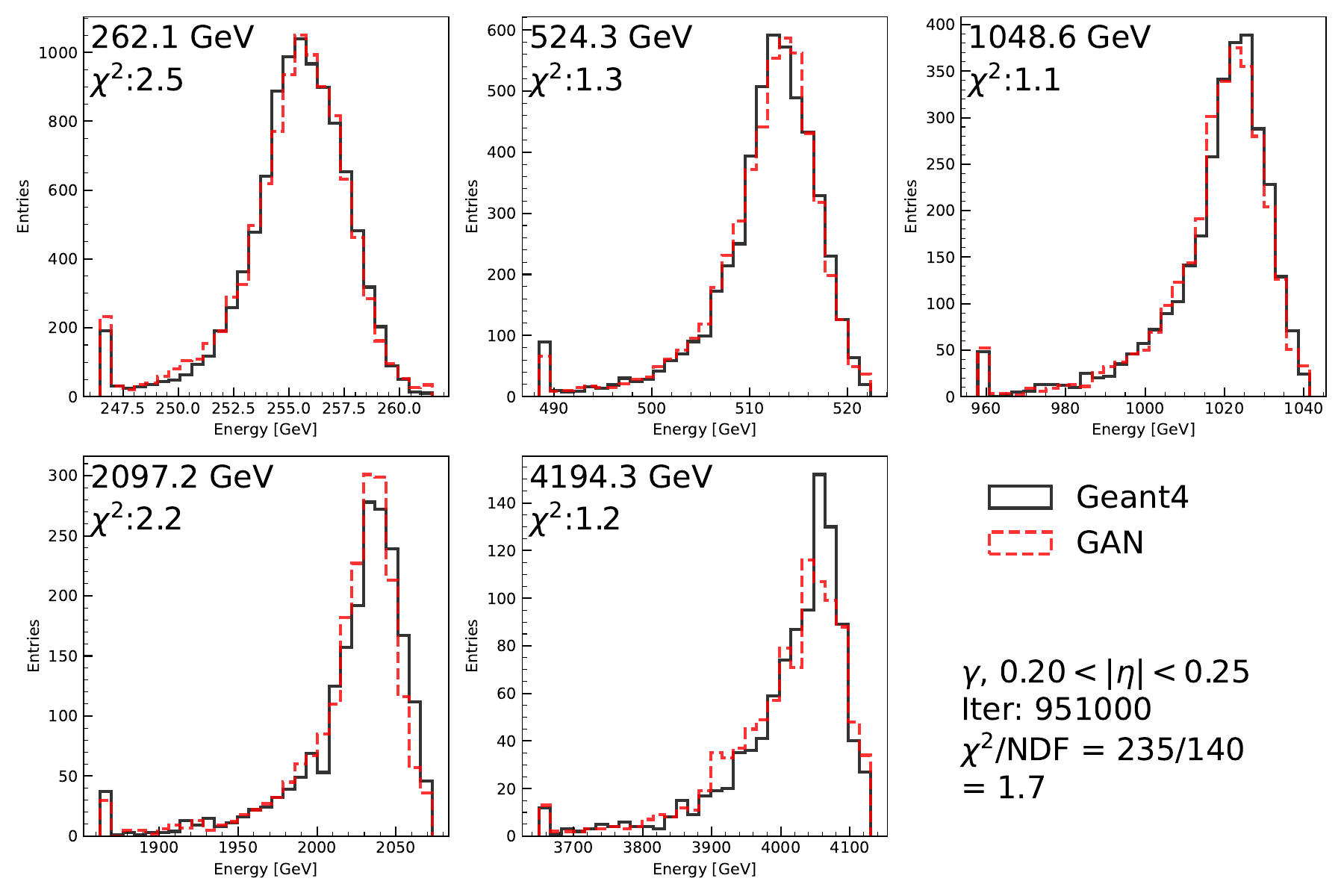}
        }
    \end{minipage}
    \hfill
    \begin{minipage}[b]{.49\linewidth}
        \subfloat[Midium-energy, 4~\GeV--262~\GeV\label{fig:totalEphotonLayerNormMid}]{
            \includegraphics[width=\textwidth]{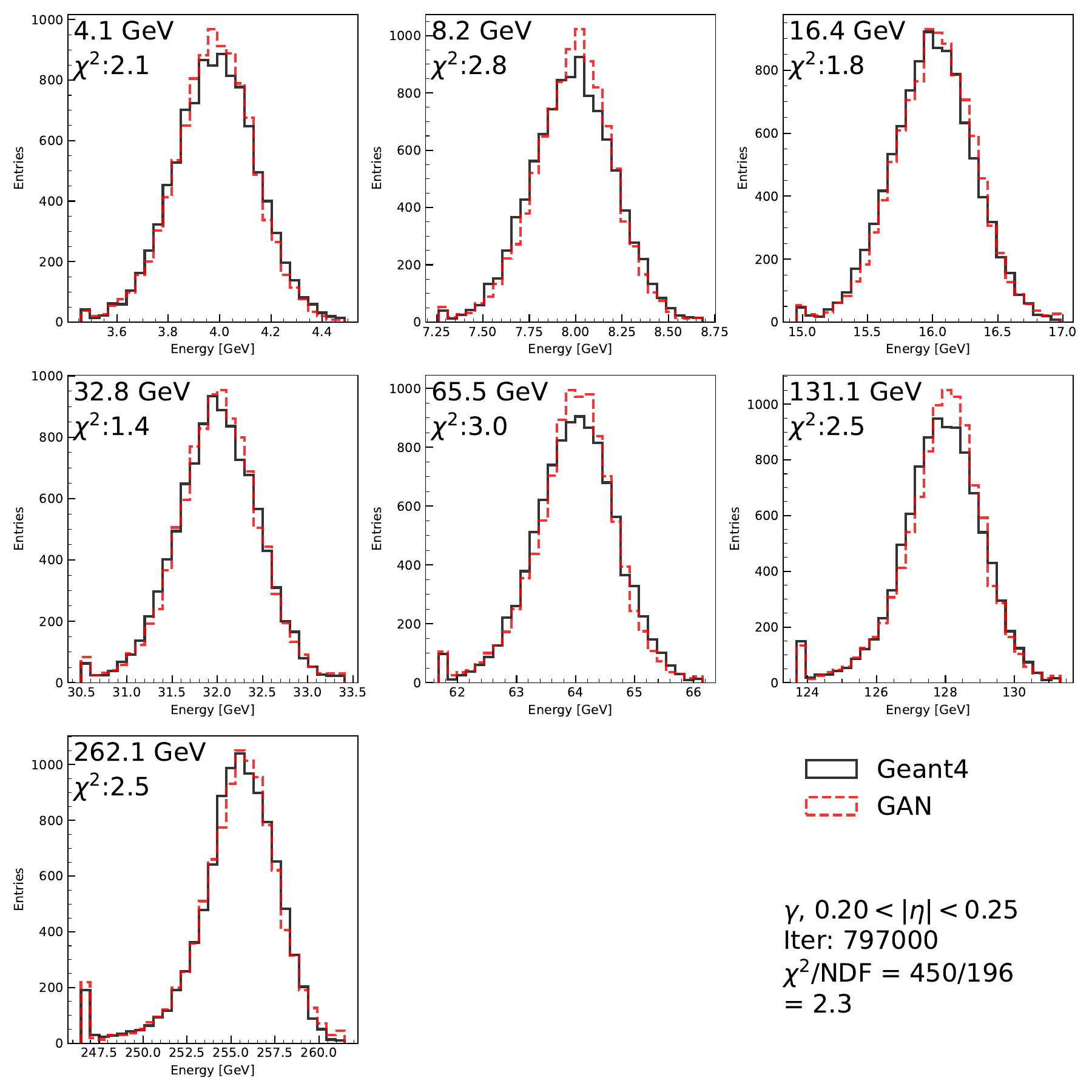}
        }
    \end{minipage}
    \caption{
        \textbf{Photon \CSG with layer-energy normalisation.}
        \captionEtotalStart for photons samples in the three momentum ranges using the layer-energy normalisation. 
        \captionEtotalEnd
        }
    \label{fig:totalEphotonLayerNorm}
\end{figure}

\begin{figure}[htp]
\centering
    \includegraphics[width=0.7\textwidth]{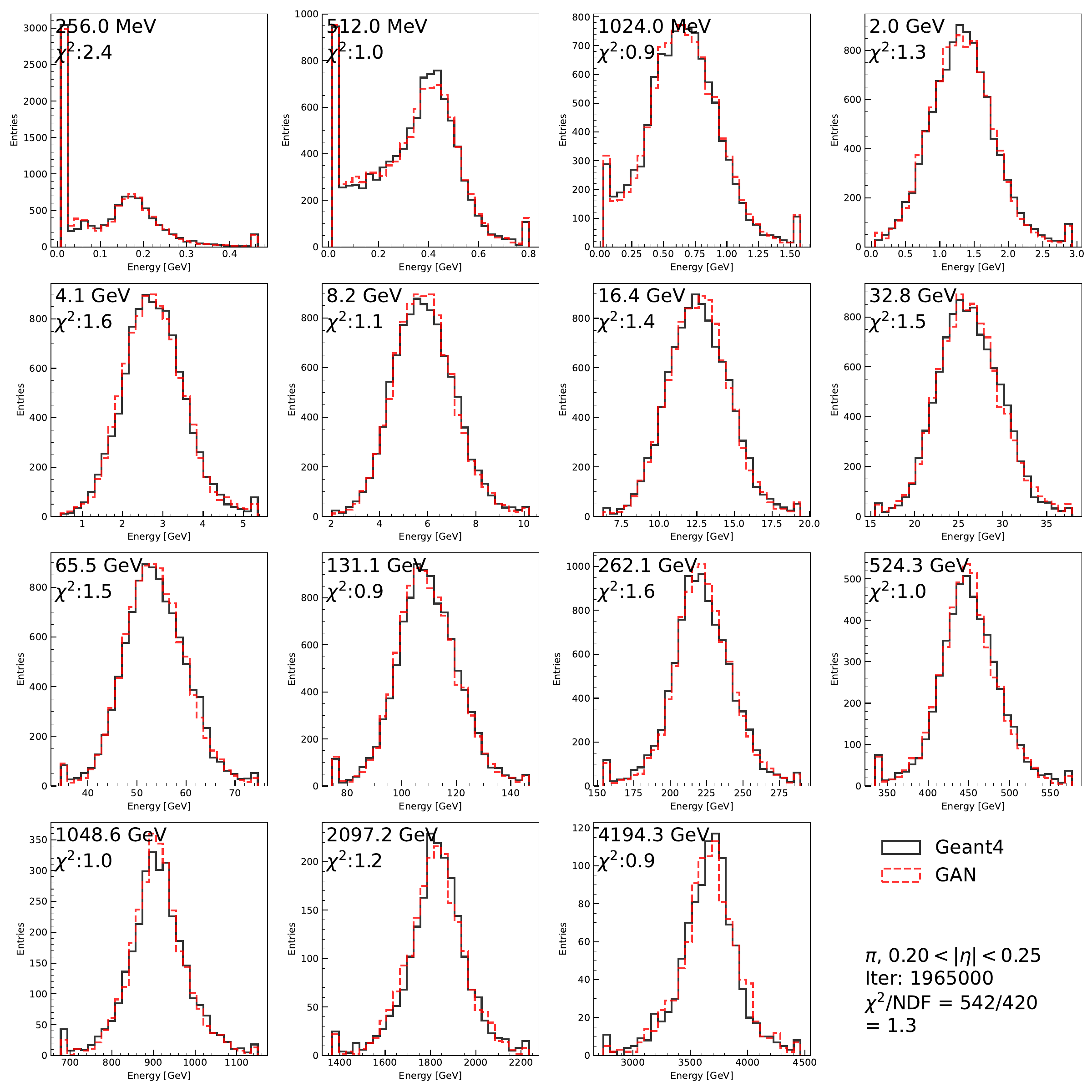}
    \caption{
        \textbf{Pion \CSG, with layer-energy normalisation.}
        \captionEtotalStart for pions in the full momentum range using the layer-energy normalisation. 
        \captionEtotalEnd
        }
    \label{fig:totalEpionLayerNorm}
\end{figure}

The results of \CSG with the layer-energy normalisation are depicted in \Figrange{\ref{fig:totalEphotonLayerNorm}}{\ref{fig:totalEpionLayerNorm}}. 
By employing layer-energy normalisation, a substantial improvement in \chisq is observed.
In particular, the performance achieved by \CSG on the pion dataset reaches a comparable performance to that of CaloFlow~\cite{Krause:2022jna} in terms of the \chisq metric.
It is worth noting that there exists a slight disparity in the definition of \chisq between the two studies.
In this study the actual number of degrees of freedom (NDF) is adopted by excluding empty bins, while the CaloFlow incorporates the total number of bins in the normalisation.
Considering the proximity of the sum of all $\chi^2$ values, we conclude that the two models reach comparable performance.

Detailed comparisons using more complex metrics, such as a classifier, are the objective of the CaloChallenge, where both models are actively participating. 
Therefore, we defer further quantitative comparisons beyond the scope of the \chisq metric to the CaloChallenge, where these considerations are appropriately addressed in a consistent and fair way for all models.

The results obtained from the 3 GANs used to simulate the full momentum range of the photon dataset are summarised in \Tab{\ref{tab:energysplitLayerNorm}}.
While \CSG does not quite match the performance level of CaloFlow, it still attains a high accuracy.
This achievement yields promise for enhancing the performance of GANs employed by ATLAS.

\begin{table}[htbp]
    \centering
    \caption{
        \textbf{\CSG, with layer-energy normalisation.}
        The ``sum''-med $\chi^2$ and NDF are the sum of these values for the GANs considered.
    }
    \begin{tabular}{cc
        S[round-mode=places, round-precision=0]
        S[round-mode=places, round-precision=0]
        S[round-mode=places, round-precision=1]
    }
        \toprule
        Name        & {momentum range}              & {$\chi^2$} & {NDF}  & {$\chi^2/\text{NDF}$} \\
        \midrule
        \rowcolor{Gray}
        Photon single GAN  & \SI{256}{\MeV}--\SI{4}{\TeV} & 1131             & 420    & 2.7 \\
        \midrule
                    & \SI{256}{\MeV}-\SI{4}{\GeV} & 160              & 140	  & 1.14 \\ 
                    & \SI{4}{\GeV}--\SI{4}{\TeV}  & 703              & 308    & 2.3 \\ 
        \rowcolor{Gray}
        Photon two GANs    &    Sum            & 863             & 448    & 1.9 \\
        \midrule
                    & [\SI{256}{\MeV}, \SI{4}{\GeV}] & 160              & 140	  & 1.1 \\ 
                    & [\SI{4}{\GeV}, \SI{262}{\GeV}] & 450              & 196    & 2.3 \\ 
                    & [\SI{262}{\GeV}, \SI{4}{\TeV}] & 235              & 140    & 1.7 \\ 
        \rowcolor{Gray}
        Photon three GANs  &     Sum           & 845             & 476    & 1.8 \\
        \midrule
        \midrule
        Pion GAN 1M & [\SI{256}{\MeV}, \SI{4}{\TeV}] & 628             & 420    & 1.5 \\
        \midrule
        Pion GAN 2M & [\SI{256}{\MeV}, \SI{4}{\TeV}] & 542             & 420    & 1.3 \\
        \bottomrule
    \end{tabular}
\label{tab:energysplitLayerNorm}
\end{table}

\FloatBarrier

\subsection{Stability of the training}

The training of a GAN, being an adversarial game between the two networks, is not a well-defined minimisation problem.
In general, a longer training should produce a better result; however, as stated previously, the optimal GAN may not always come from the final iteration of the training.
Therefore, the behaviour of \chisq over the saved iterations is examined.
The progression of \chisq throughout iterations is visually represented in \Fig{\ref{fig:chi2photon}} and \Fig{\ref{fig:chi2pion}} for photons and pions, respectively.
Evidently, the \chisq values for both individual energies and the aggregate shows a decreasing trend as a function of the iteration and reach a relatively stable value toward the end of the training, indicating the model is progressively improving during training.
It is interesting to observe that the best overall model, i.e.\ the one that produces the best total \chisq, may not coincide with the optimal iteration for a specific momentum.
The present selection methodology procedure is a balance between performance across all momentum points.
In general, the final selection is driven by momentum points that are relatively less accurately modelled, as these contribute the most to the overall total \chisq.

In the case of pions, a notable distinction in \chisq is observed across various momentum points.
Higher momentum points tend to converge more quickly towards a stable solution, while lower momentum points have a relatively slower convergence in \chisq. 
The longer training used for the pions helps to achieve a good convergence for all energies.

\def\captionchi{The first $N-1$ panels show the individual \chisq for a momentum energy and the last panel shows the total \chisq.
The selected iteration is indicated in red dot while the orange dots in the individual cases indicate the lowest \chisq when only considering that momentum.}
\begin{figure}[htb]
\centering
    \begin{minipage}[b]{.49\linewidth}
        \subfloat[low-momentum, 256~\MeV--4~\GeV]{
            \includegraphics[width=\textwidth]{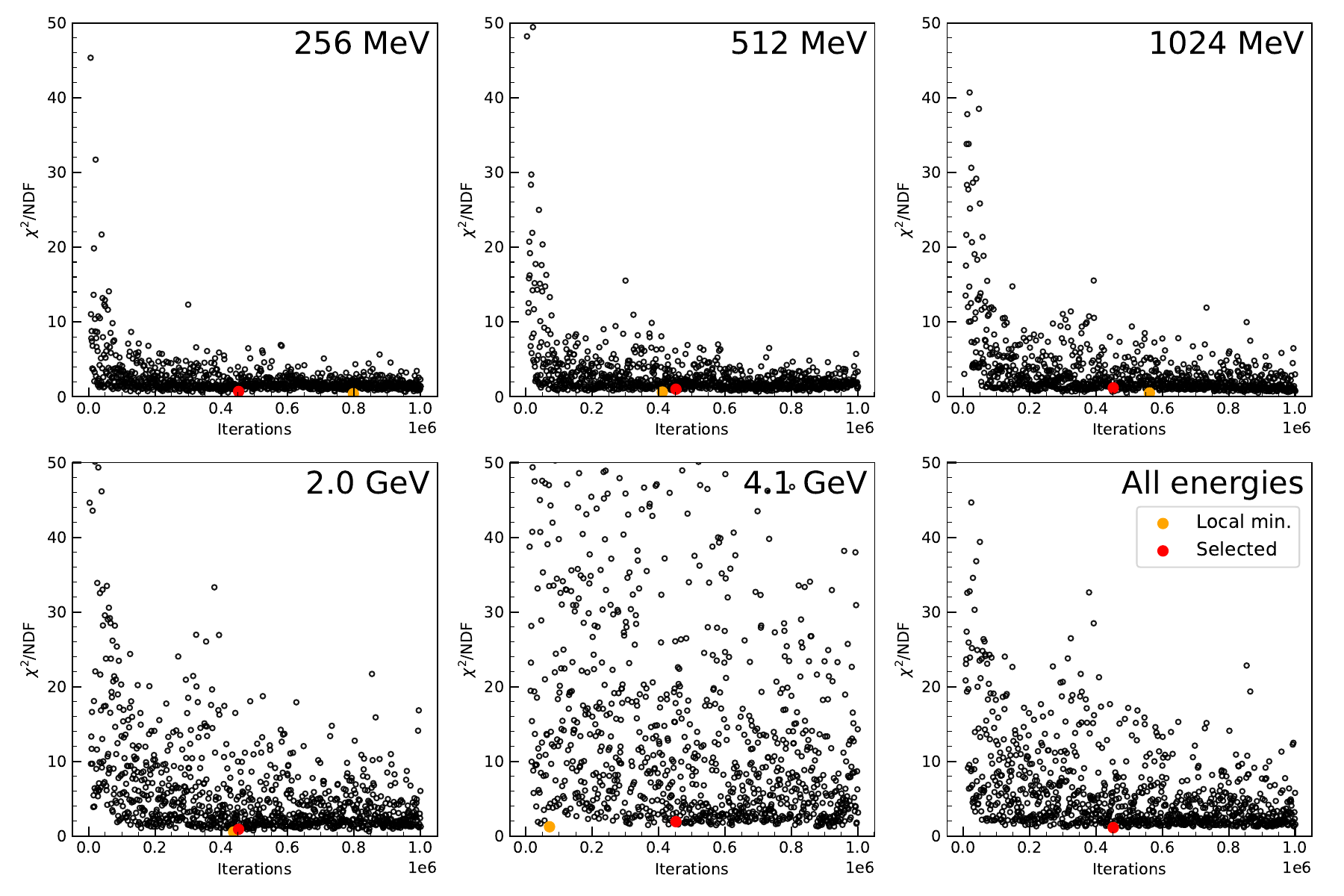}
        } \\
        \subfloat[high-momentum, 262~\GeV--4~\TeV]{
            \includegraphics[width=\textwidth]{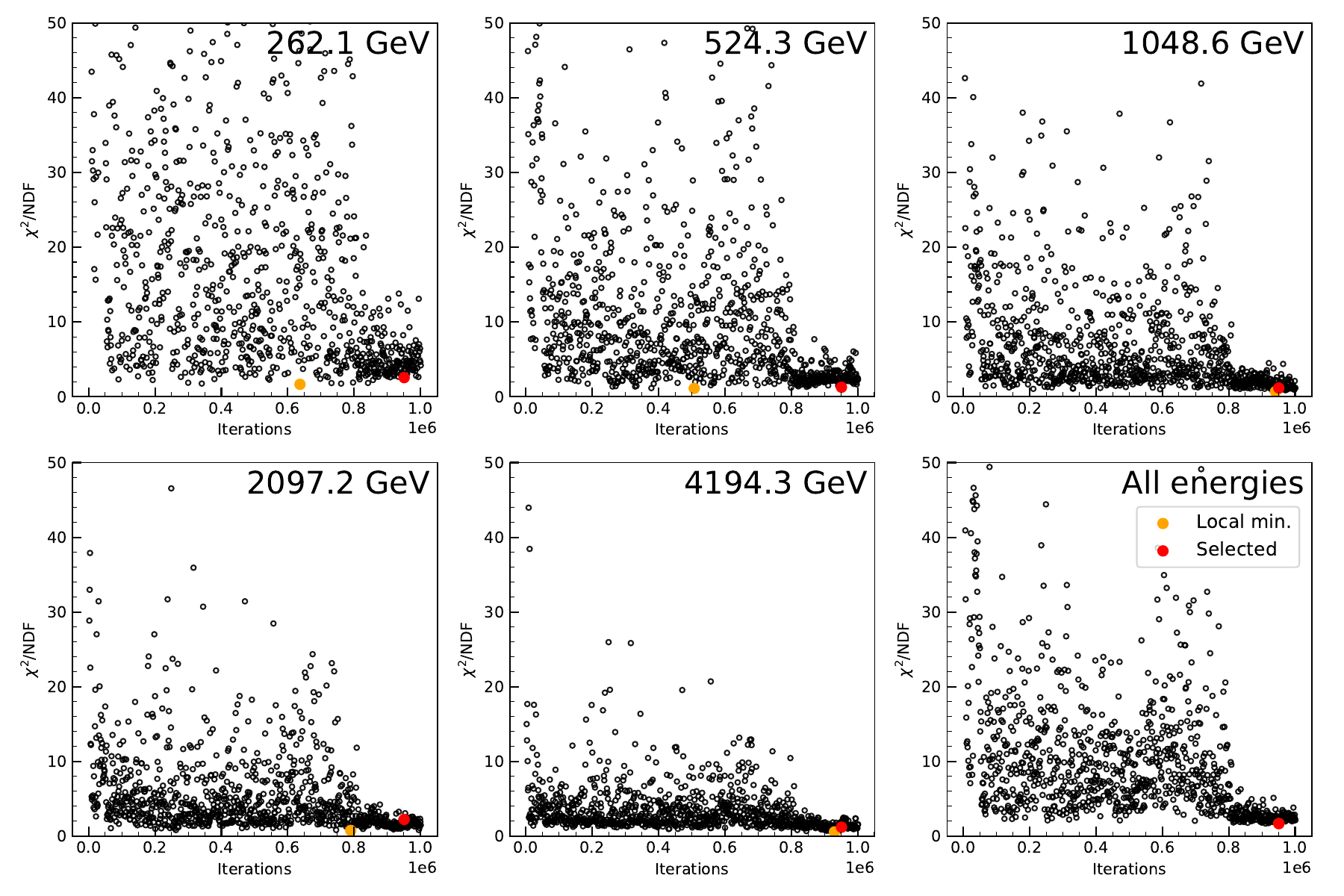}
        }
    \end{minipage}
    \hfill
    \begin{minipage}[b]{.49\linewidth}
        \subfloat[Midium-energy, 4~\GeV--262~\GeV]{
            \includegraphics[width=\textwidth]{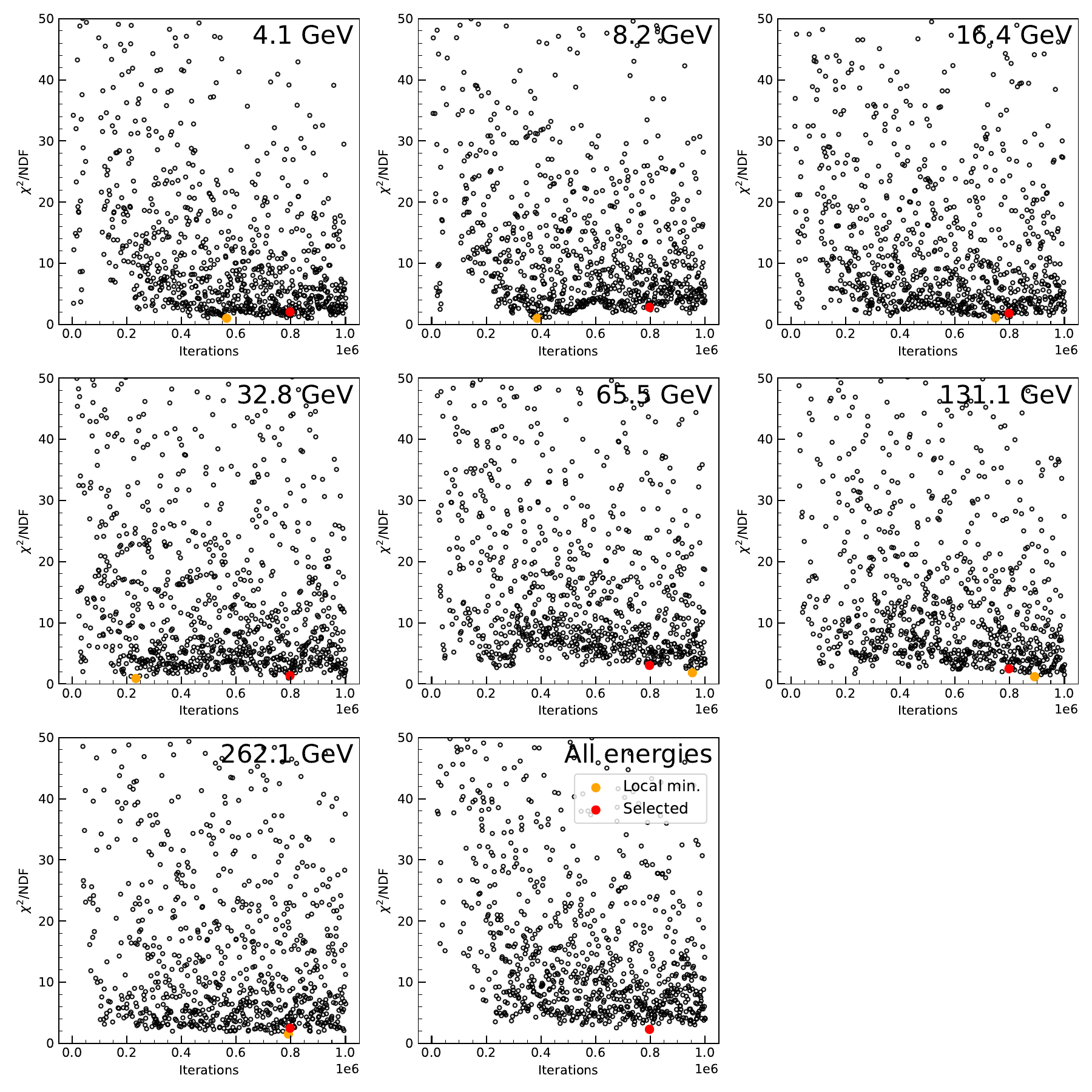}
        }
    \end{minipage}
    \caption{
        \textbf{Photon \CSG, with layer-energy normalisation.}
        Evolution of the individual and total $\chi^2$ as a function of iteration in photons.
        \captionchi
        }
    \label{fig:chi2photon}
\end{figure}

\begin{figure}[htb]
\centering
    \includegraphics[width=0.7\textwidth]{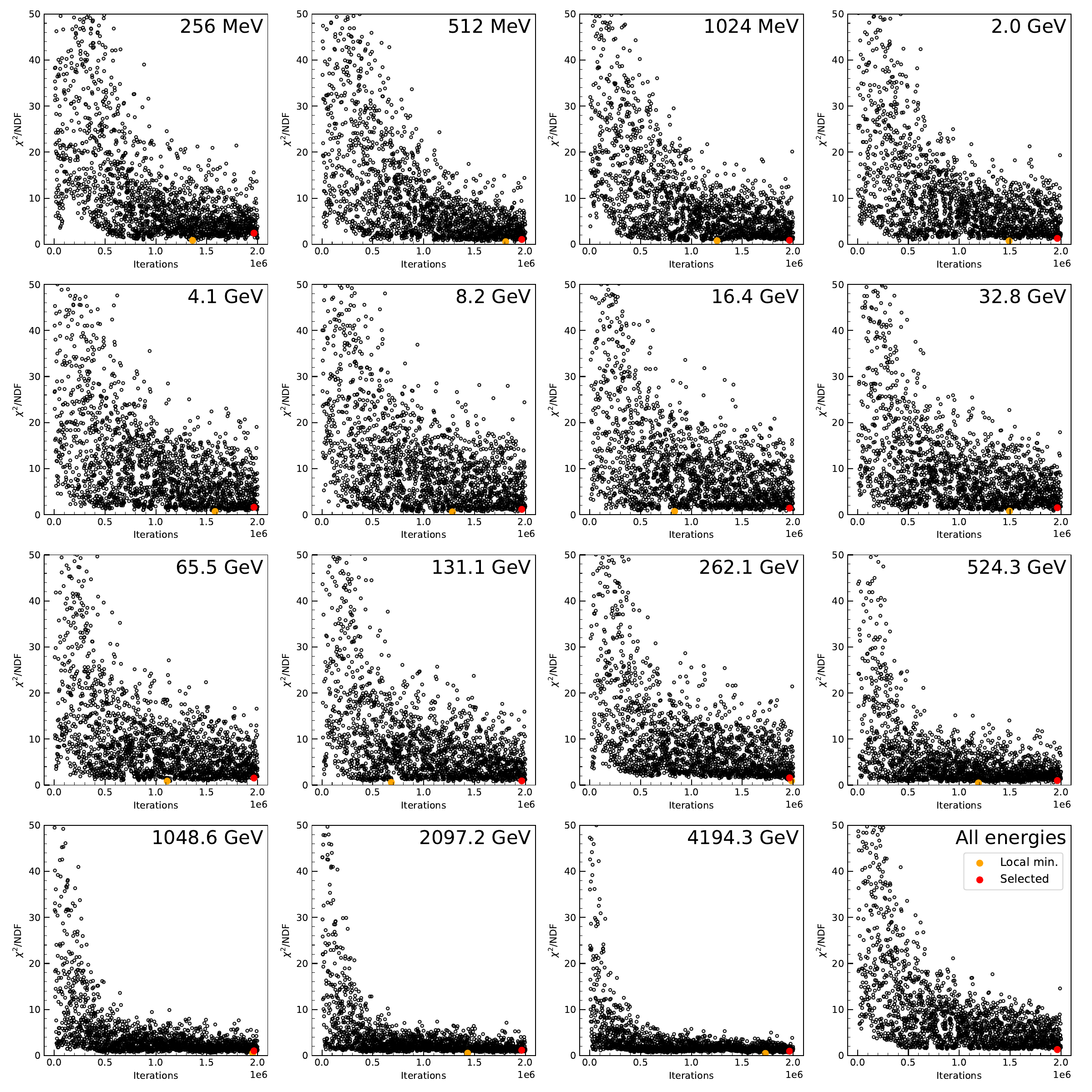}
    \caption{
        \textbf{Pion \CSG, with layer-energy normalisation.}
        Evolution of the individual and total $\chi^2$ as a function of iteration in pions.
        \captionchi
    }
    \label{fig:chi2pion}
\end{figure}

\FloatBarrier

\subsection{Energy per layer}

\begin{figure}[htp]
\centering
    \includegraphics[width=0.3\textwidth]{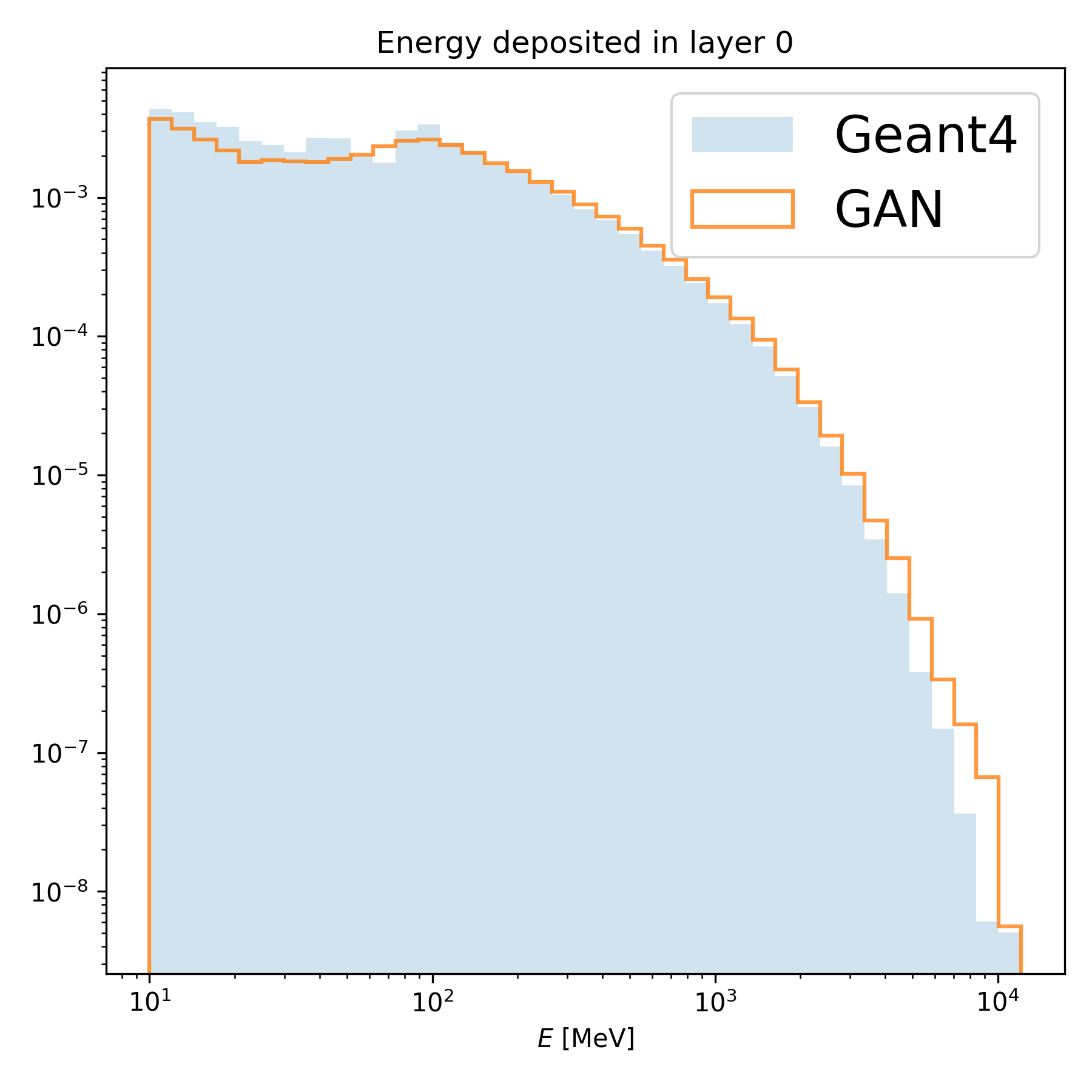}
    \includegraphics[width=0.3\textwidth]{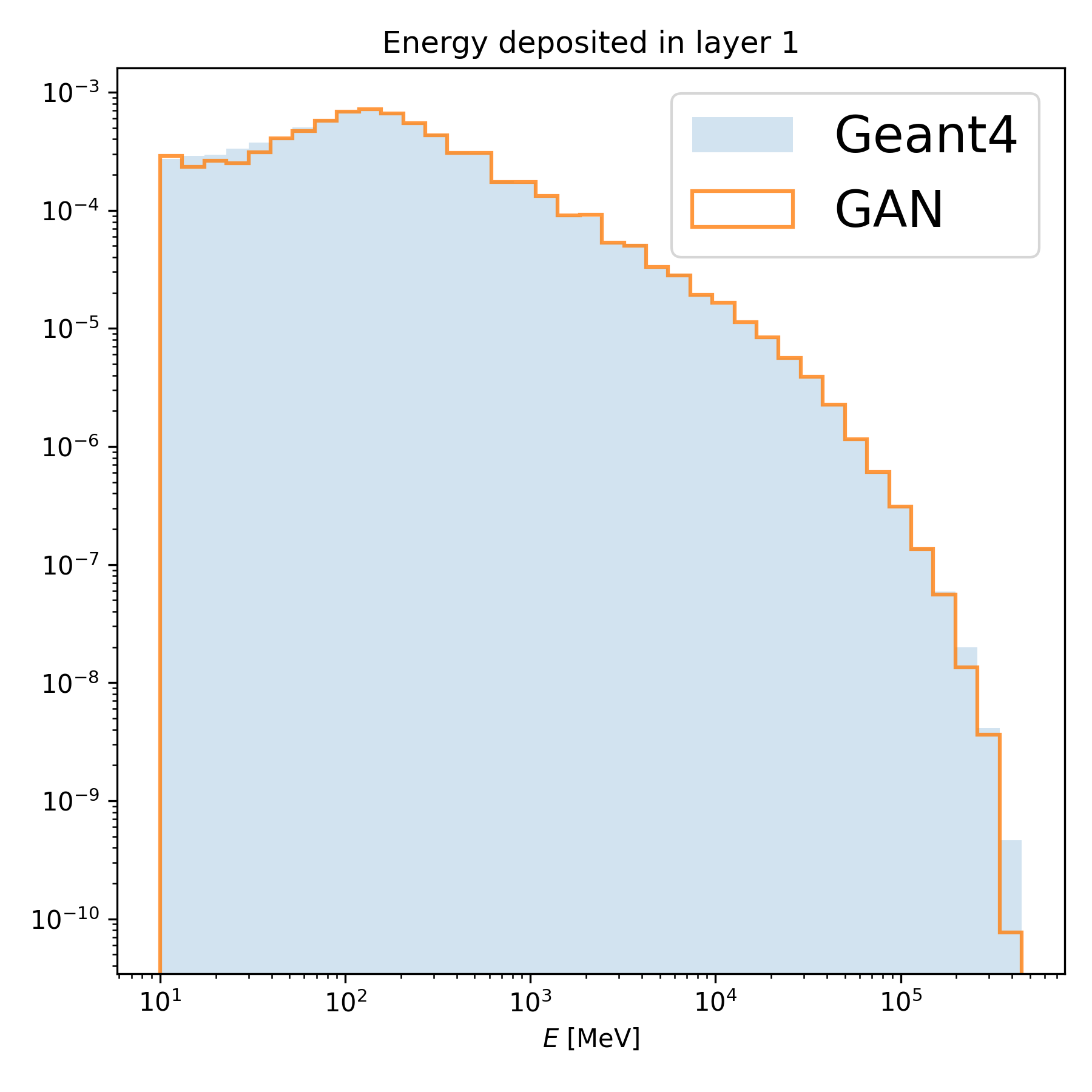}
    \includegraphics[width=0.3\textwidth]{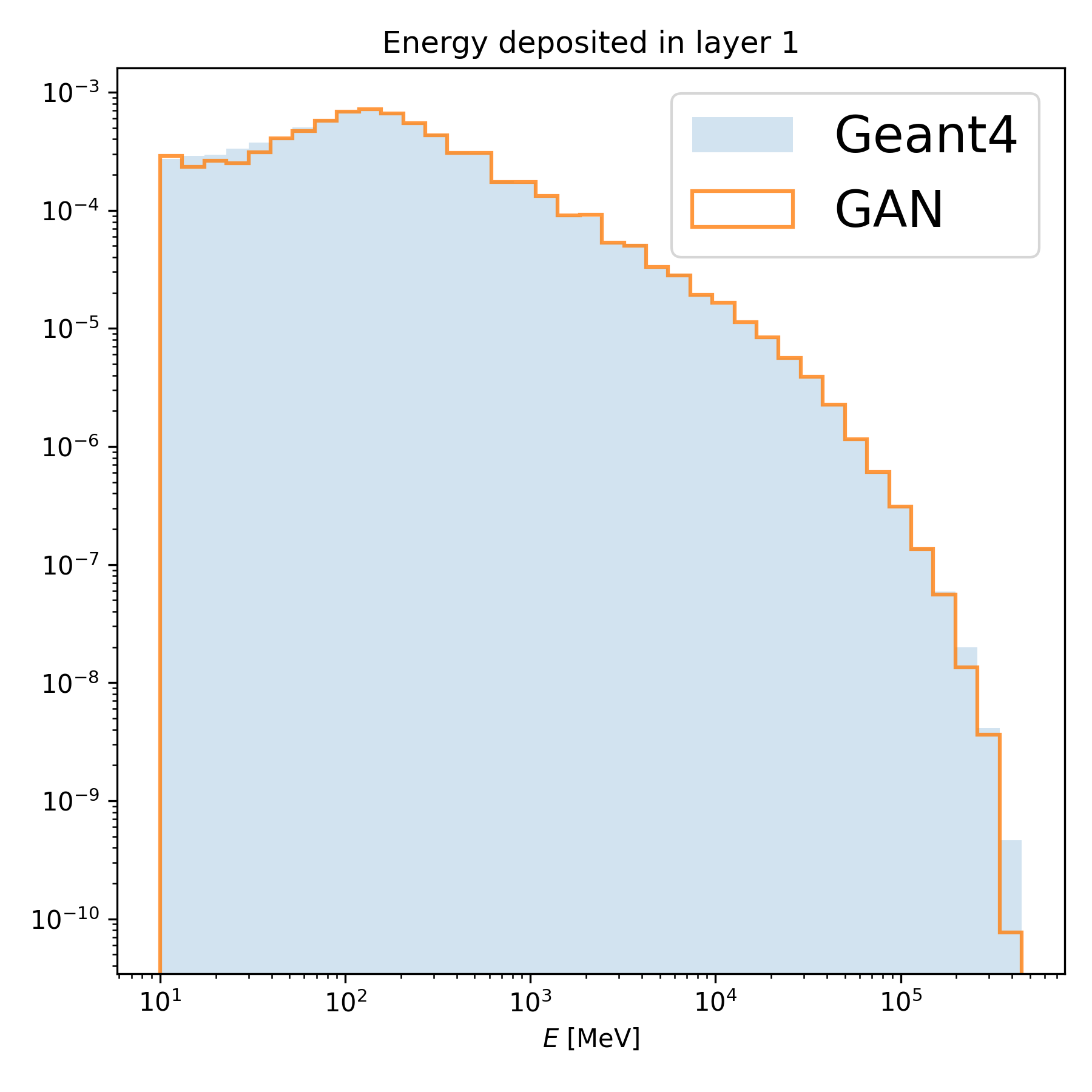} \\
    \includegraphics[width=0.3\textwidth]{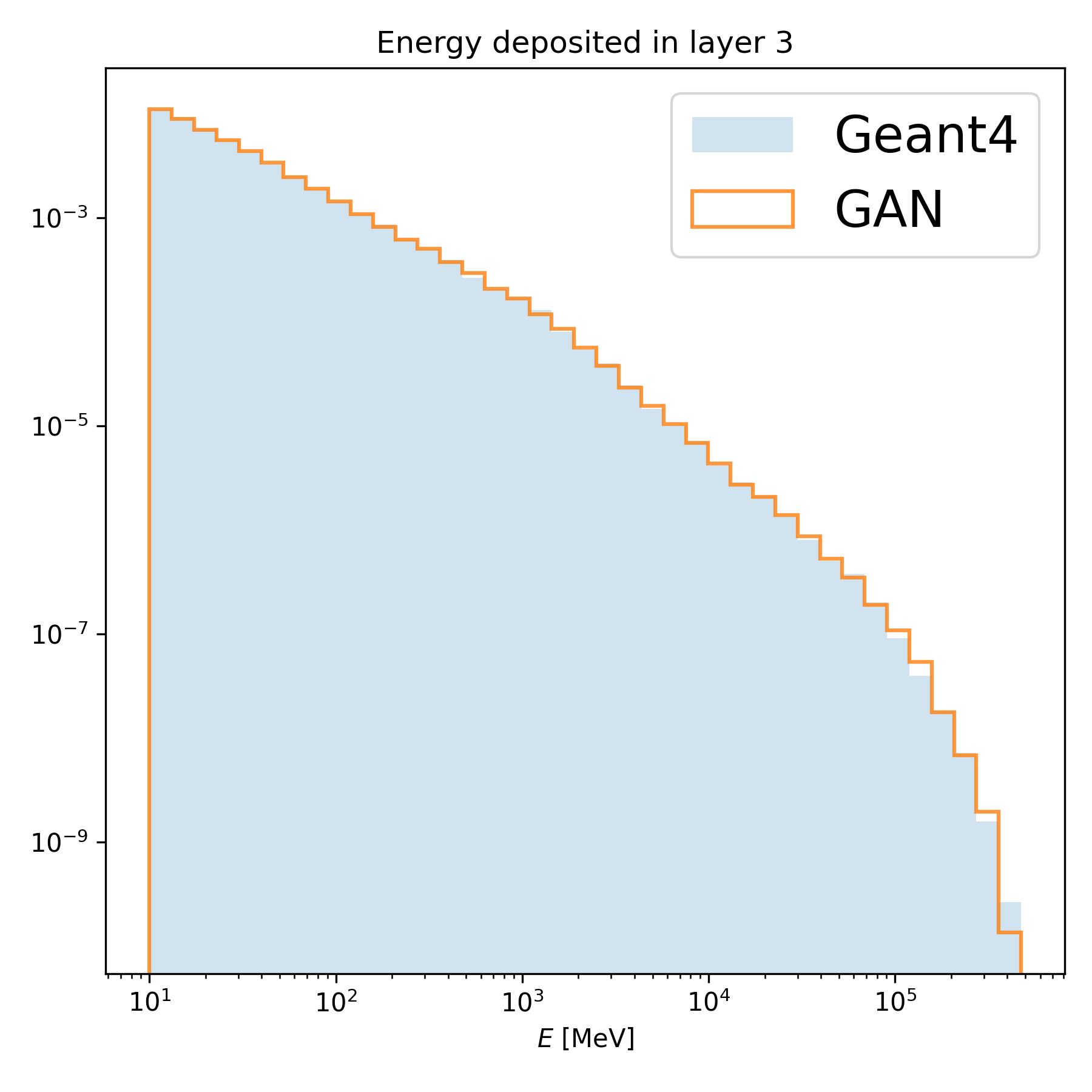}
    \includegraphics[width=0.3\textwidth]{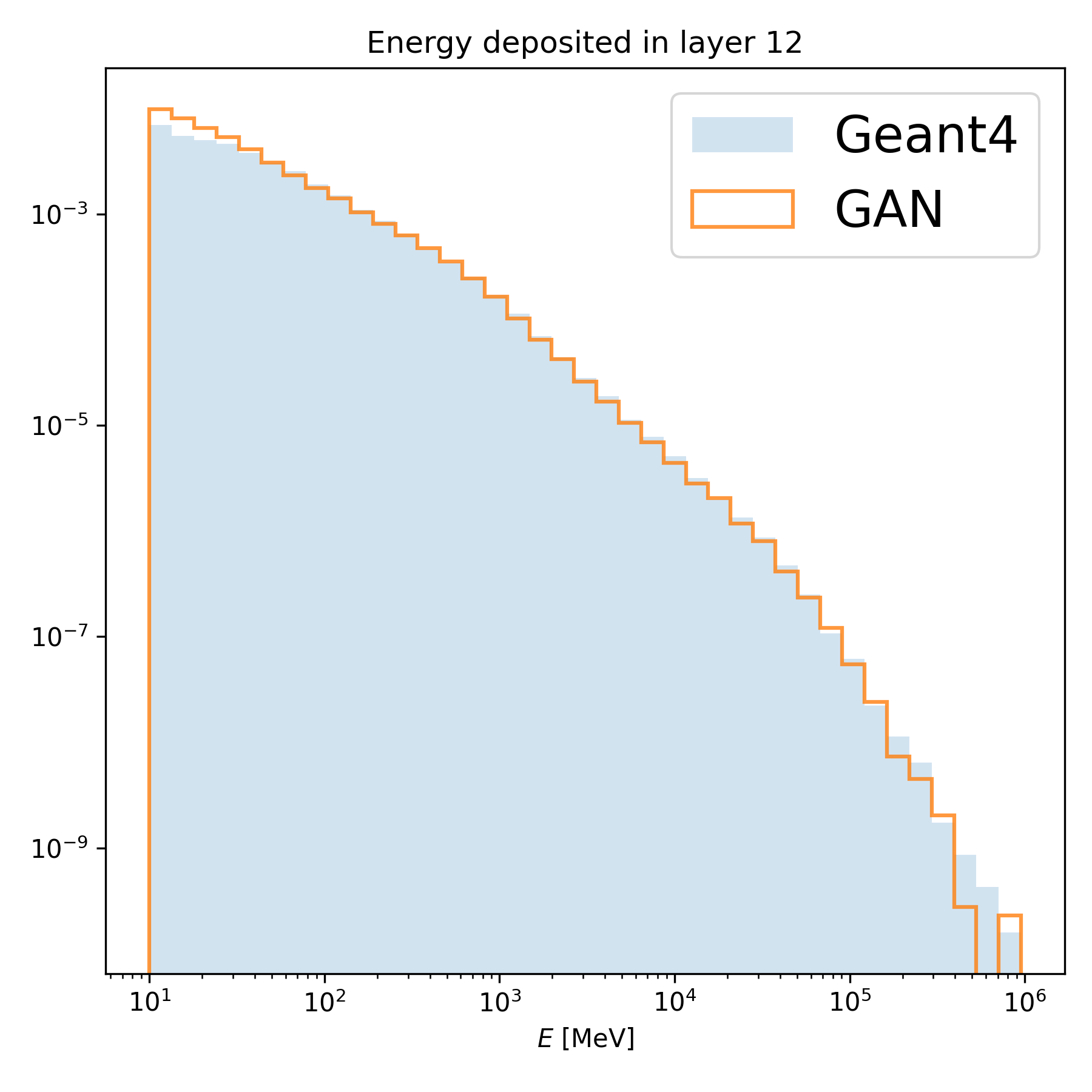}
    \caption{
        Energy deposit in each calorimeter layer for photons, summed over all incident energies.
    }
    \label{fig:layerphoton}
\end{figure}

\begin{figure}[htp]
\centering
    \includegraphics[width=0.3\textwidth]{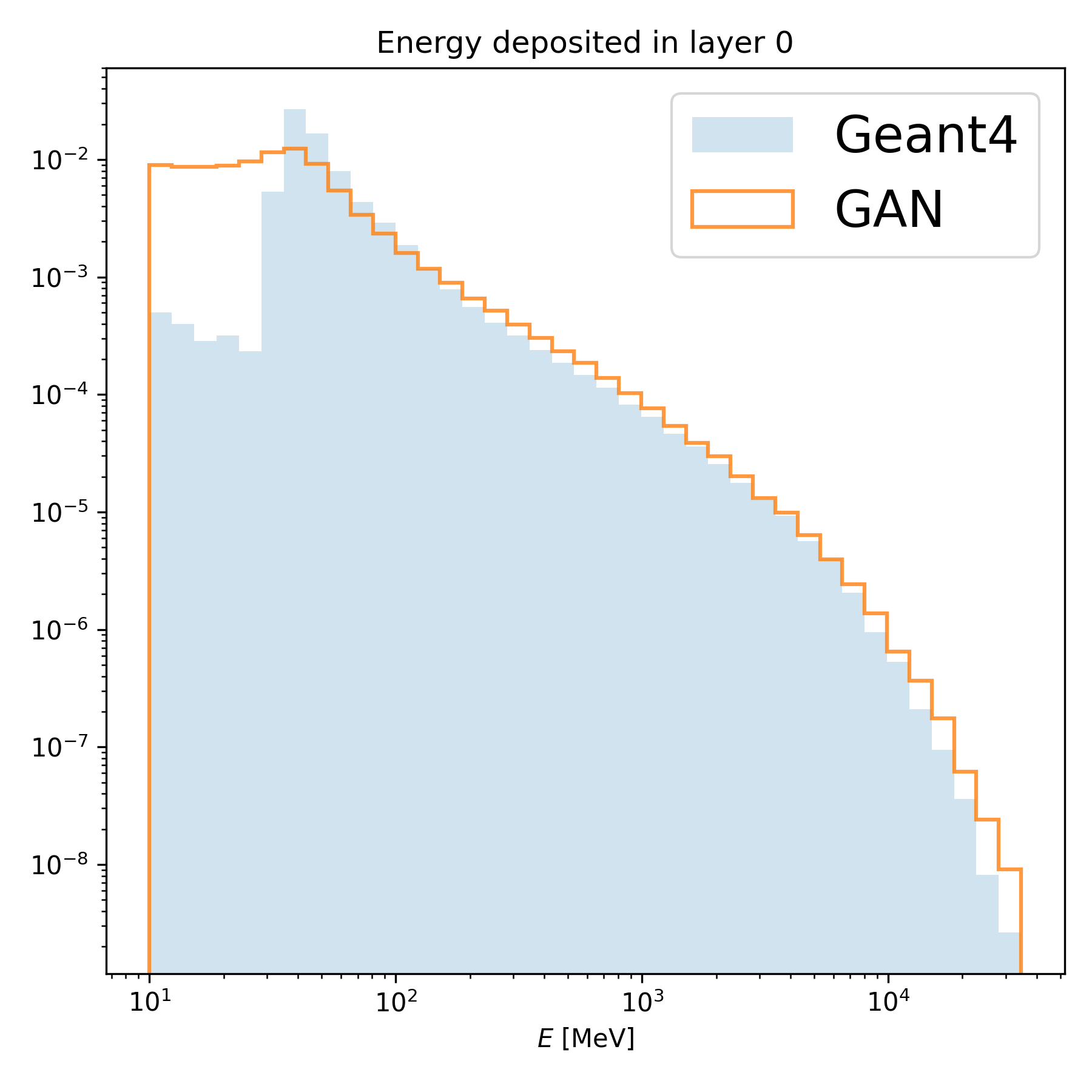}
    \includegraphics[width=0.3\textwidth]{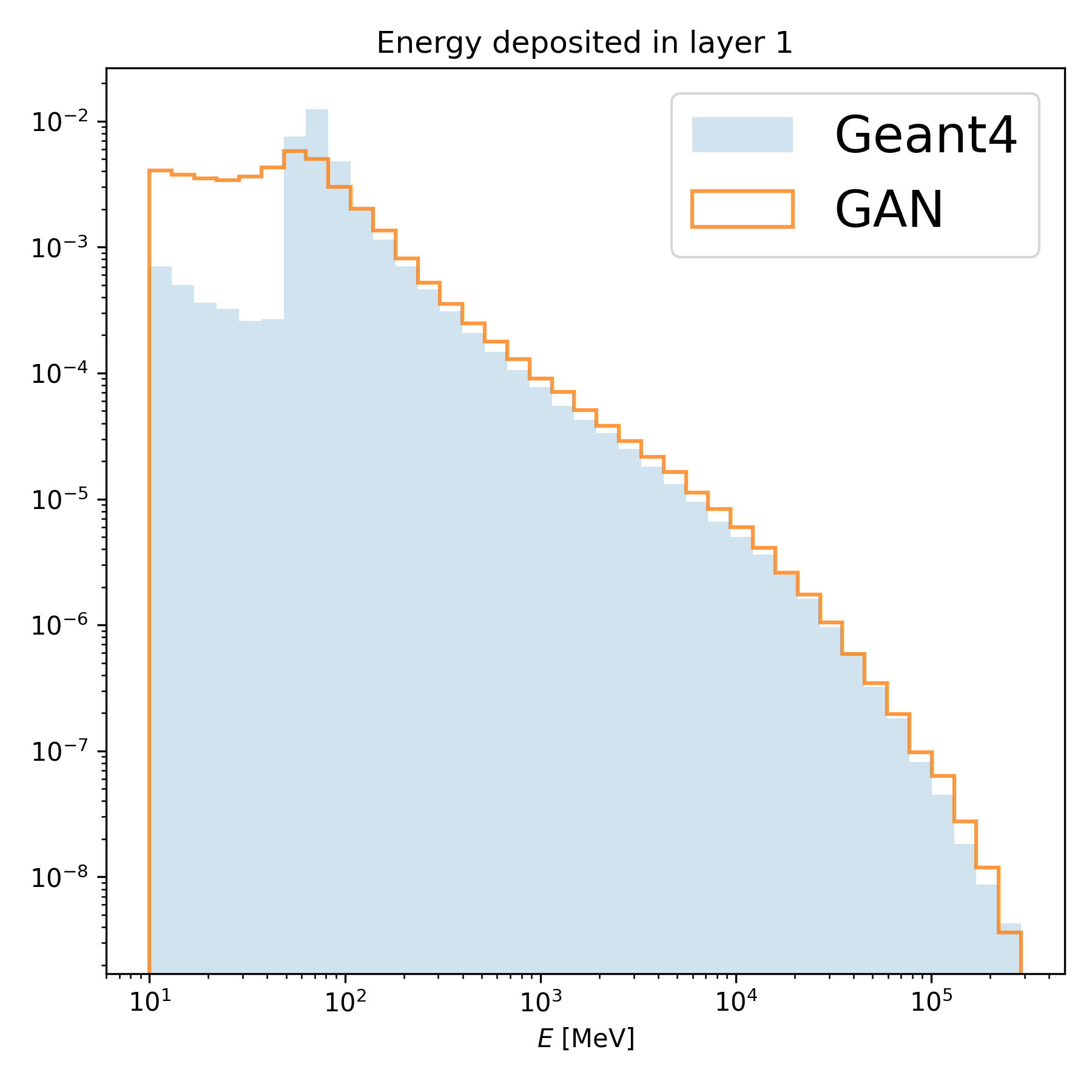}
    \includegraphics[width=0.3\textwidth]{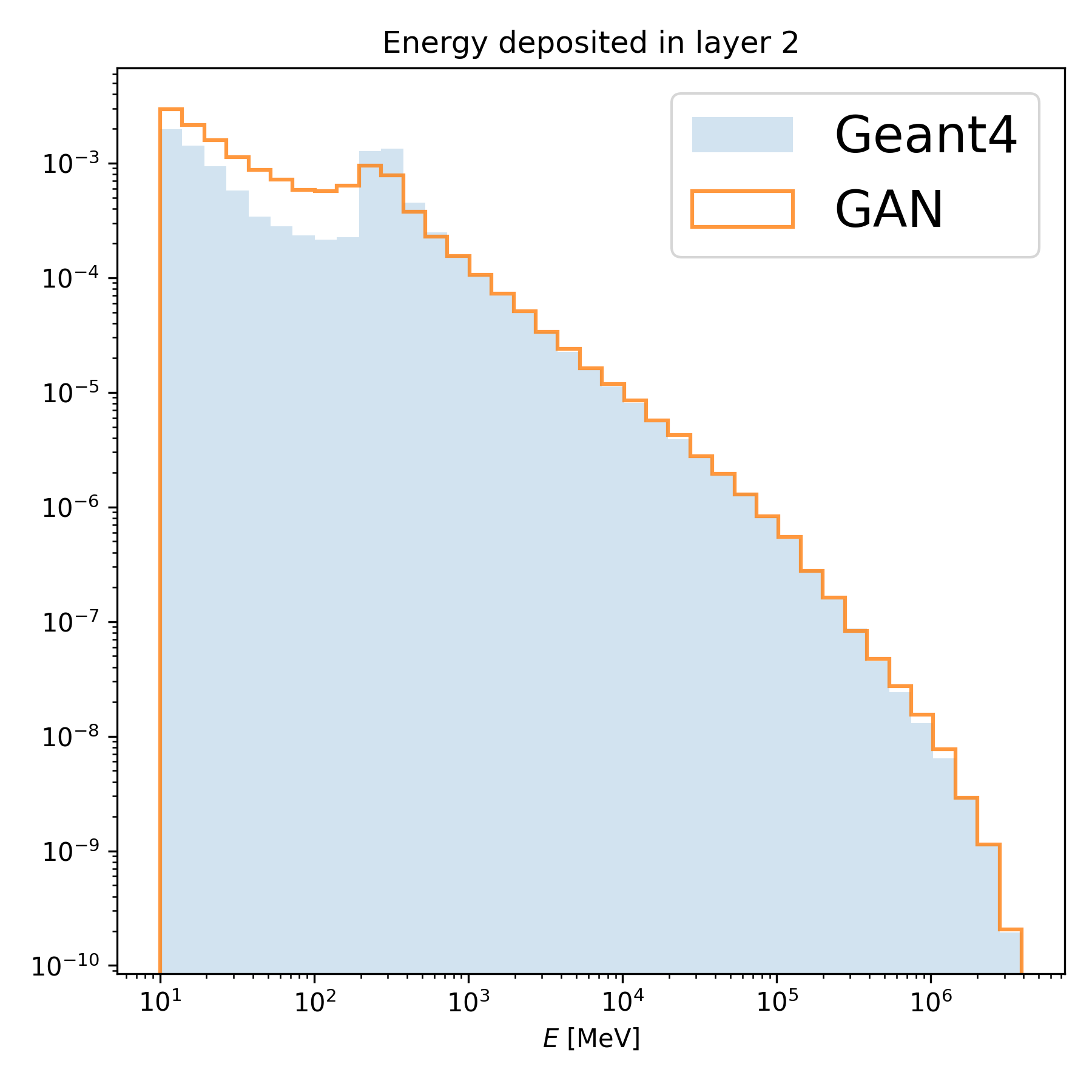} \\
    \includegraphics[width=0.3\textwidth]{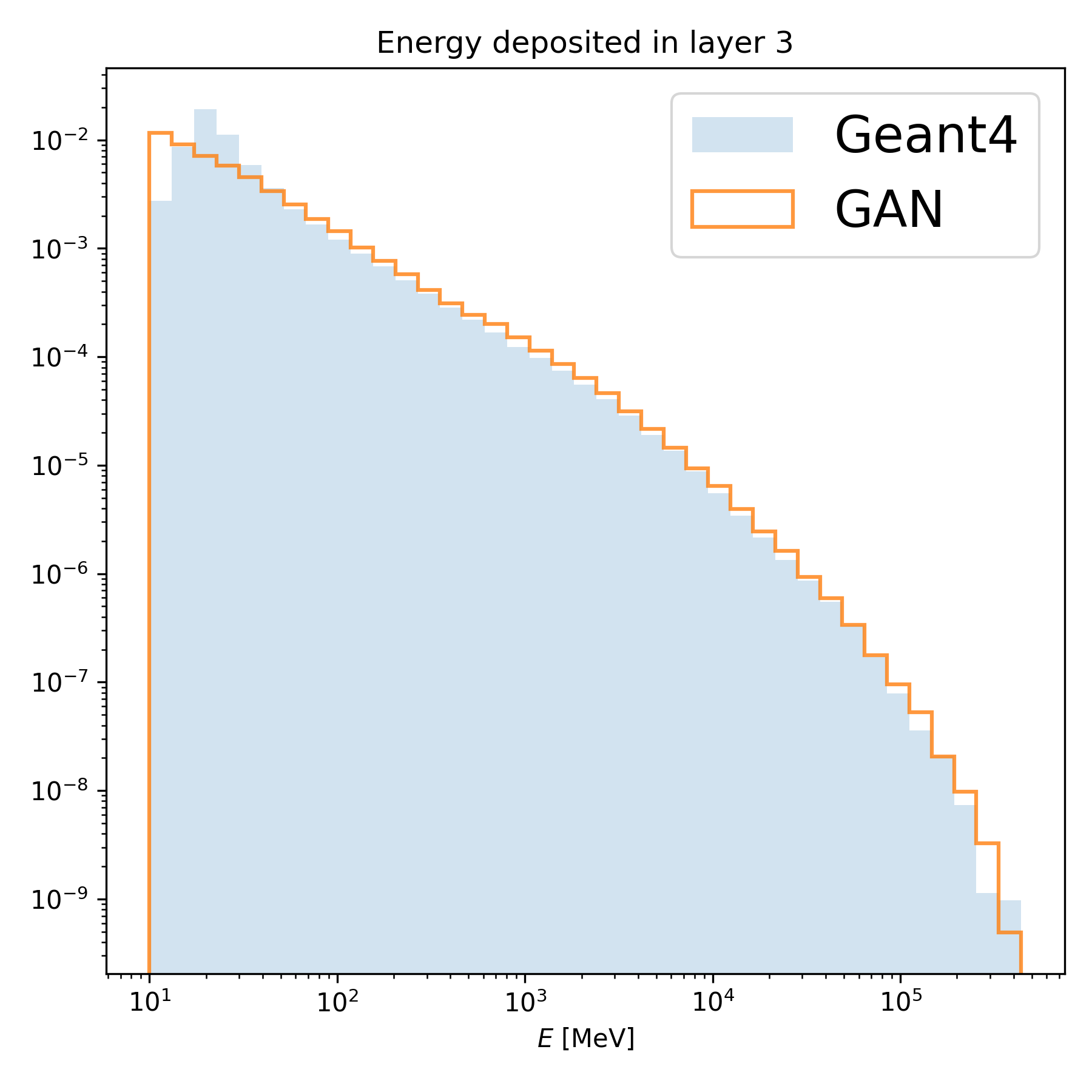}
    \includegraphics[width=0.3\textwidth]{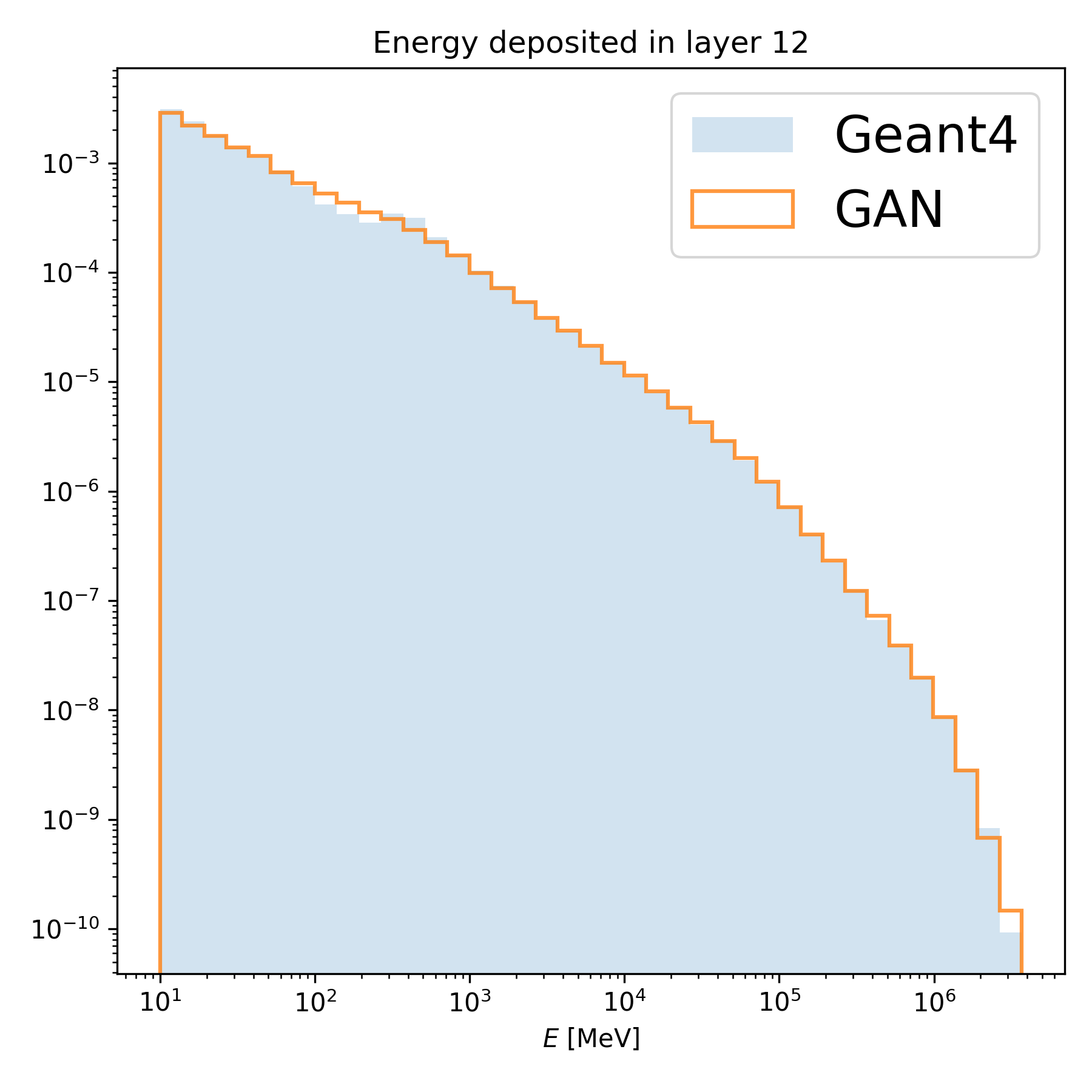}
    \includegraphics[width=0.3\textwidth]{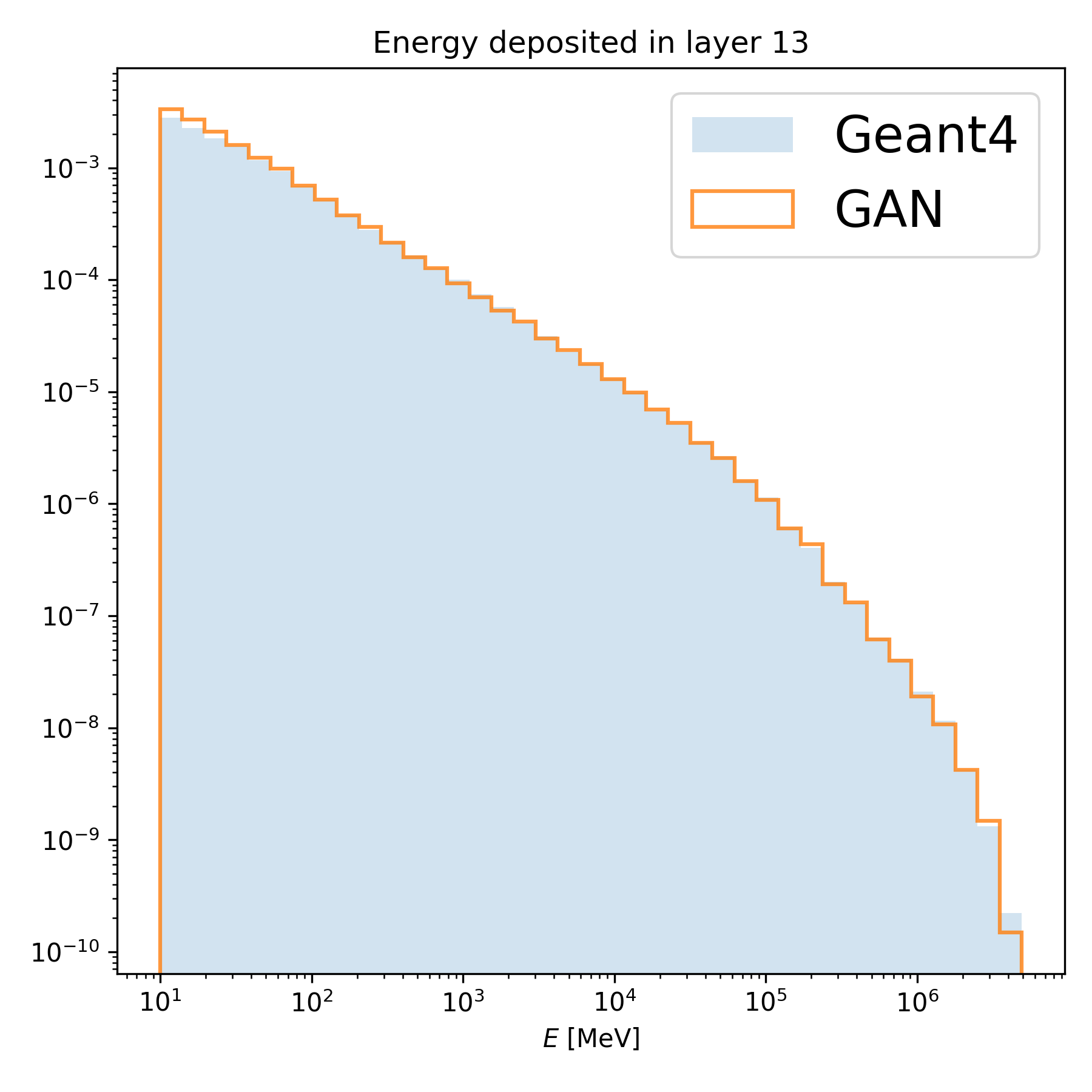} \\
    \includegraphics[width=0.3\textwidth]{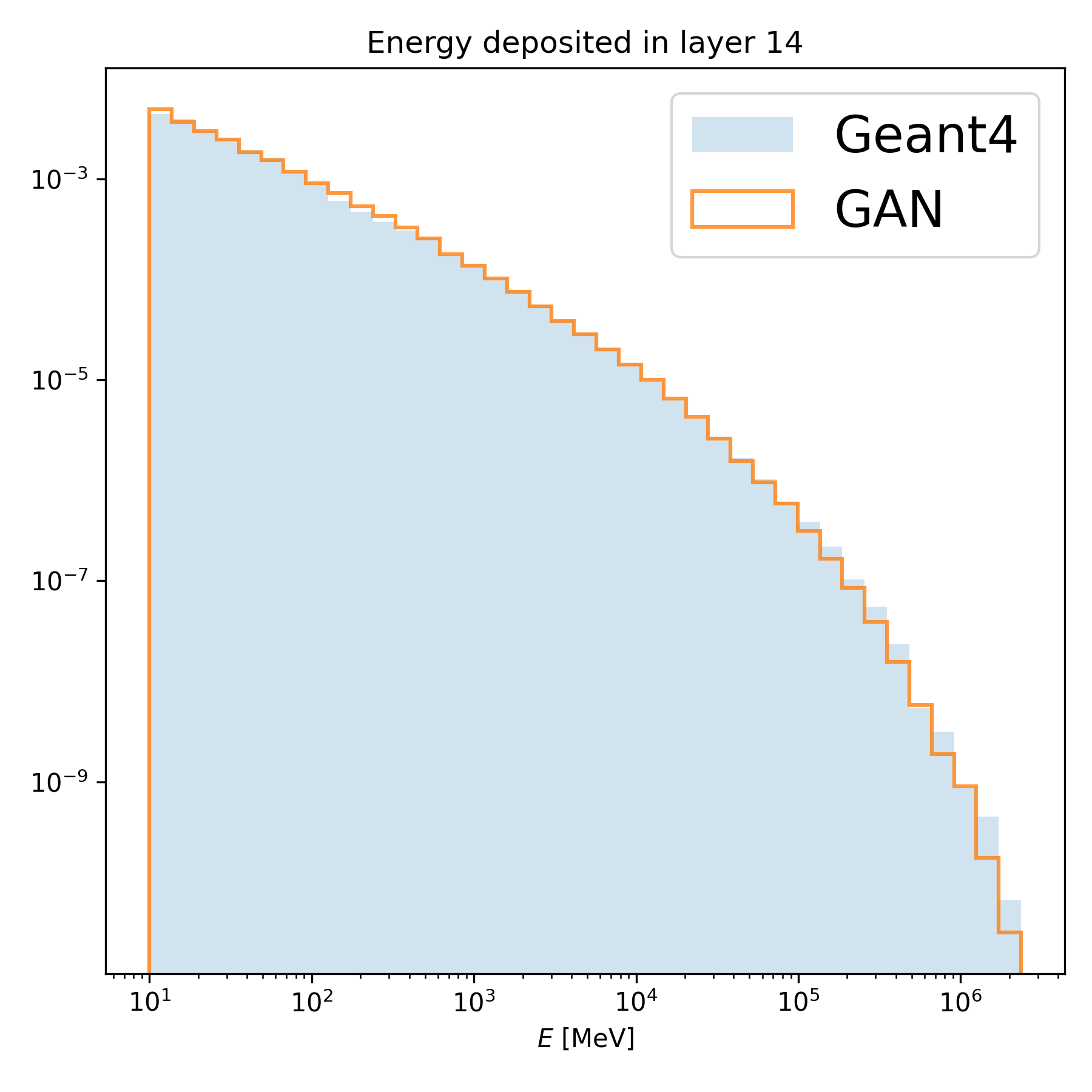}
    \caption{
        Energy deposit in each calorimeter layer for pions, summed over all incident energies.
    }
    \label{fig:layerpion}
\end{figure}

High fidelity of the model require the GAN to reproduce not only the total energy distribution, but also the energy distributions inside the showers
To validate this point, the energy distributions across each calorimeter layer are checked and shown in \Fig{\ref{fig:layerphoton}} and \Fig{\ref{fig:layerpion}}.
Good agreement with the \GEANT samples is observed when samples of all incident momentum points are merged together.
.

In the case of photons depicted in \Fig{\ref{fig:layerphoton}}, layer~0 poses a slight challenge for modelling, as the GANs slightly overshoot the intended energy deposits.
It is important to note that this effect is relatively minor, considering the logarithmic scale of the plots.
In fact, it affects less than 1\% of the events.
Moreover, the effect is limited for overall physics performance, as ATLAS electron and photon reconstructions primarily rely on layers~1 and 2.

For pions in \Fig{\ref{fig:layerpion}}, a bimodal structure emerges in several layers for momenta below \SI{1}{\GeV}, which is a feature that GANs encounter difficulty in reproducing.
This phenomenon arises from situations where pions initially deposit a small, consistent amount of energy in the layers before the hadronic shower starts.
This energy is compatible with deposits of a minimum ionising particle (MIP).
While the GANs struggle to accurately model this aspect, this discrepancy is unlikely to impact physics outcomes either.
The good agreement in the deeper layers and the high-momentum tails of the distribution serves as a strong indicator that \CSG is capable of delivering good performance andfavourable outcomes in physics applications.

\FloatBarrier

\subsection{Shower shapes}
An additional validation to assess the performance of \CSG involves examining the shape of the showers across various layers.
This requires evaluating the position of the shower centre and its width within each layer, in both the $\phi$ and $\eta$ directions.
The shower centre is defined as follows:

\begin{equation}
    \label{eqn:centroid}
    \langle\eta_{l}\rangle = \frac{\sum_{i}\left(E_i \odot H\right)}{E_l}\quad 
    \text{and} \quad 
    \langle\phi_{l}\rangle = \frac{\sum_{i}\left(E_i \odot F\right)}{E_l},
\end{equation}
where $E_i$ represents the energy in the $i$-th voxel in layer $l$, and $H$, $F$ are the position along $\eta$ and $\phi$ direction, respectively.
Additionally, $E_l = \sum_{i} E_i$ represents the energy in layer $l$.
The symbol $\odot$ corresponds to the Hadamard product, while $\sum$ denotes a summation across all elements in layer $l$.
The widths are defined as:
\begin{equation}
    \label{eqn:centroid_widths}
    \sigma^\eta_{l} = \sqrt{\frac{\sum_{i}\left(E_i \odot H^2\right)}{E_l}-\left(\langle\eta_{l}\rangle\right)^2}\quad
    \text{and}\quad 
    \sigma^\phi_{l} = \sqrt{\frac{\sum_{i}\left(E_i \odot F^2\right)}{E_l}-\left(\langle\phi_{l}\rangle\right)^2}
\end{equation}
Note that these shapes are meaningful to compute in layers featuring multiple bins in the angular ($\alpha$) direction and become not-well-defined in layers that possess only one bin along the angular direction.

\begin{figure}[htp]
\centering
    \includegraphics[width=0.3\textwidth]{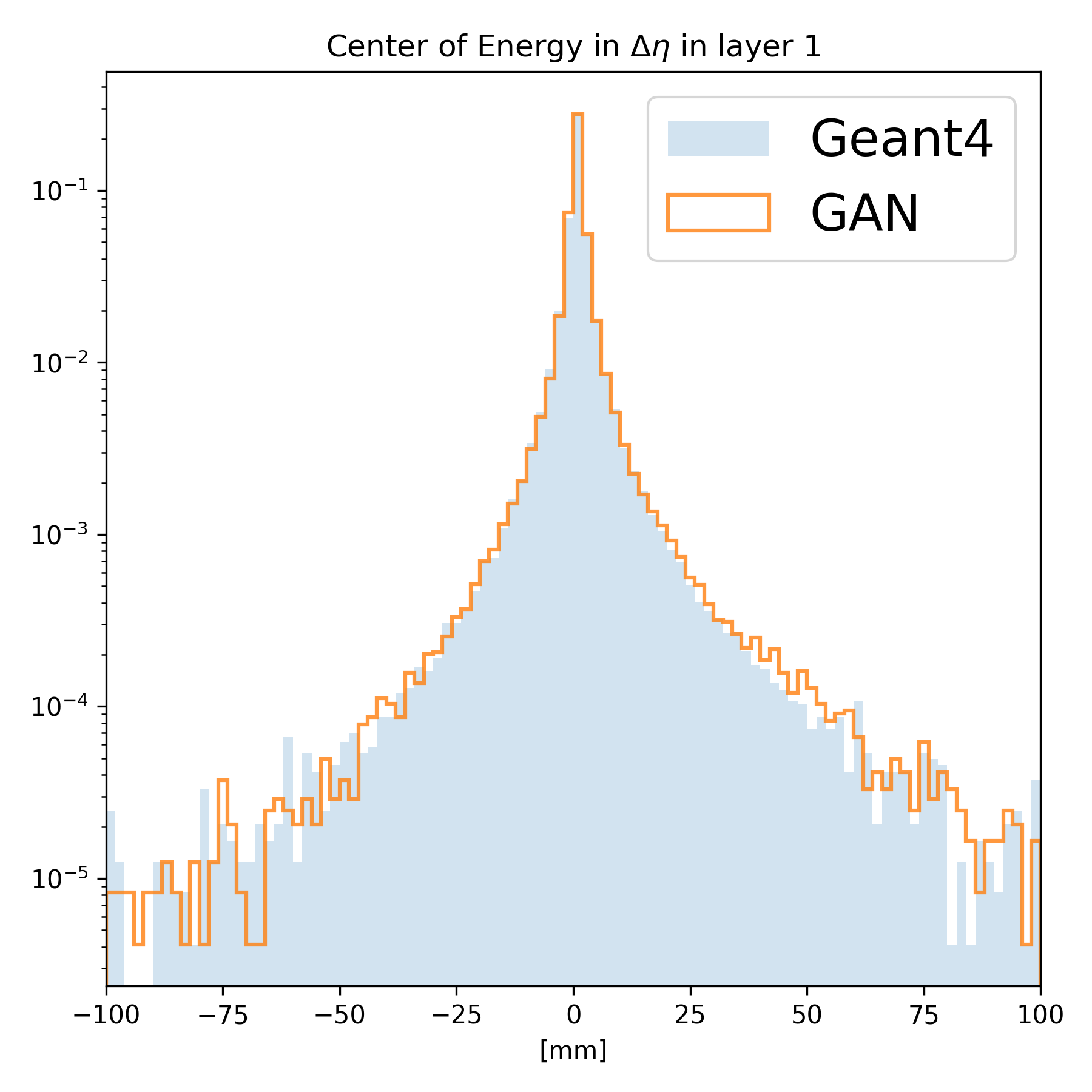}
    \includegraphics[width=0.3\textwidth]{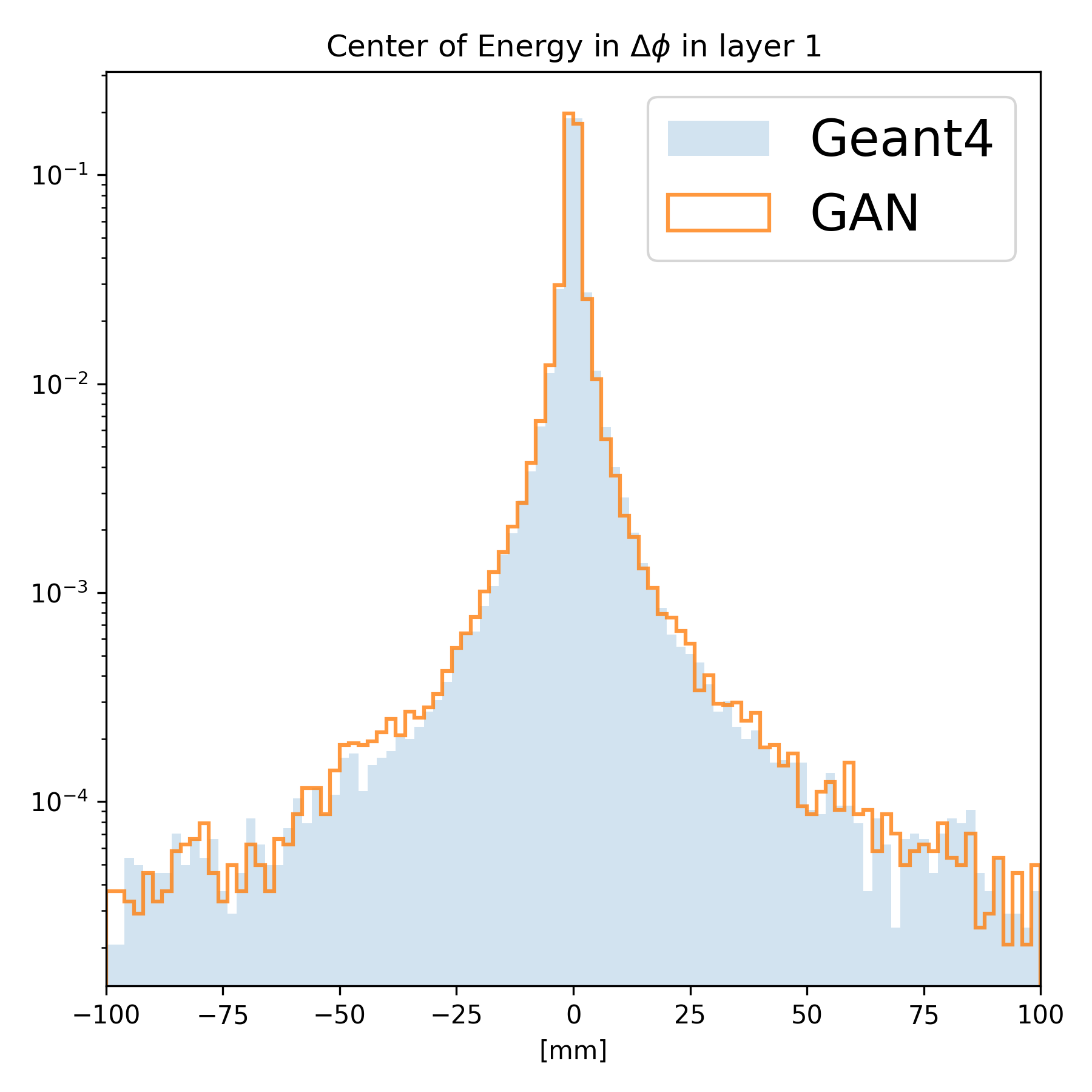} \\
    \includegraphics[width=0.3\textwidth]{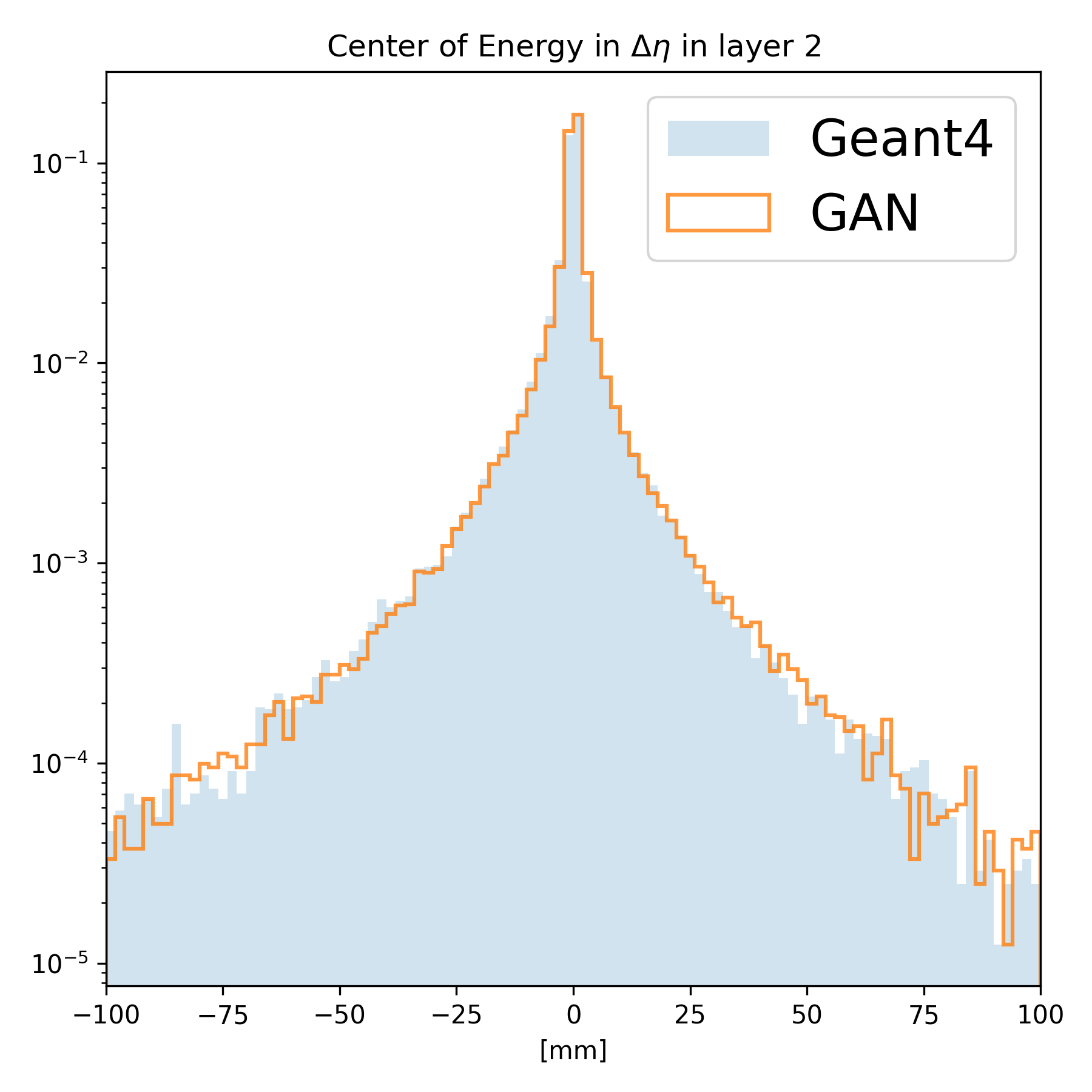} 
    \includegraphics[width=0.3\textwidth]{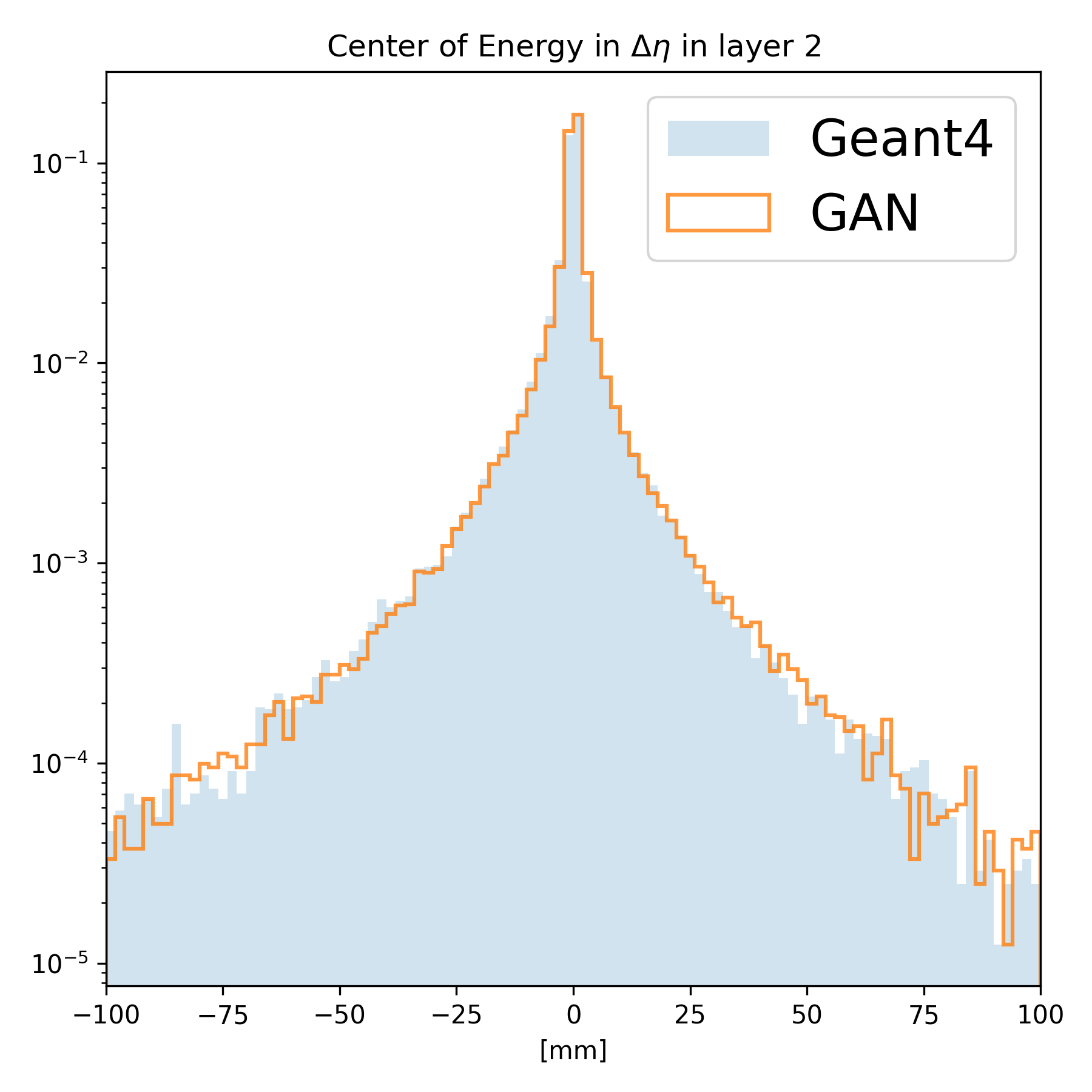} \\
    \caption{
        \textbf{Photon \CSG with layer-energy normalisation}: shower centre position compared to \GEANT simulation (solid area).
        All energies accumulated in layers 1 (top) and 2 (bottom) along $\eta$ (left) and $\phi$ (right).
    }
    \label{fig:centrephoton}
\end{figure}

\begin{figure}[htp]
\centering
    \includegraphics[width=0.3\textwidth]{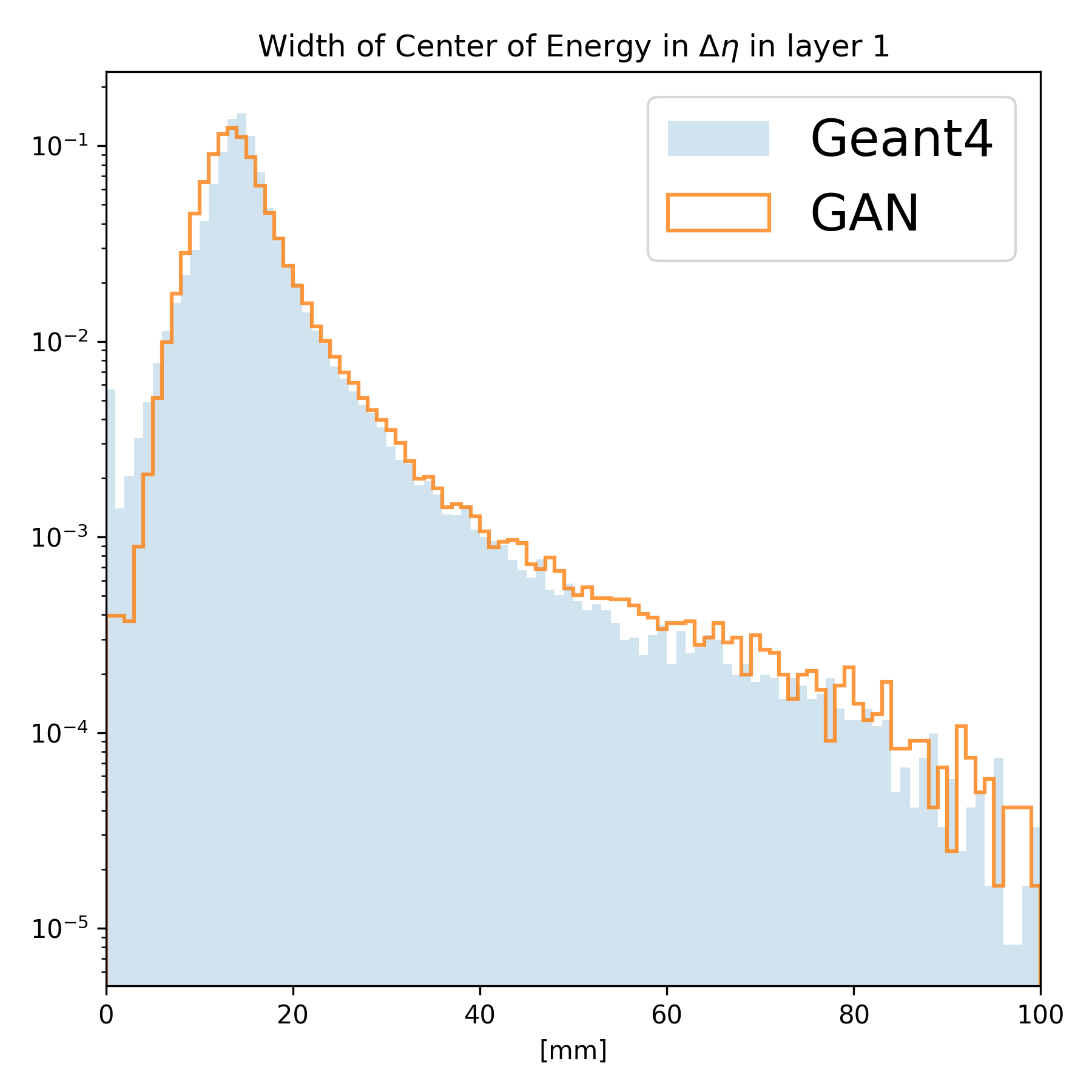}
    \includegraphics[width=0.3\textwidth]{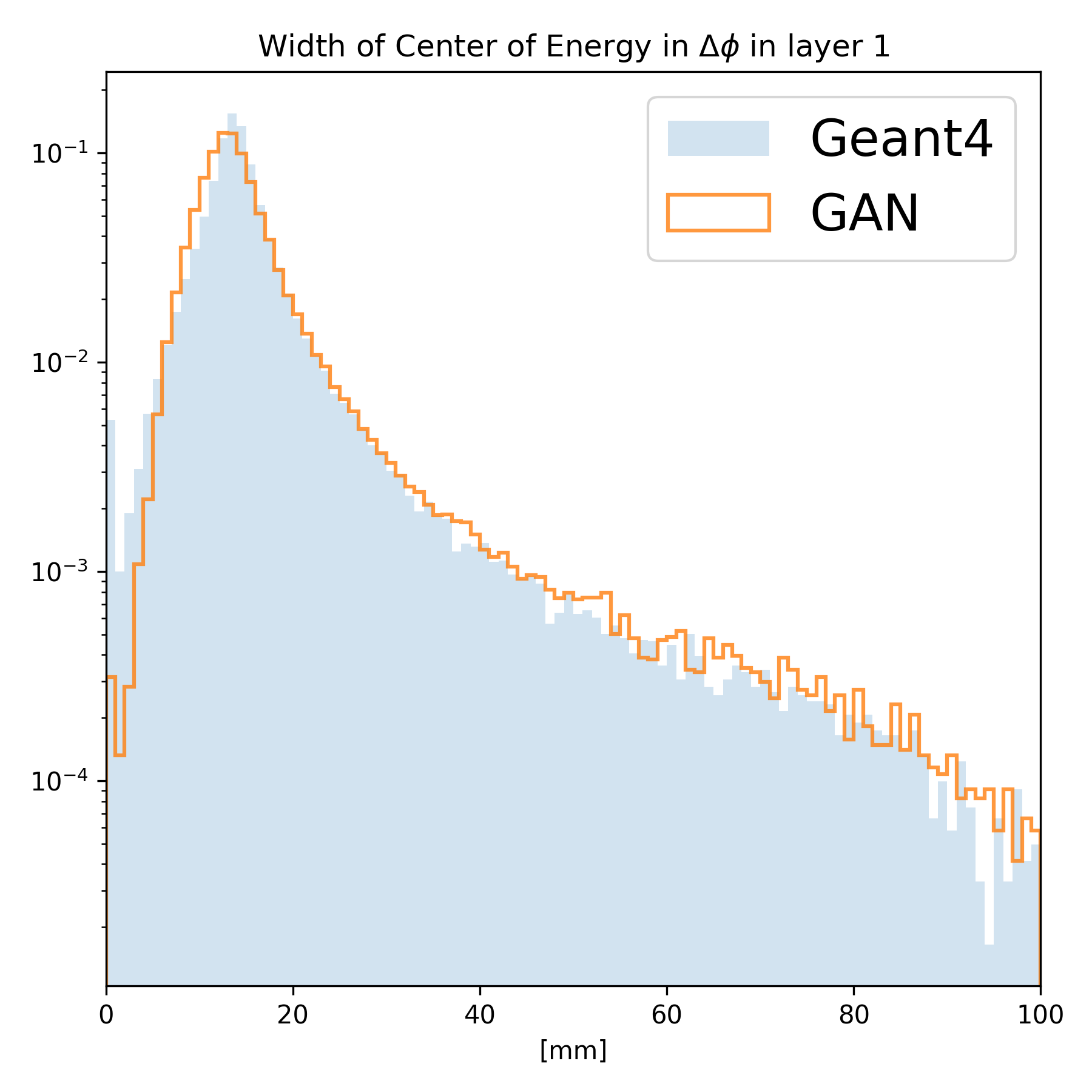} \\
    \includegraphics[width=0.3\textwidth]{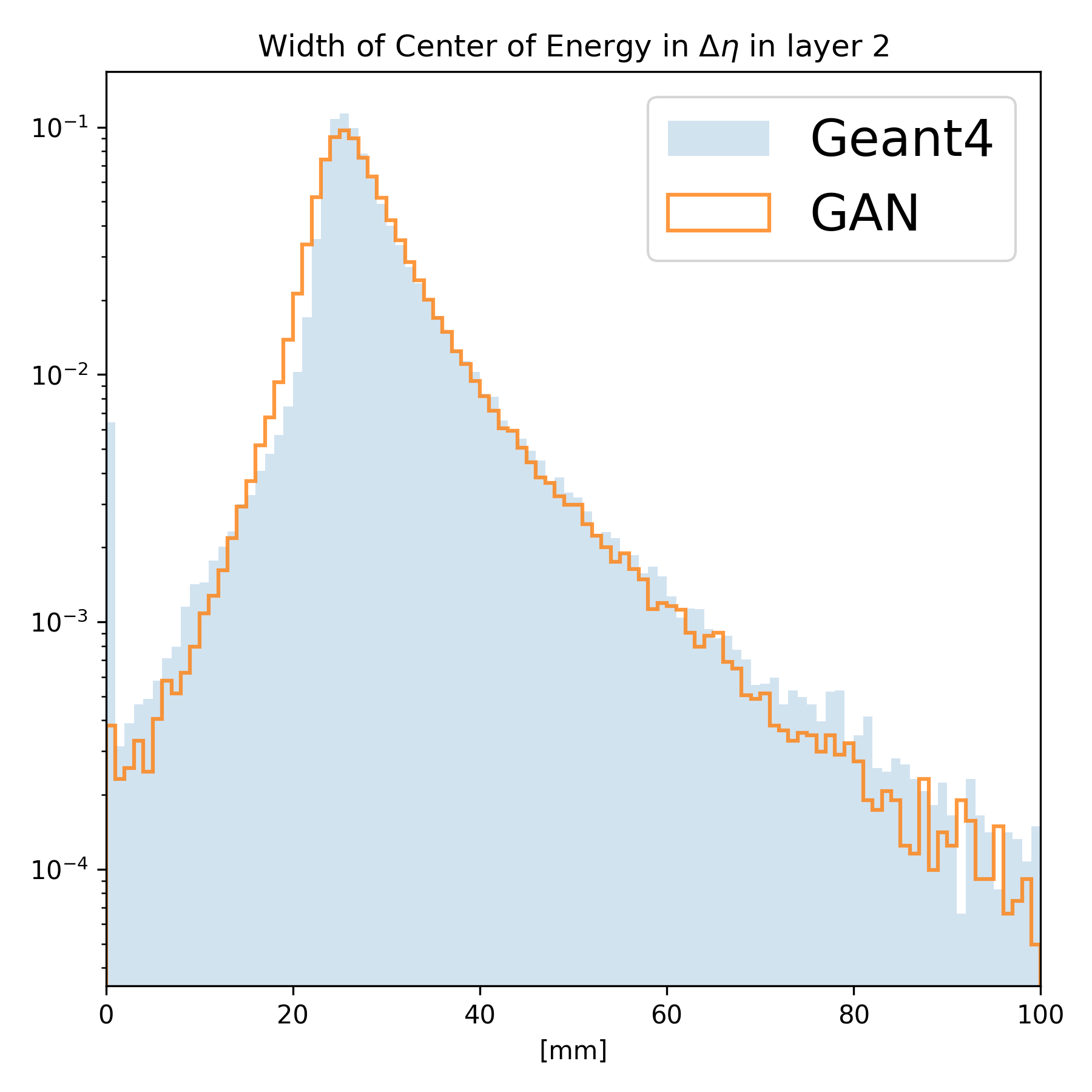} 
    \includegraphics[width=0.3\textwidth]{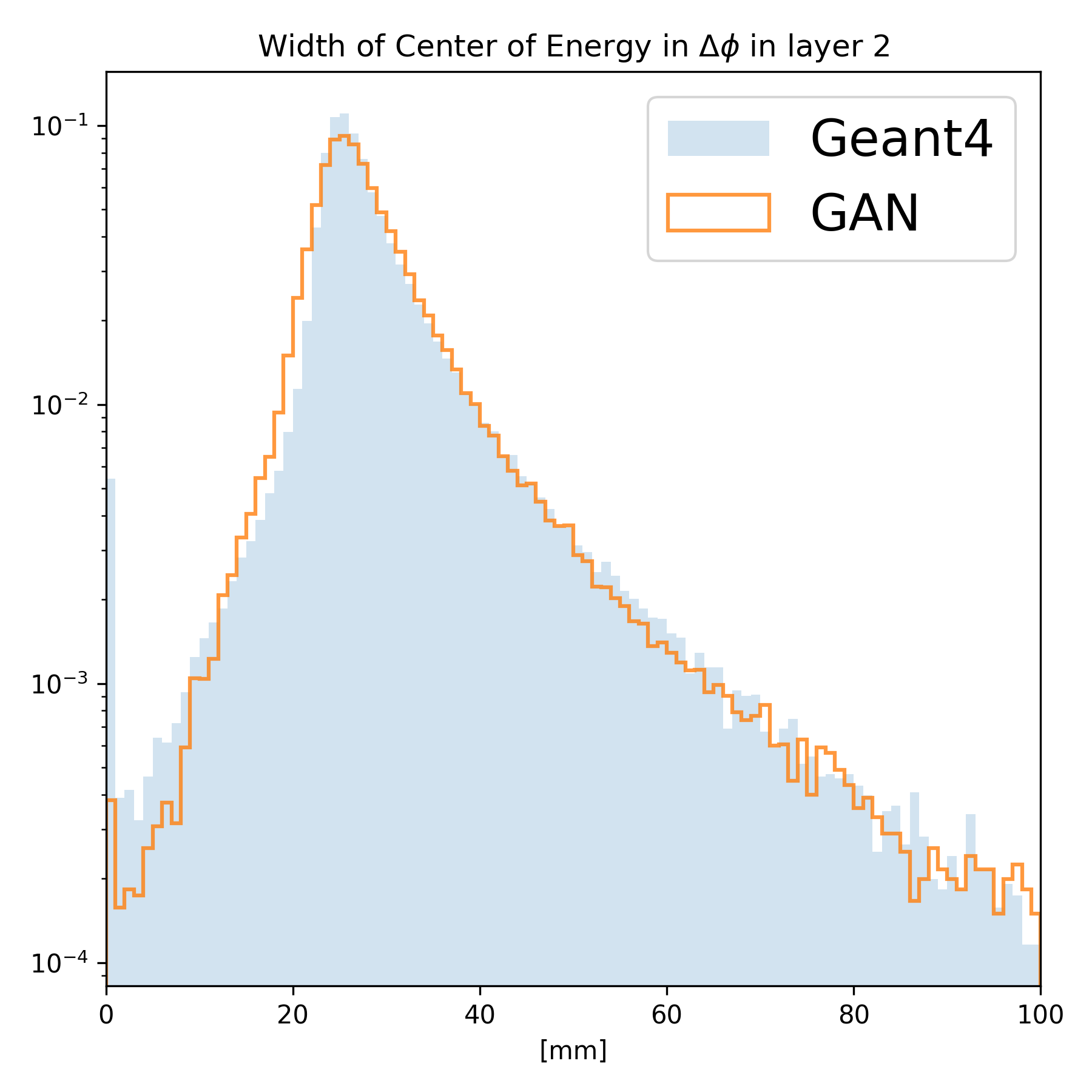} \\
    \caption{
        \textbf{Photon \CSG with layer-energy normalisation}: shower width compared to \GEANT simulation (solid area).
        All energies accumulated in layers 1 (top) and 2 (bottom) along $\eta$ (left) and $\phi$ (right).
    }
    \label{fig:widthphoton}
\end{figure}

\begin{figure}[htp]
\centering
    \includegraphics[width=0.3\textwidth]{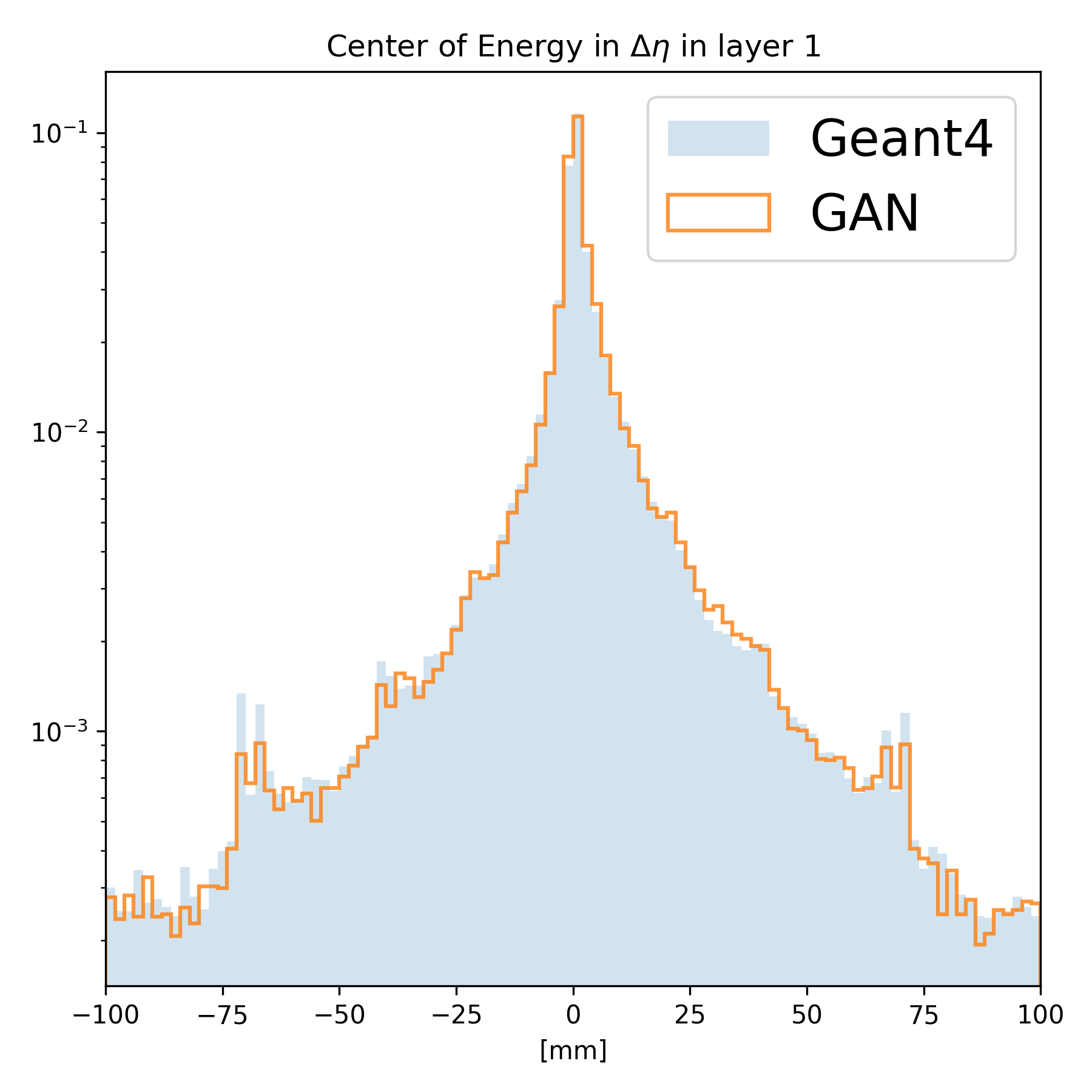}
    \includegraphics[width=0.3\textwidth]{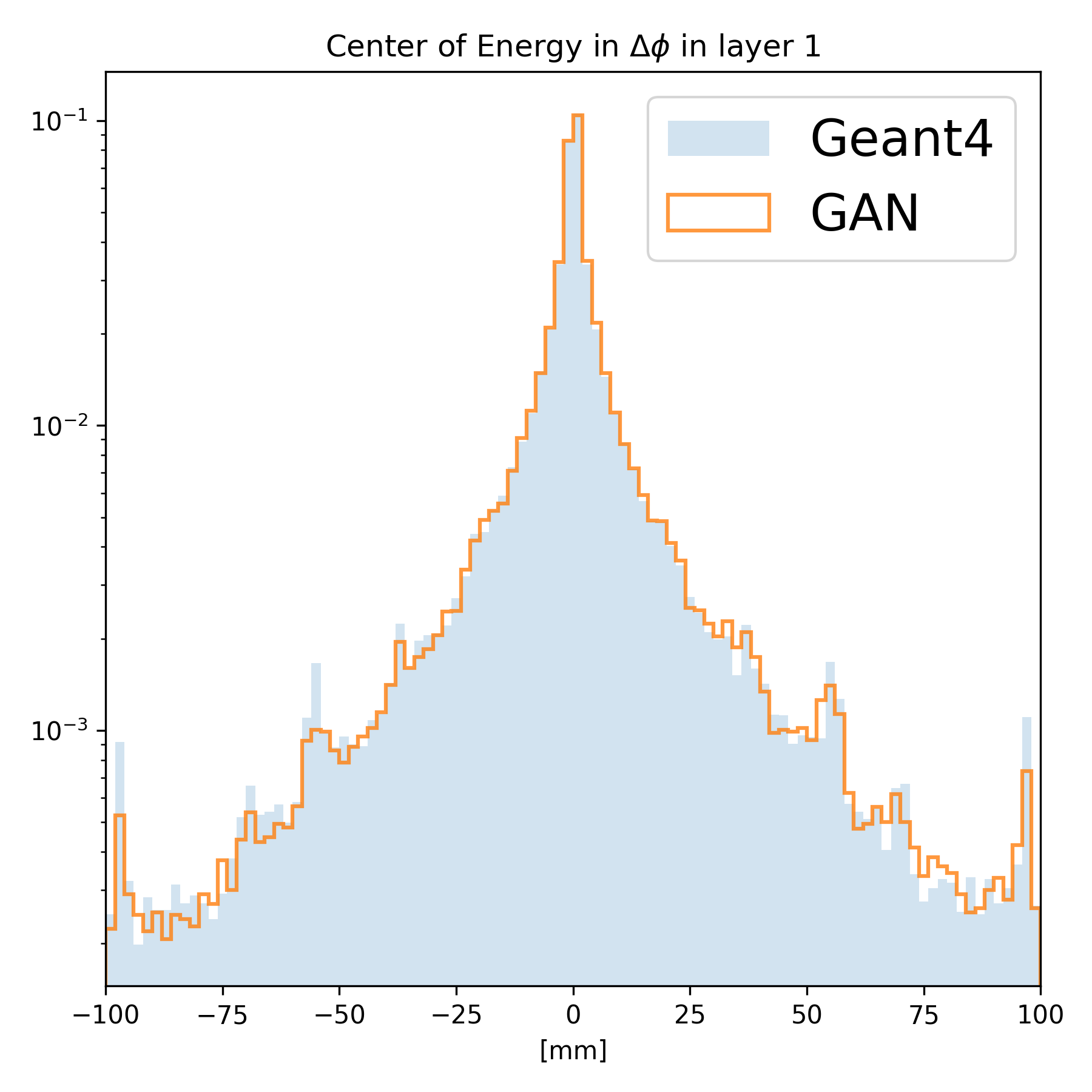} \\ 
    \includegraphics[width=0.3\textwidth]{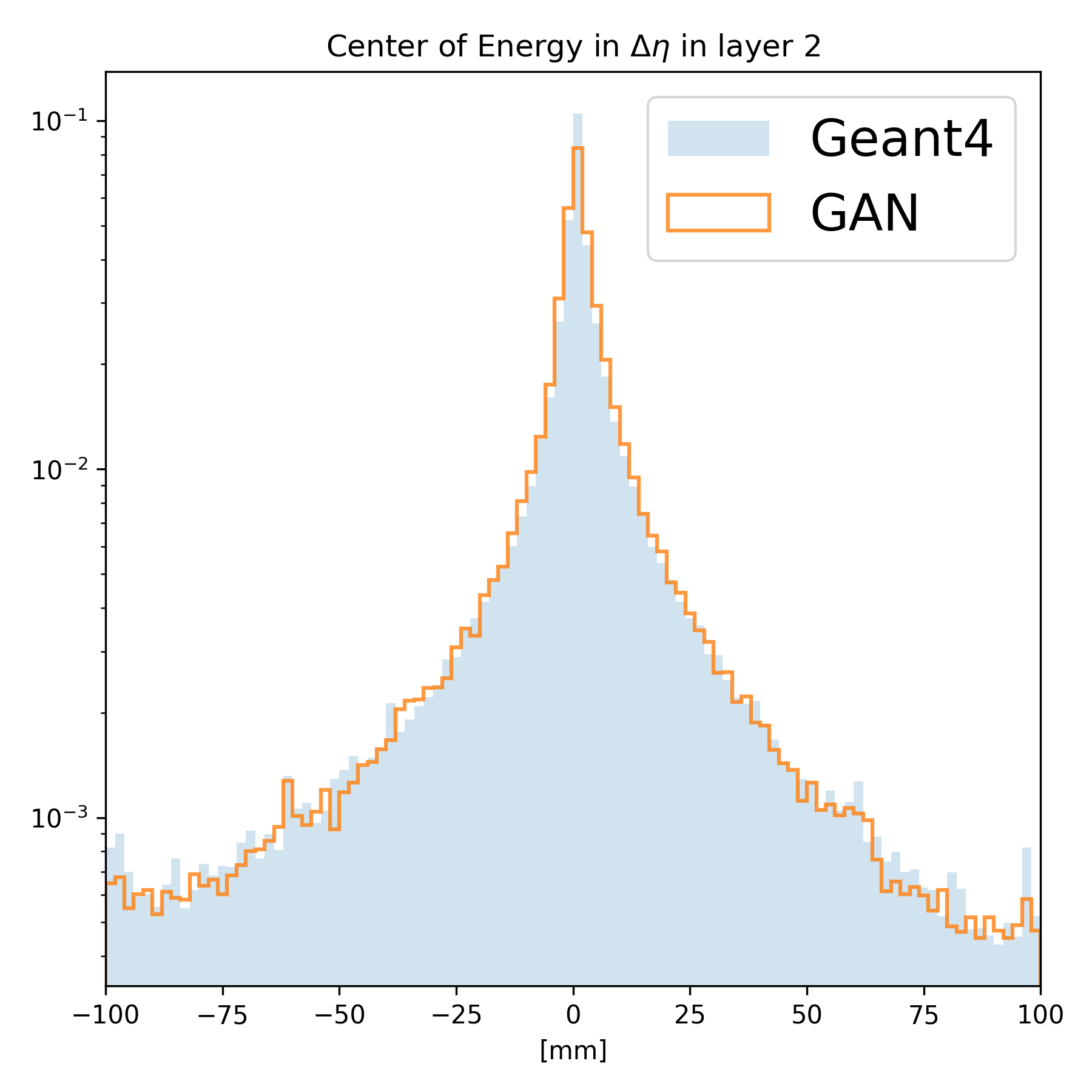}
    \includegraphics[width=0.3\textwidth]{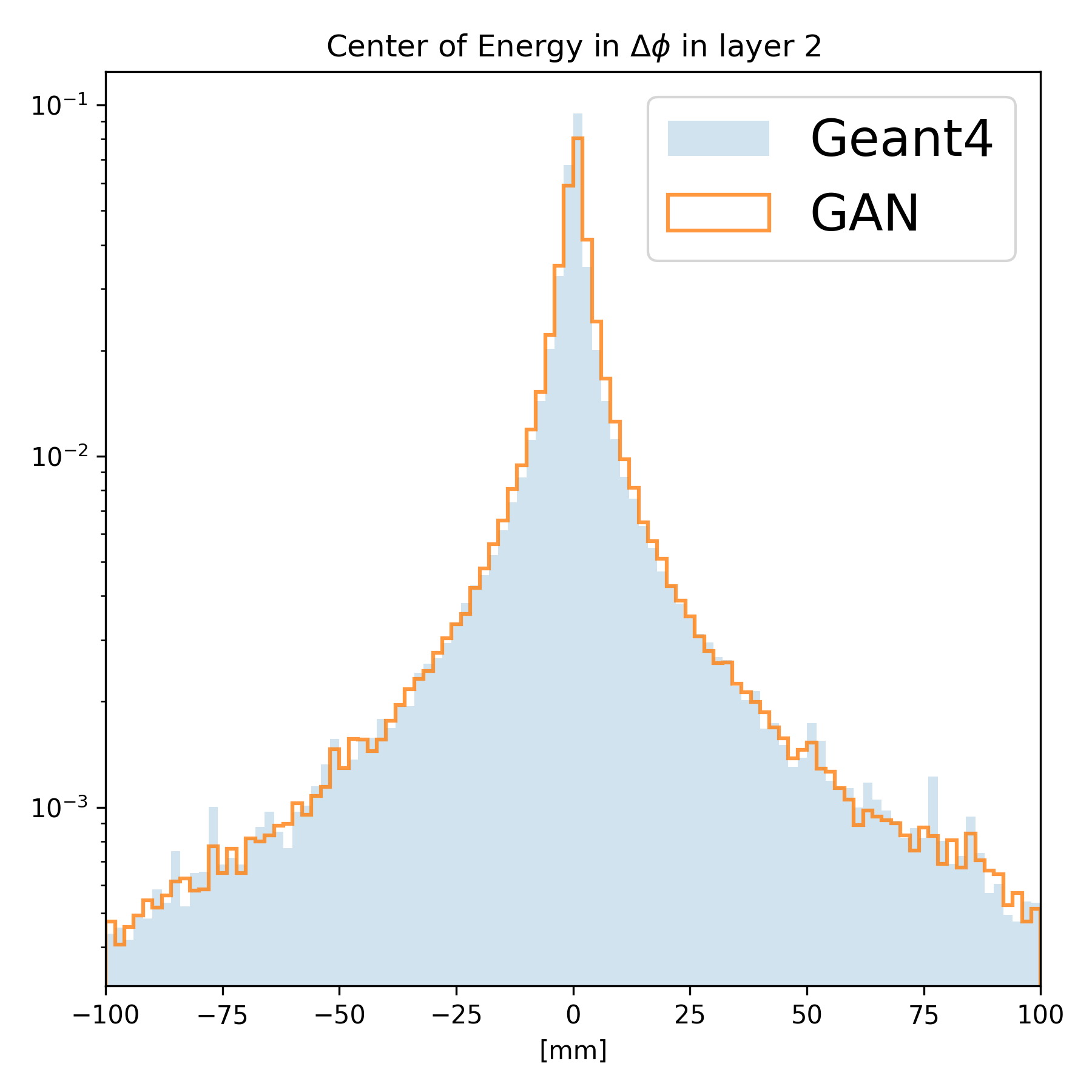} \\ 
    \includegraphics[width=0.3\textwidth]{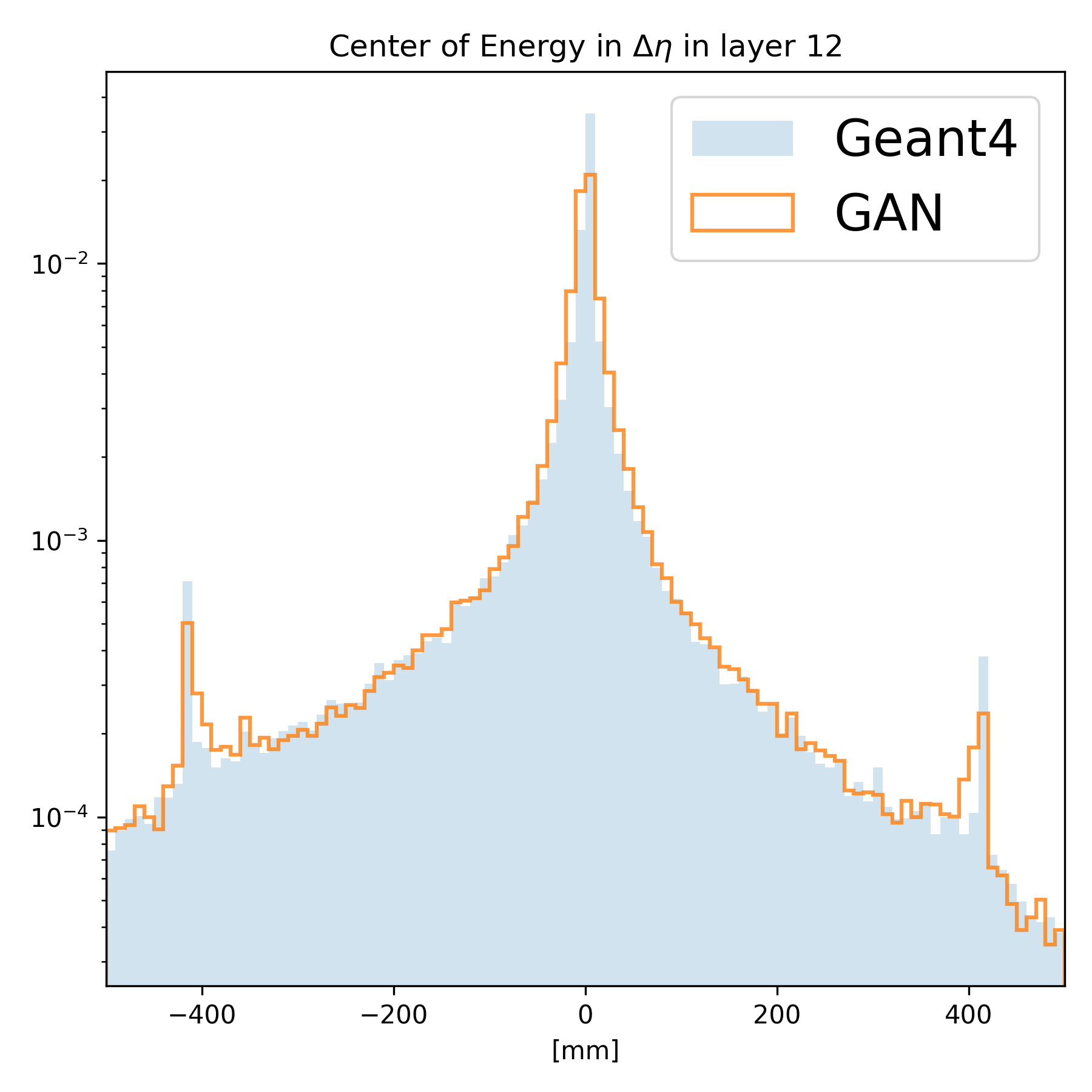}
    \includegraphics[width=0.3\textwidth]{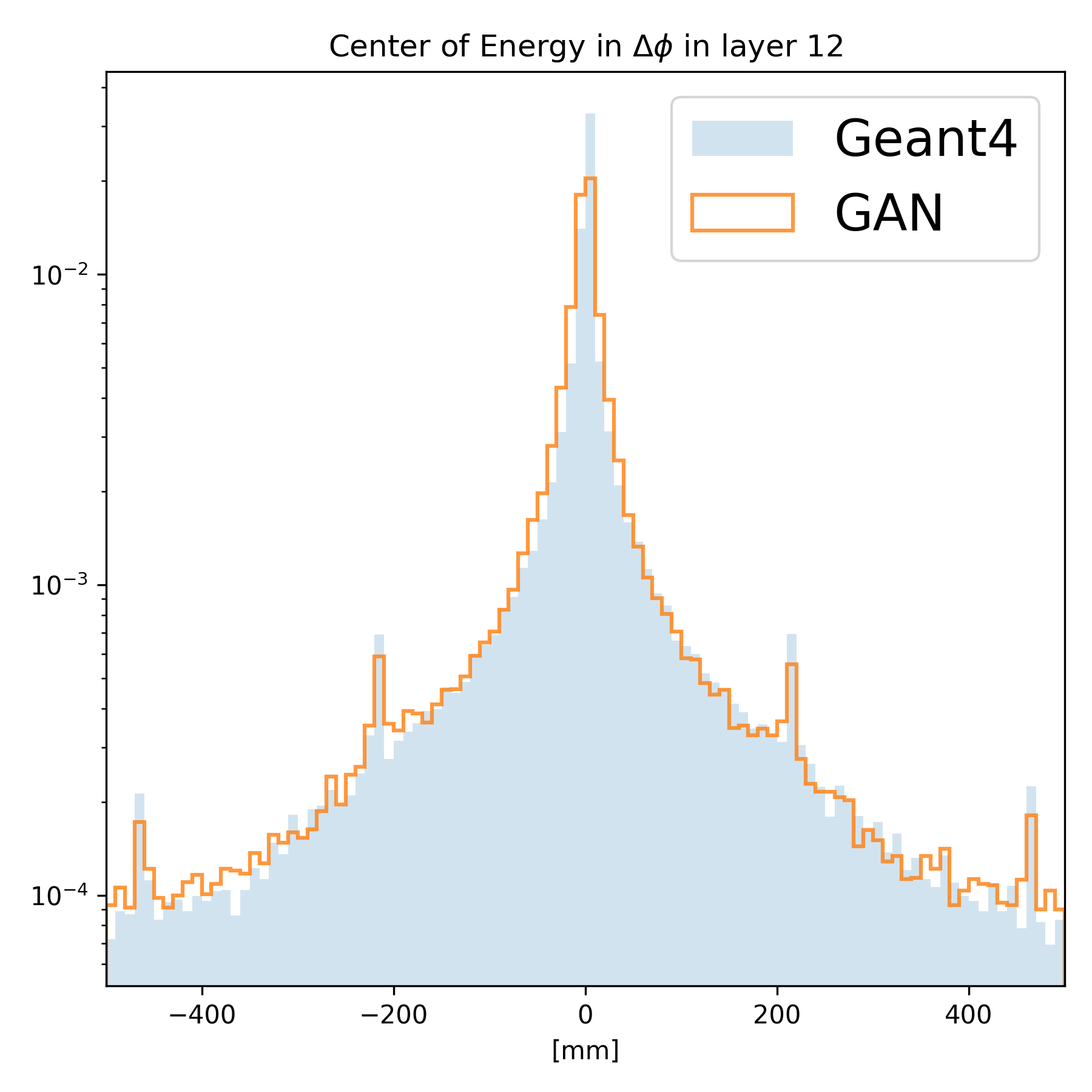} \\
    \includegraphics[width=0.3\textwidth]{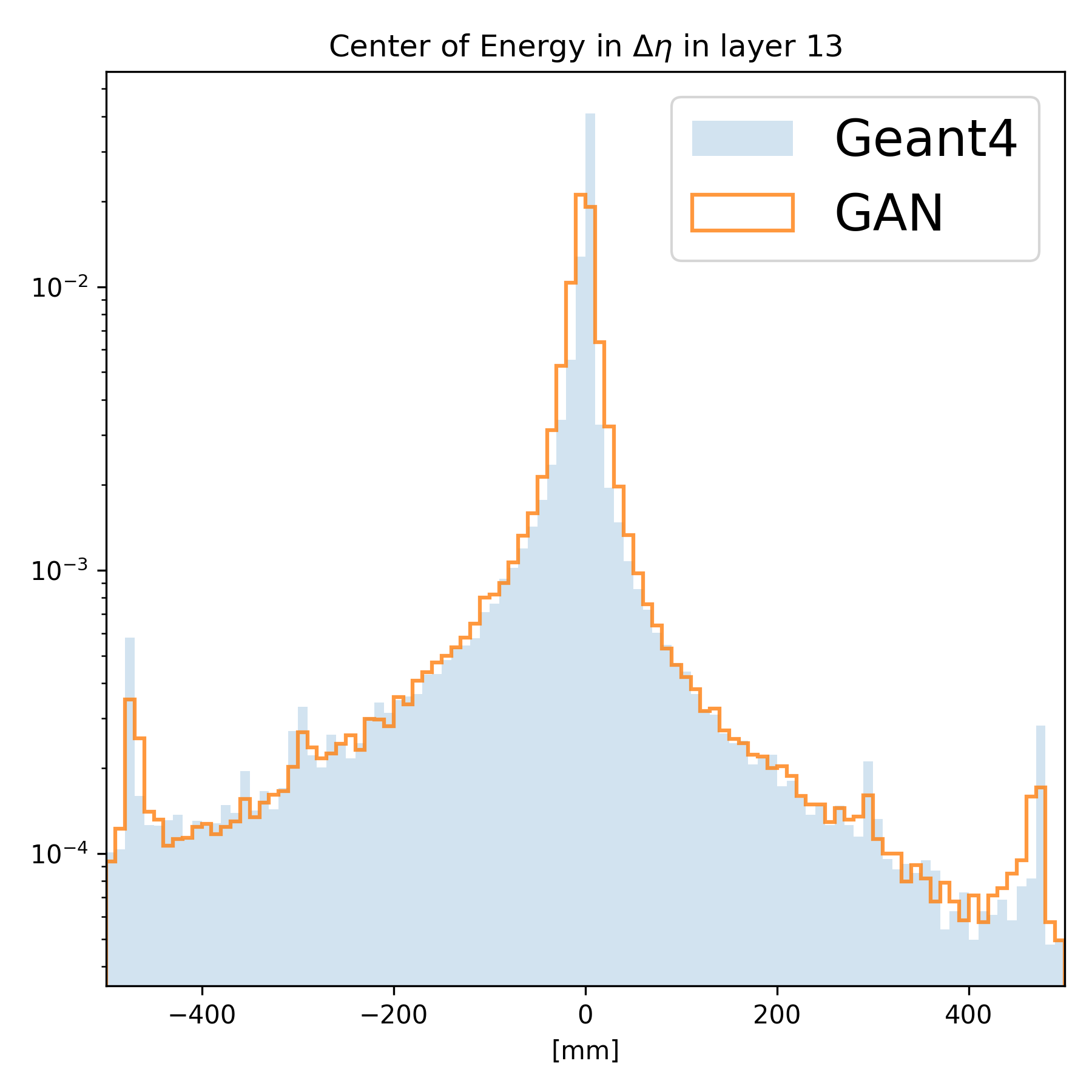} 
    \includegraphics[width=0.3\textwidth]{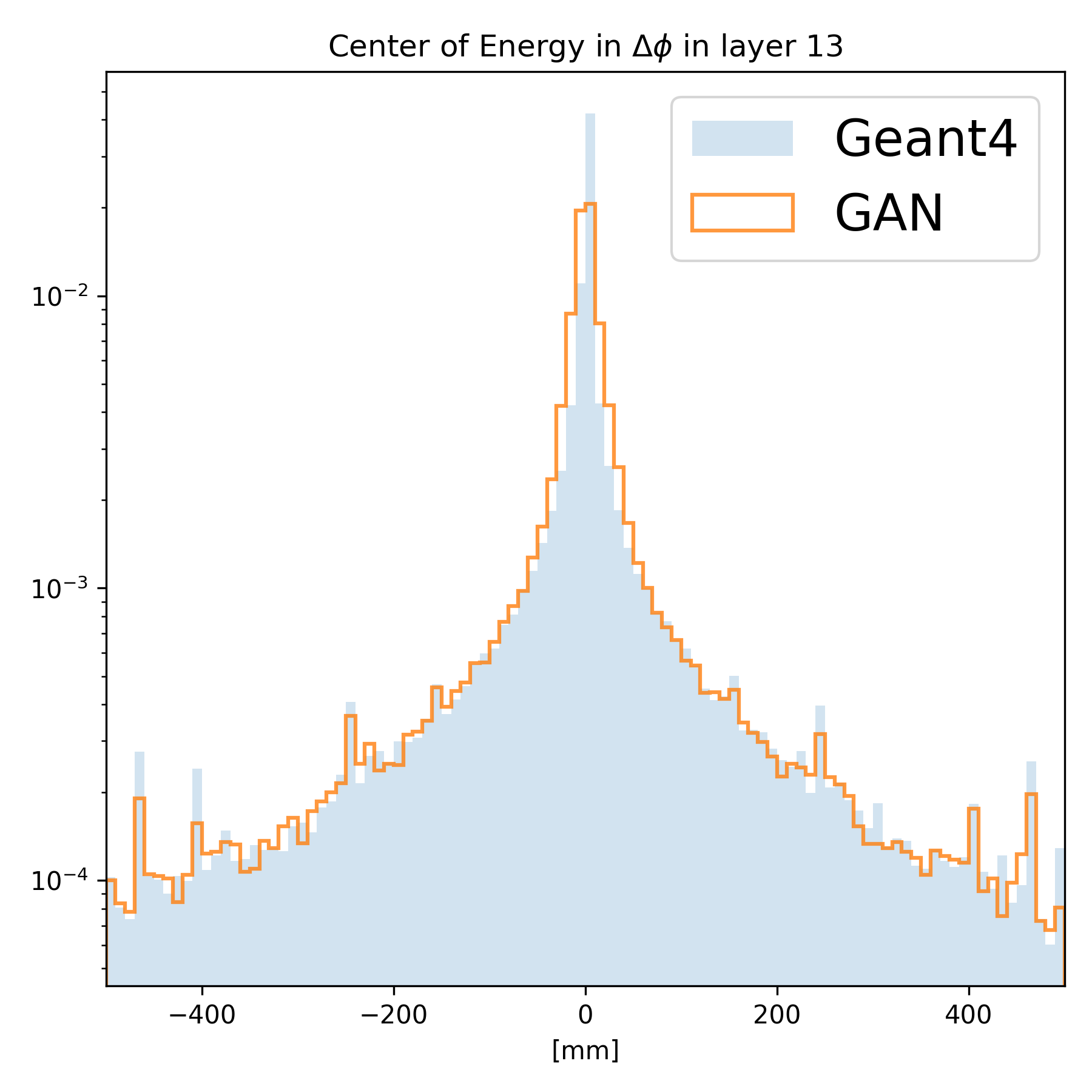}
    \caption{
        \textbf{Pion \CSG with layer-energy normalisation}: shower width compared to \GEANT simulation (solid area).
        All energies accumulated in layers 1 (first row), 2 (second row), 12 (third line) and 13 (bottom) along $\eta$ (left) and $\phi$ (right).
    }
    \label{fig:centrepion}
\end{figure}

\begin{figure}[htp]
\centering
    \includegraphics[width=0.3\textwidth]{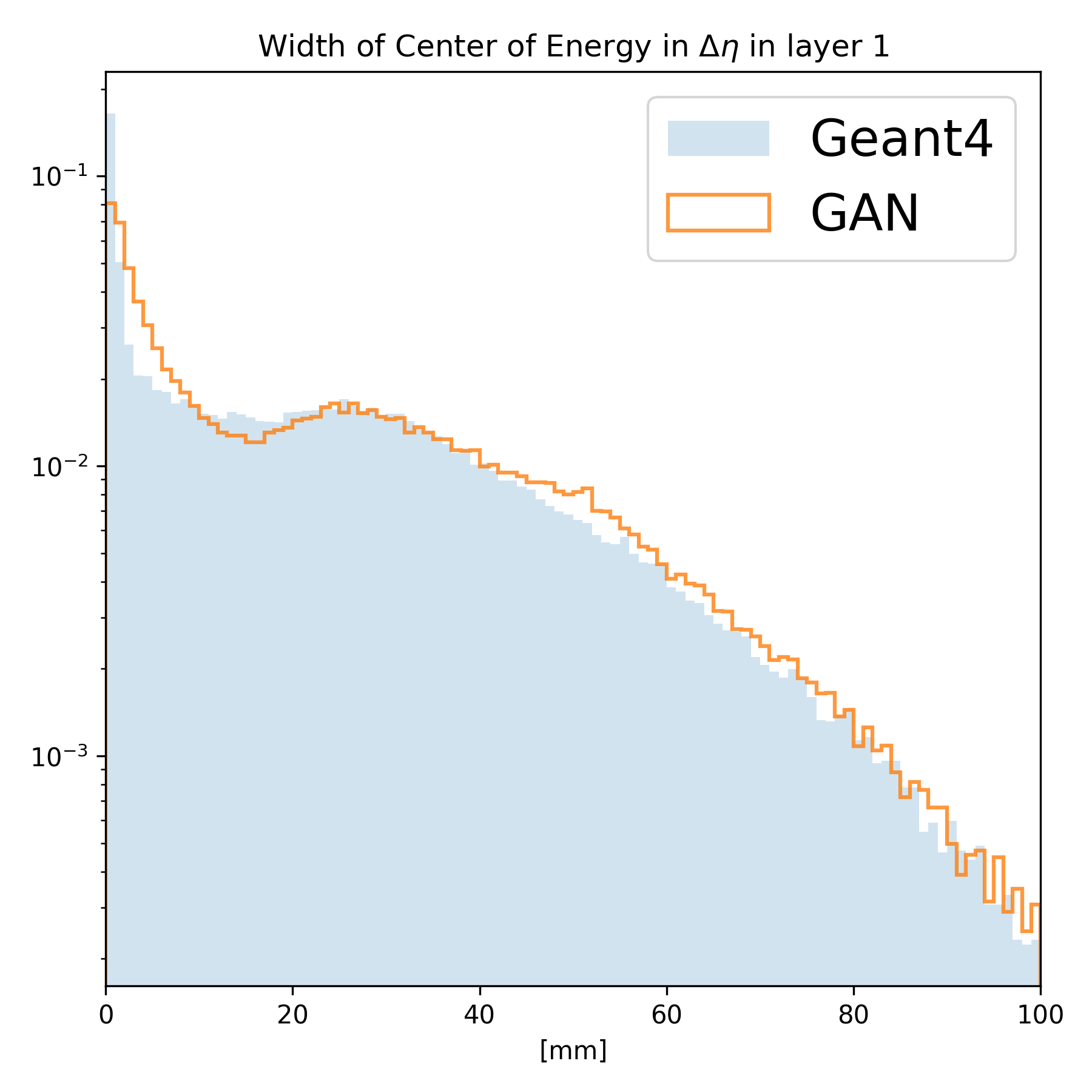}
    \includegraphics[width=0.3\textwidth]{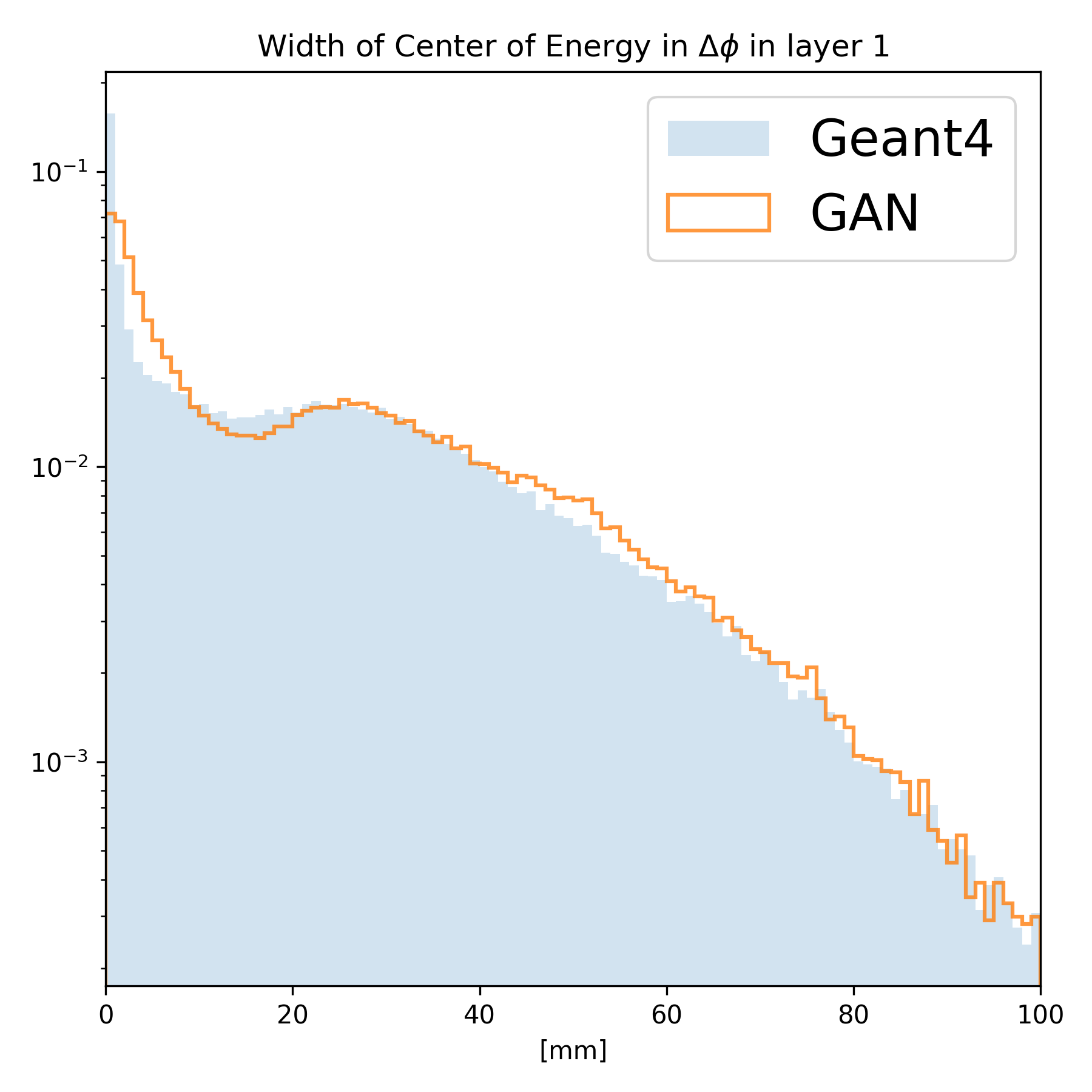} \\
    \includegraphics[width=0.3\textwidth]{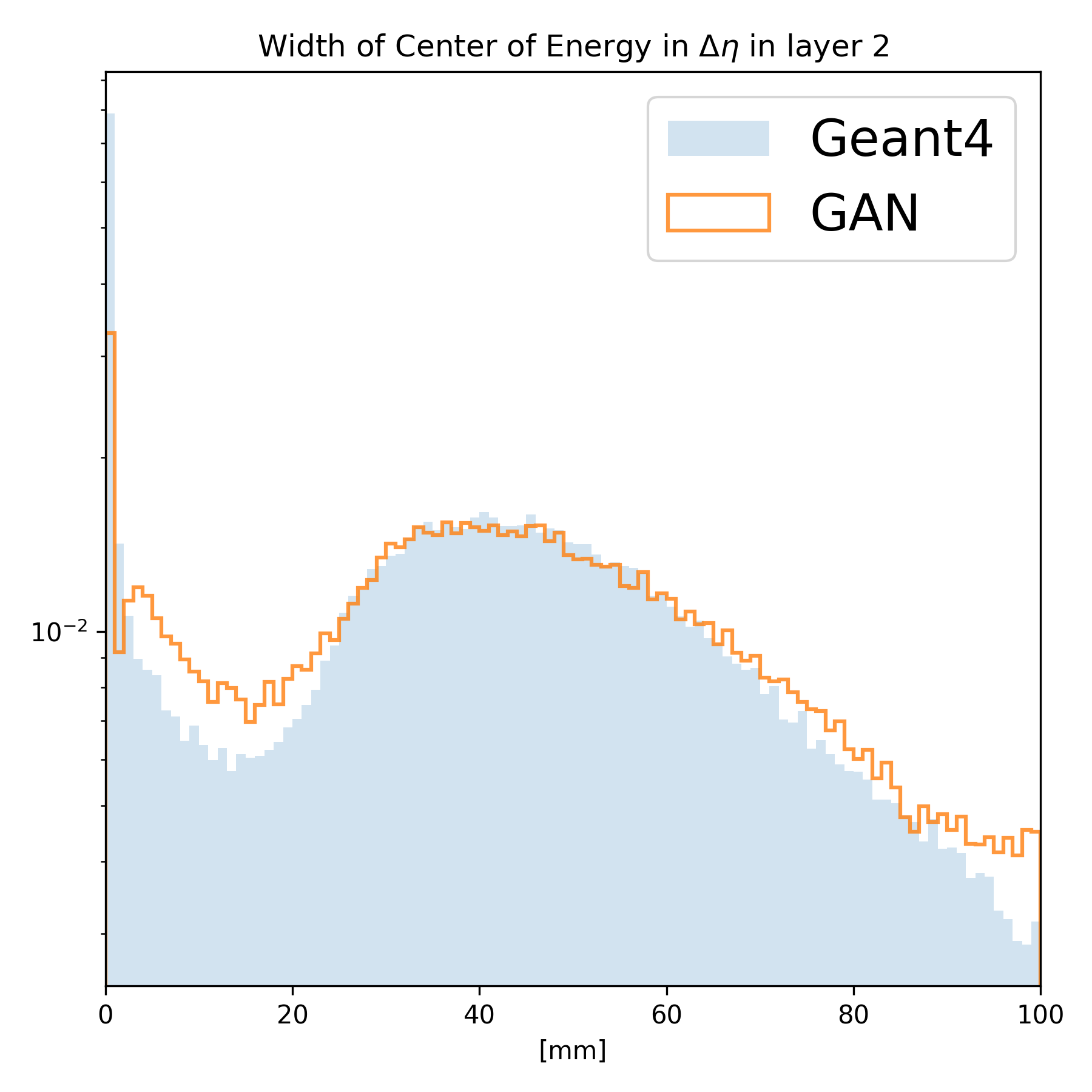} 
    \includegraphics[width=0.3\textwidth]{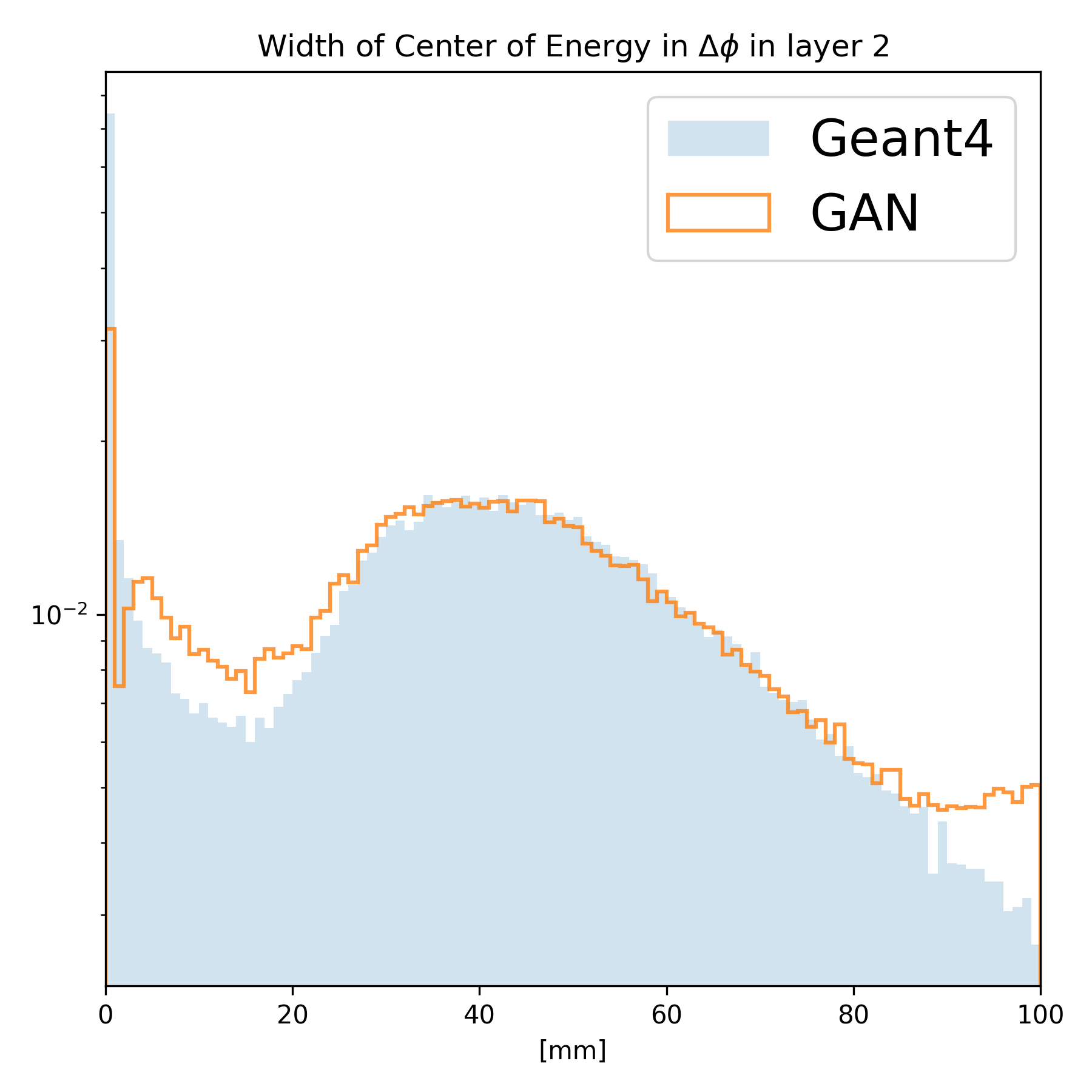} \\
    \includegraphics[width=0.3\textwidth]{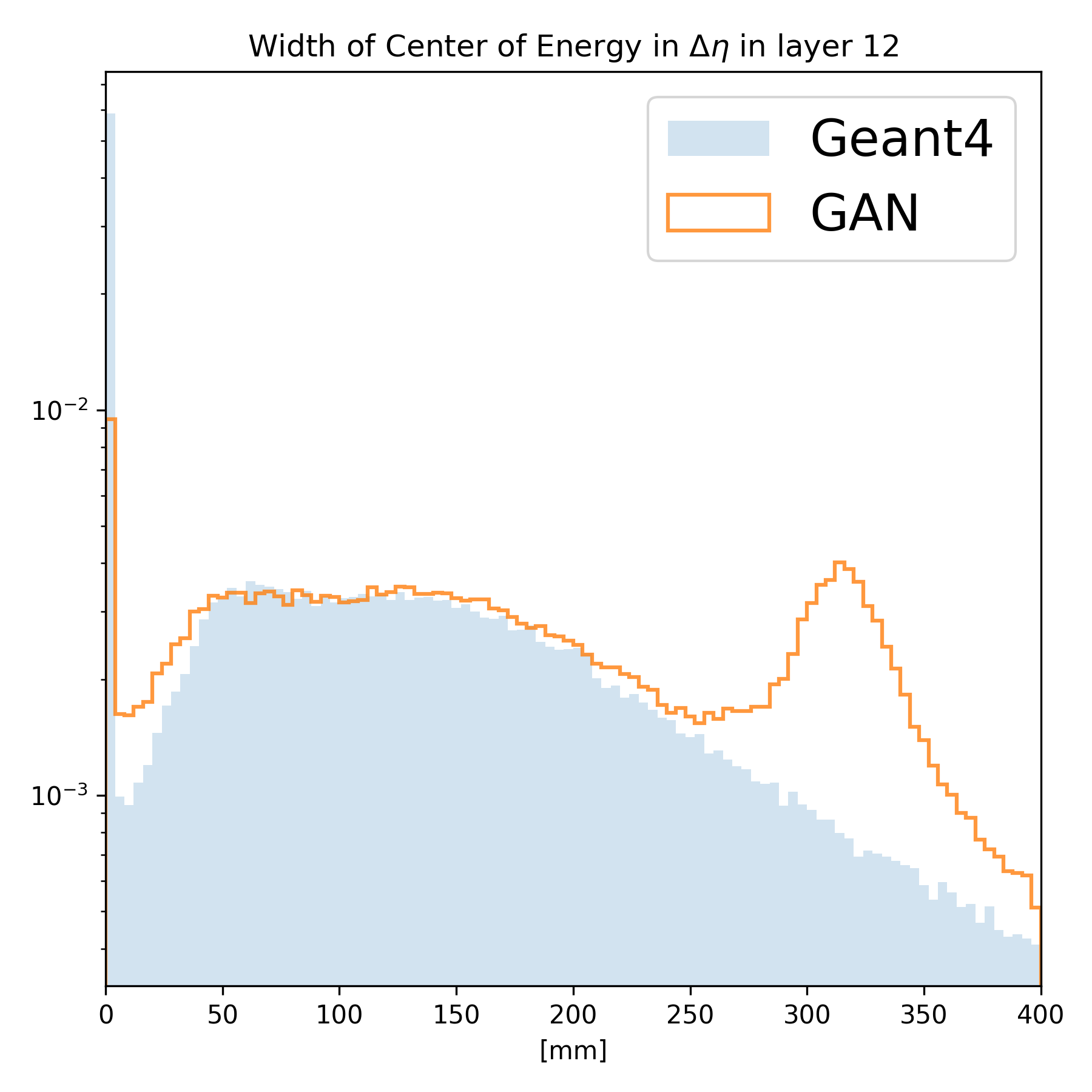}
    \includegraphics[width=0.3\textwidth]{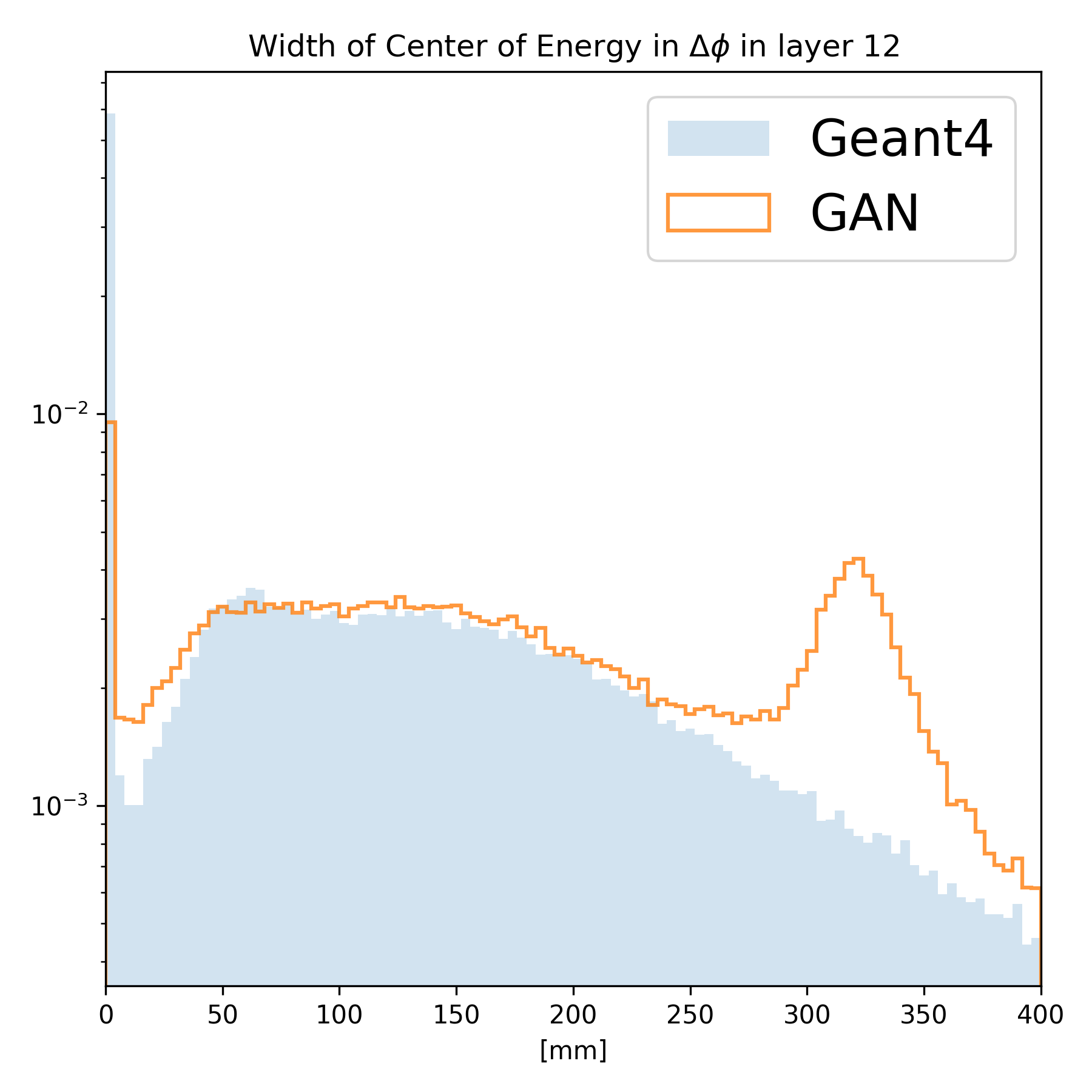} \\
    \includegraphics[width=0.3\textwidth]{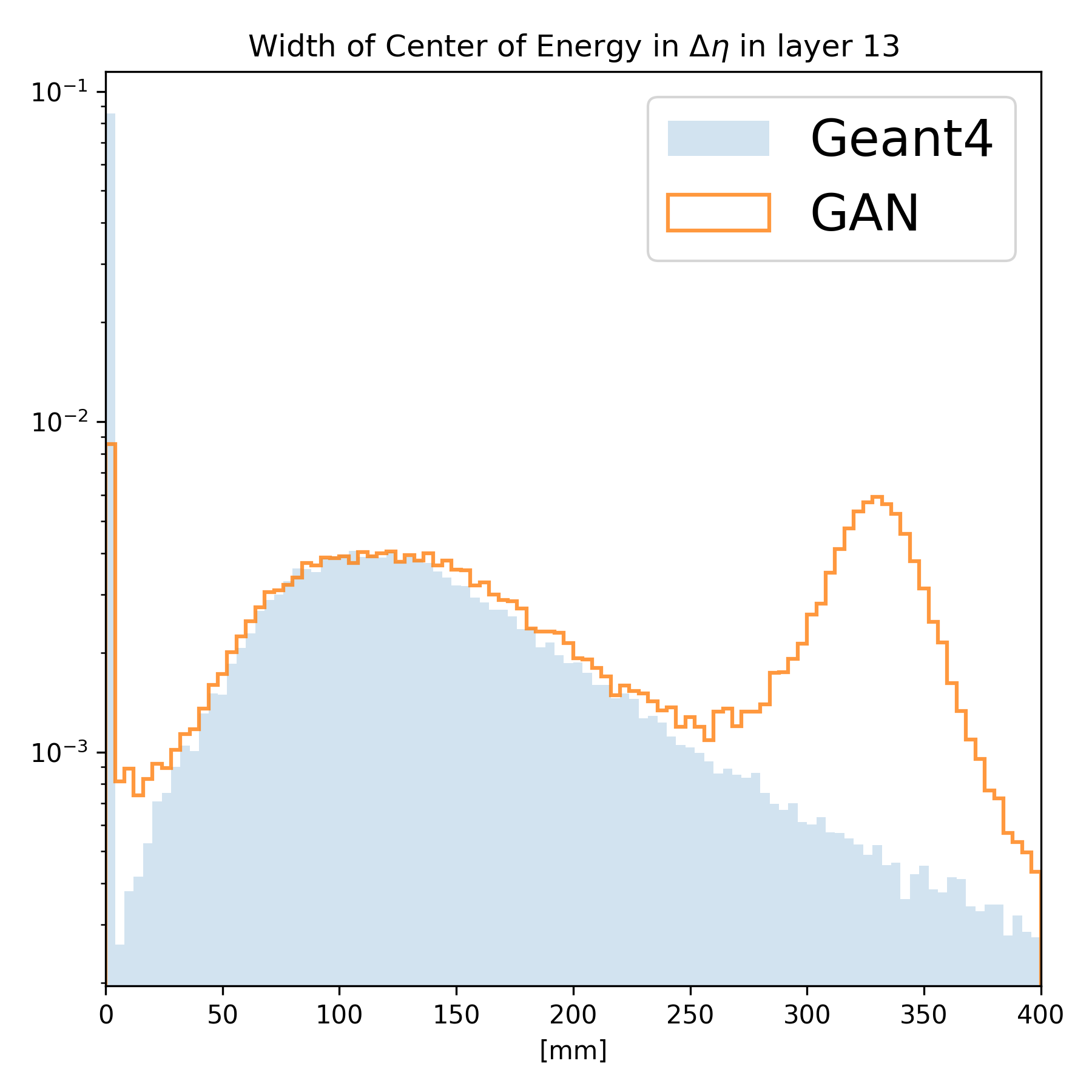} 
    \includegraphics[width=0.3\textwidth]{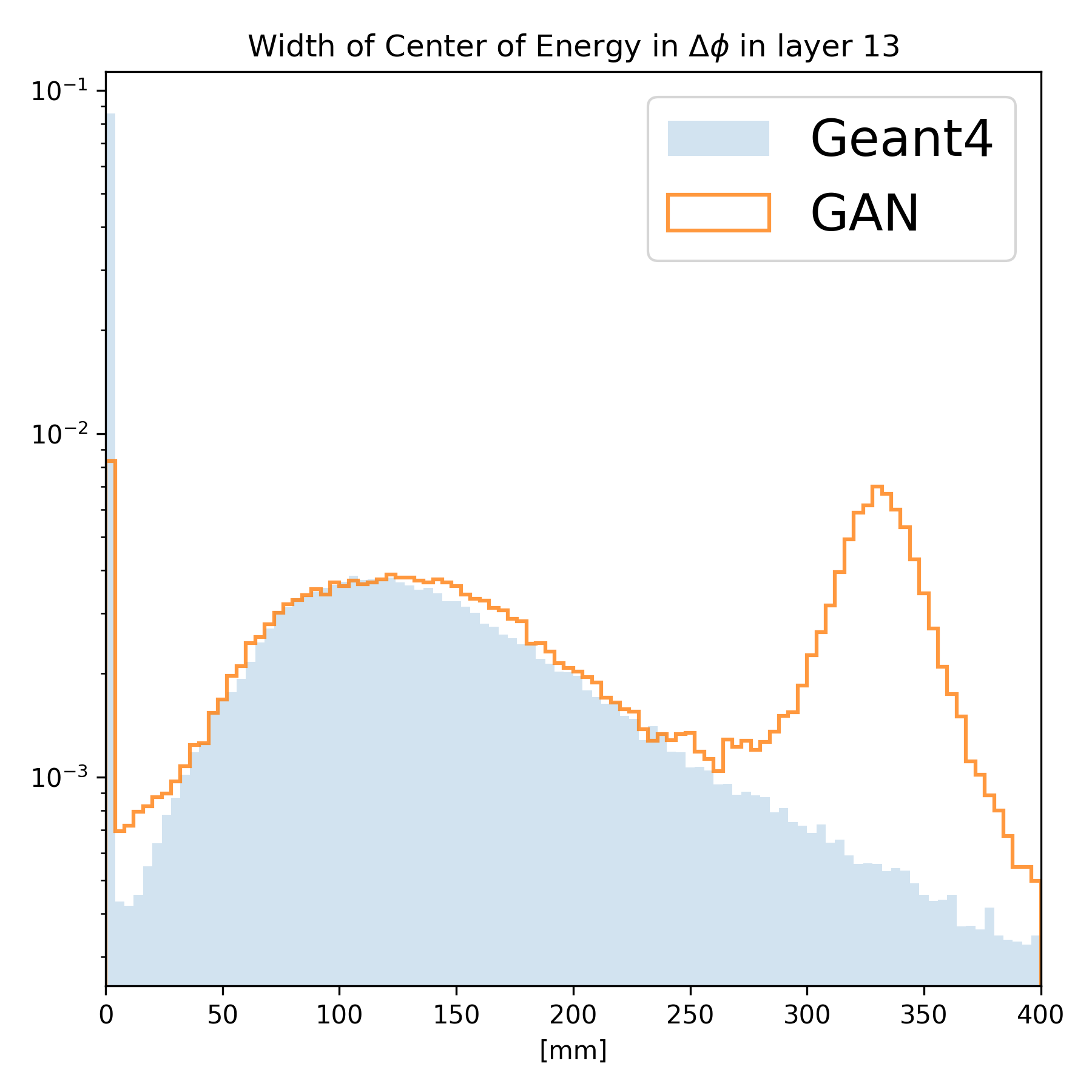} \\
    \caption{
        \textbf{Pion \CSG with layer-energy normalisation}: shower width compared to \GEANT simulation (solid area).
        All energies accumulated in layers 1 (first row) and 2 (second row) 12 (third line) and 13 (bottom) along $\eta$ (left) and $\phi$ (right).
    }
    \label{fig:widthpion}
\end{figure}

The distributions of shape characteristics from \CSG are depicted in \Figrange{\ref{fig:centrephoton}}{\ref{fig:widthphoton}} and \Figrange{\ref{fig:centrepion}}{\ref{fig:widthpion}} for photons and pions, respectively.
These distributions include all incident momenta.
In general, the showers generated by \CSG effectively replicate the attributes found in the \GEANT distribution of shower centres.

\begin{figure}[htp]
\centering
    \includegraphics[width=0.3\textwidth]{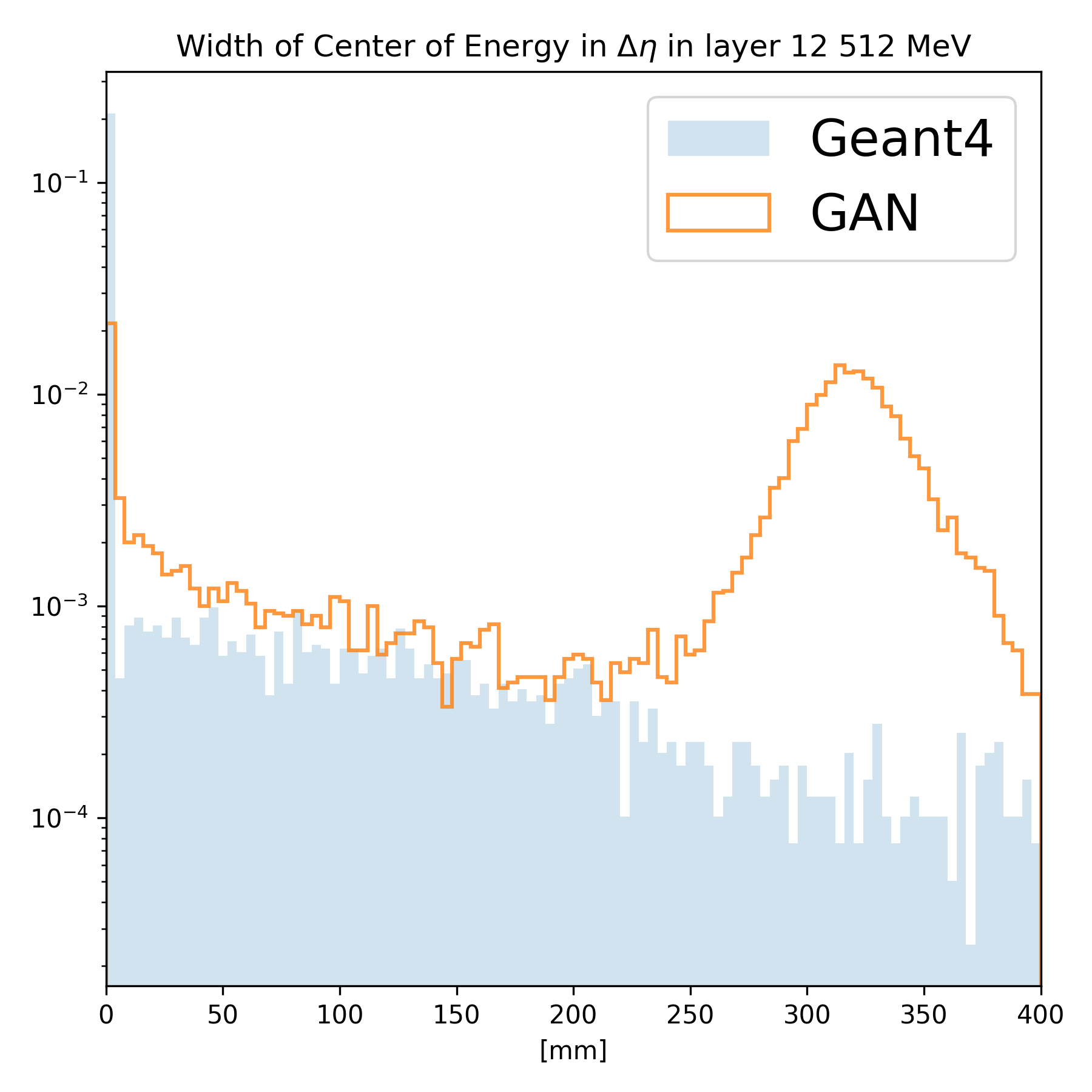}
    \includegraphics[width=0.3\textwidth]{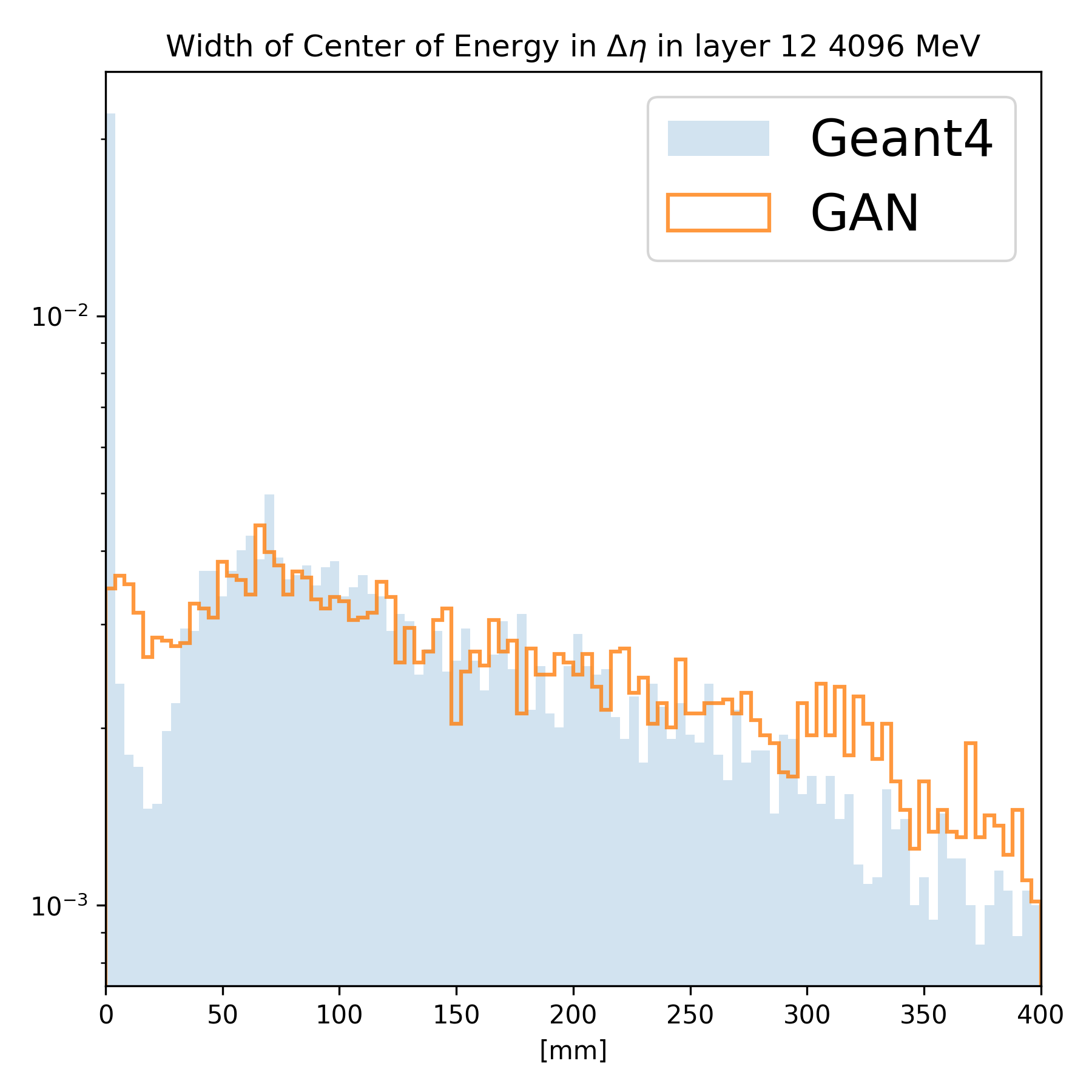}
    \includegraphics[width=0.3\textwidth]{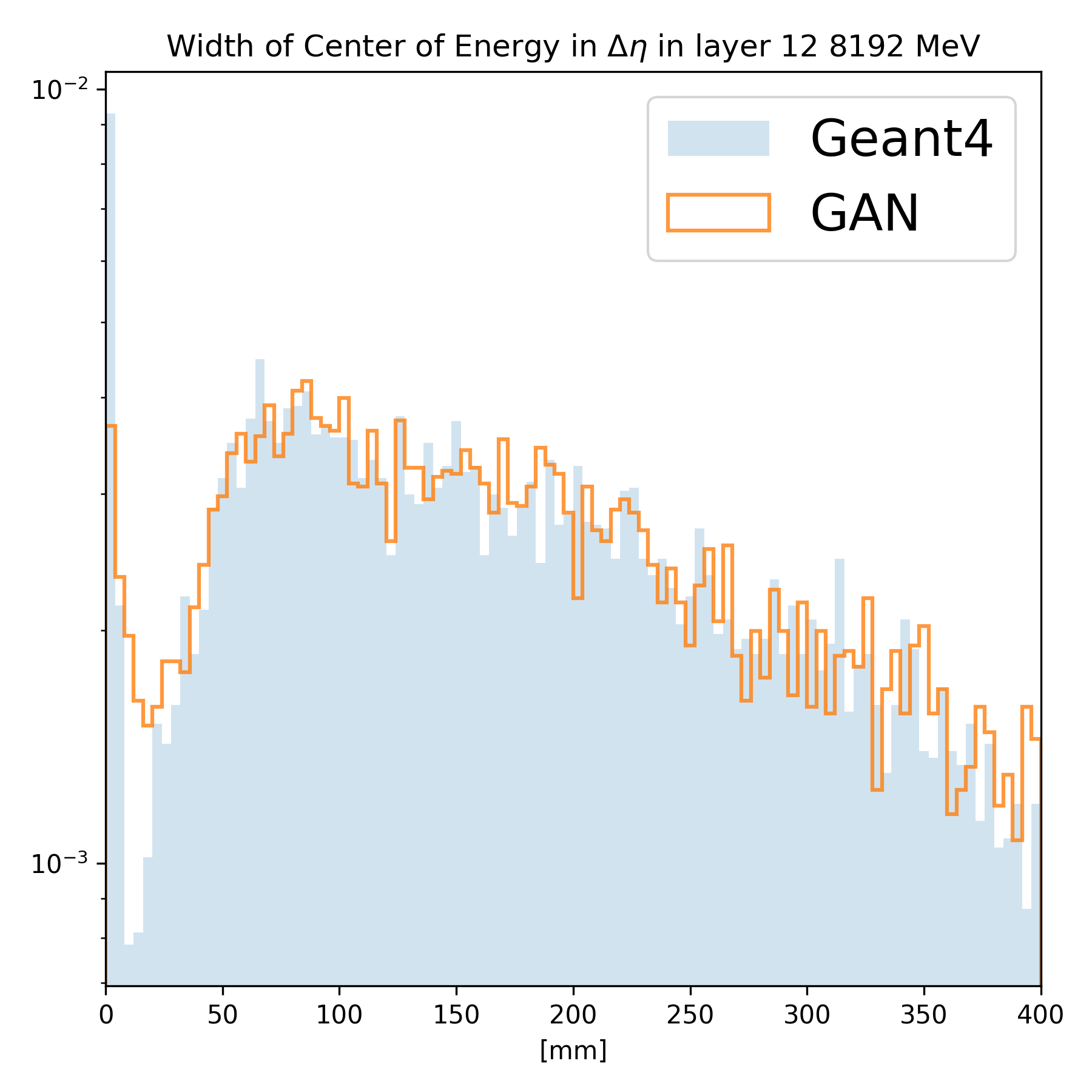} \\
    \includegraphics[width=0.3\textwidth]{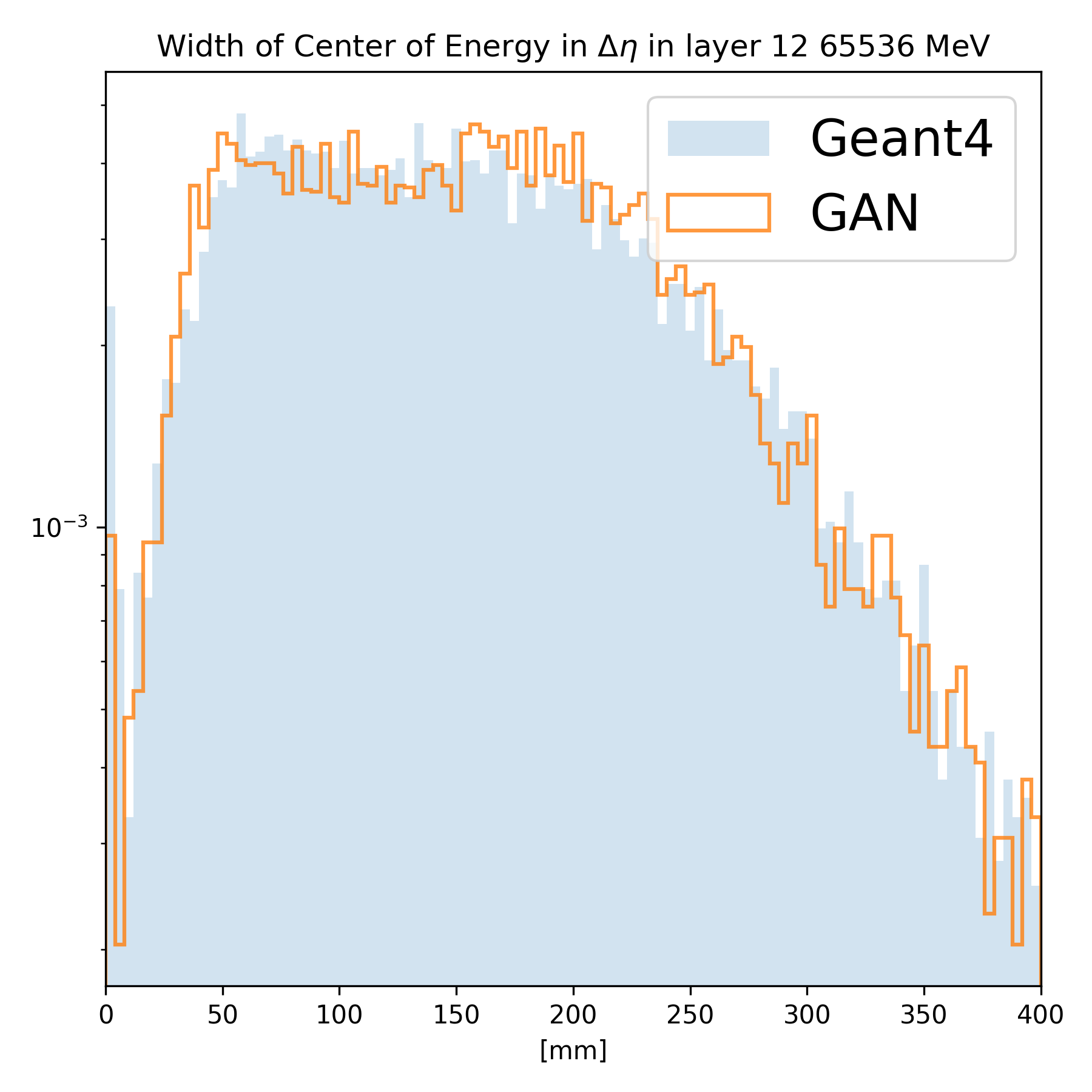}
    \includegraphics[width=0.3\textwidth]{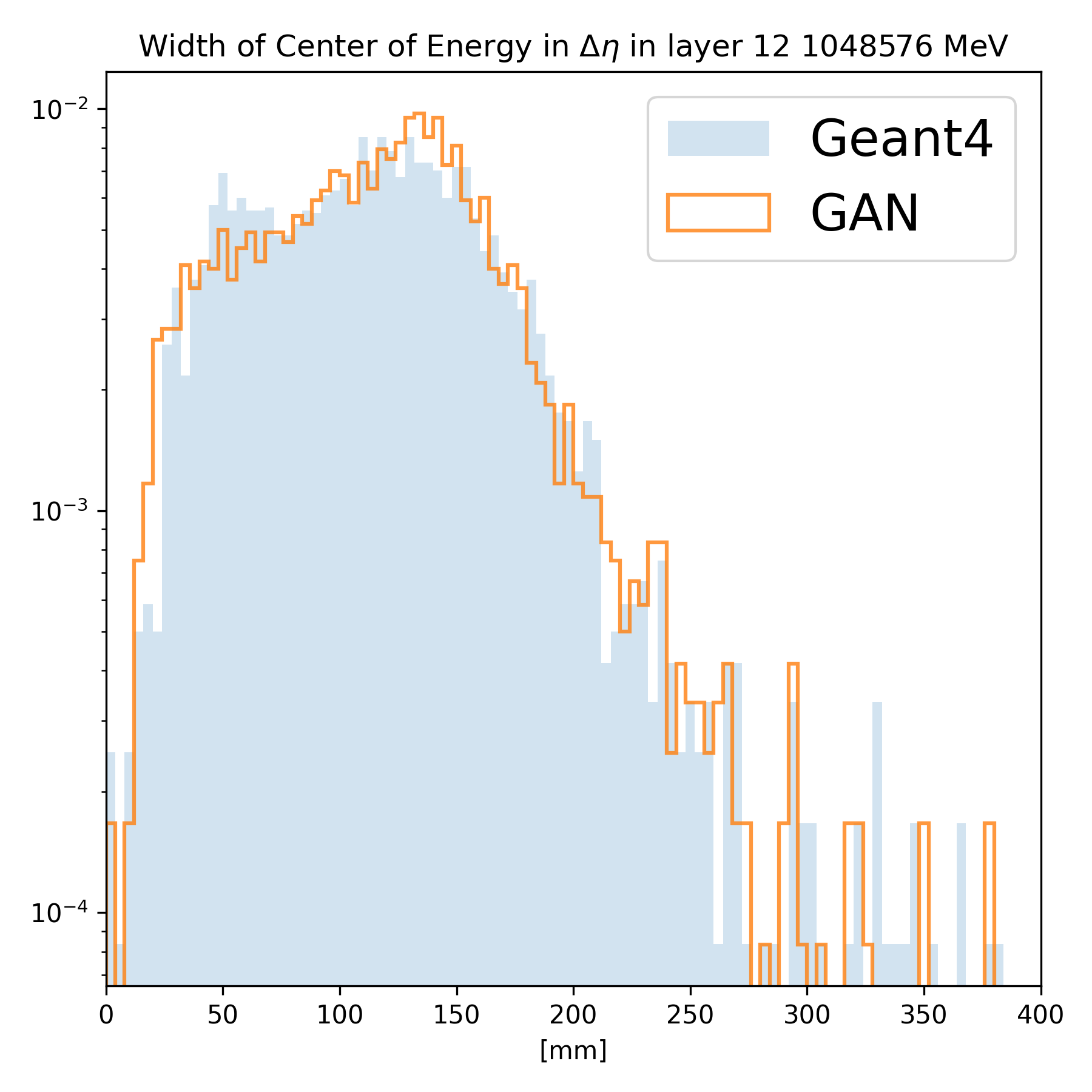} \\
    \includegraphics[width=0.3\textwidth]{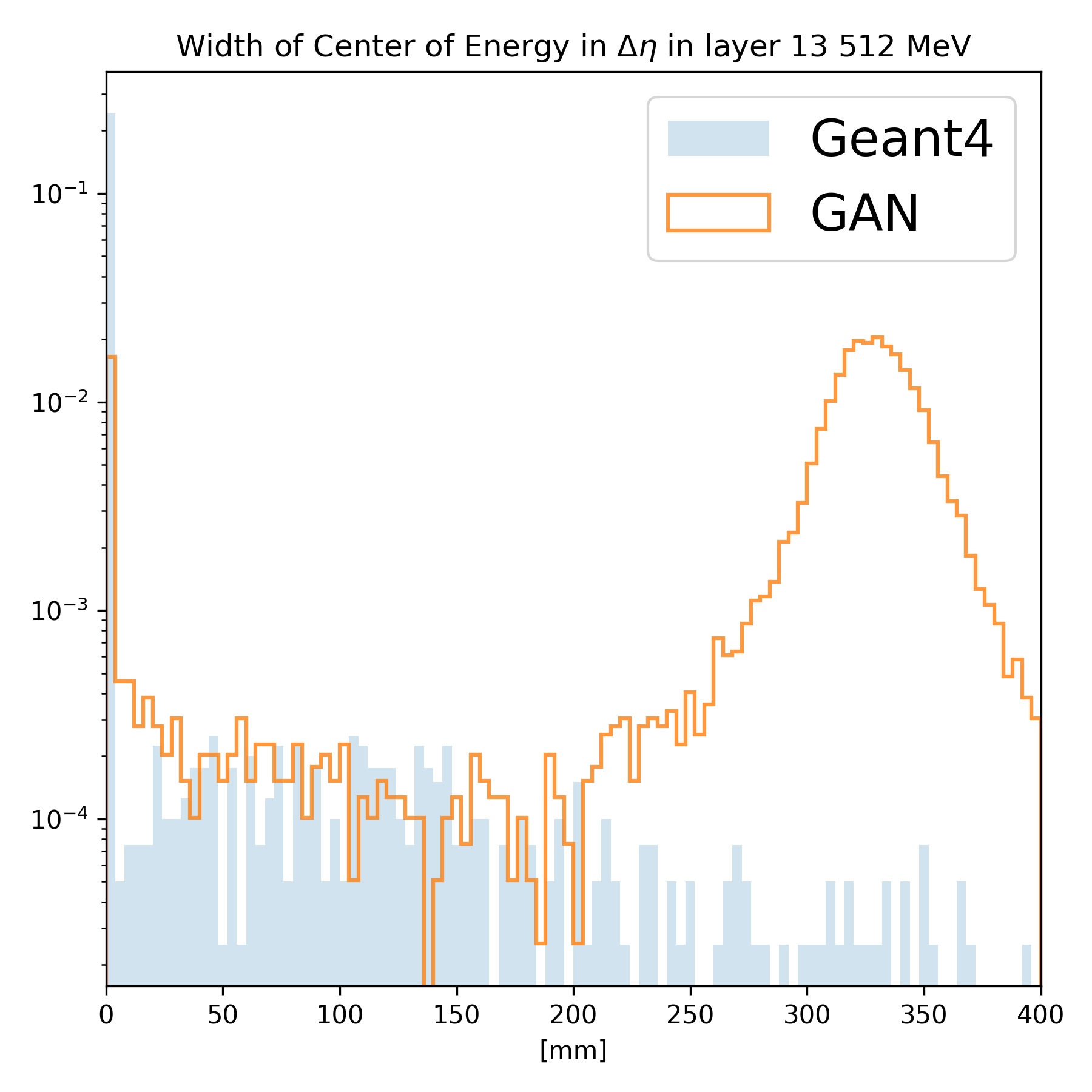}
    \includegraphics[width=0.3\textwidth]{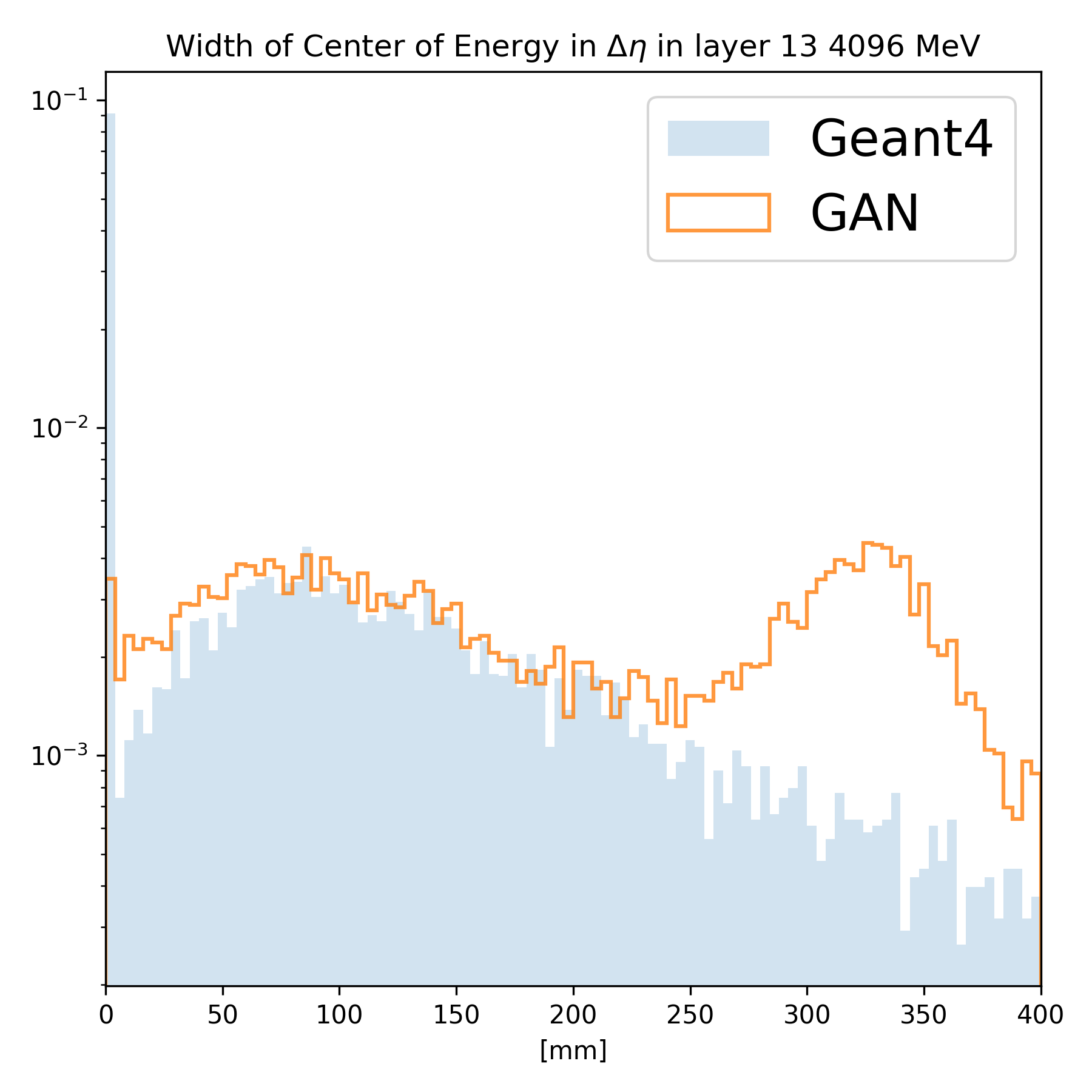}
    \includegraphics[width=0.3\textwidth]{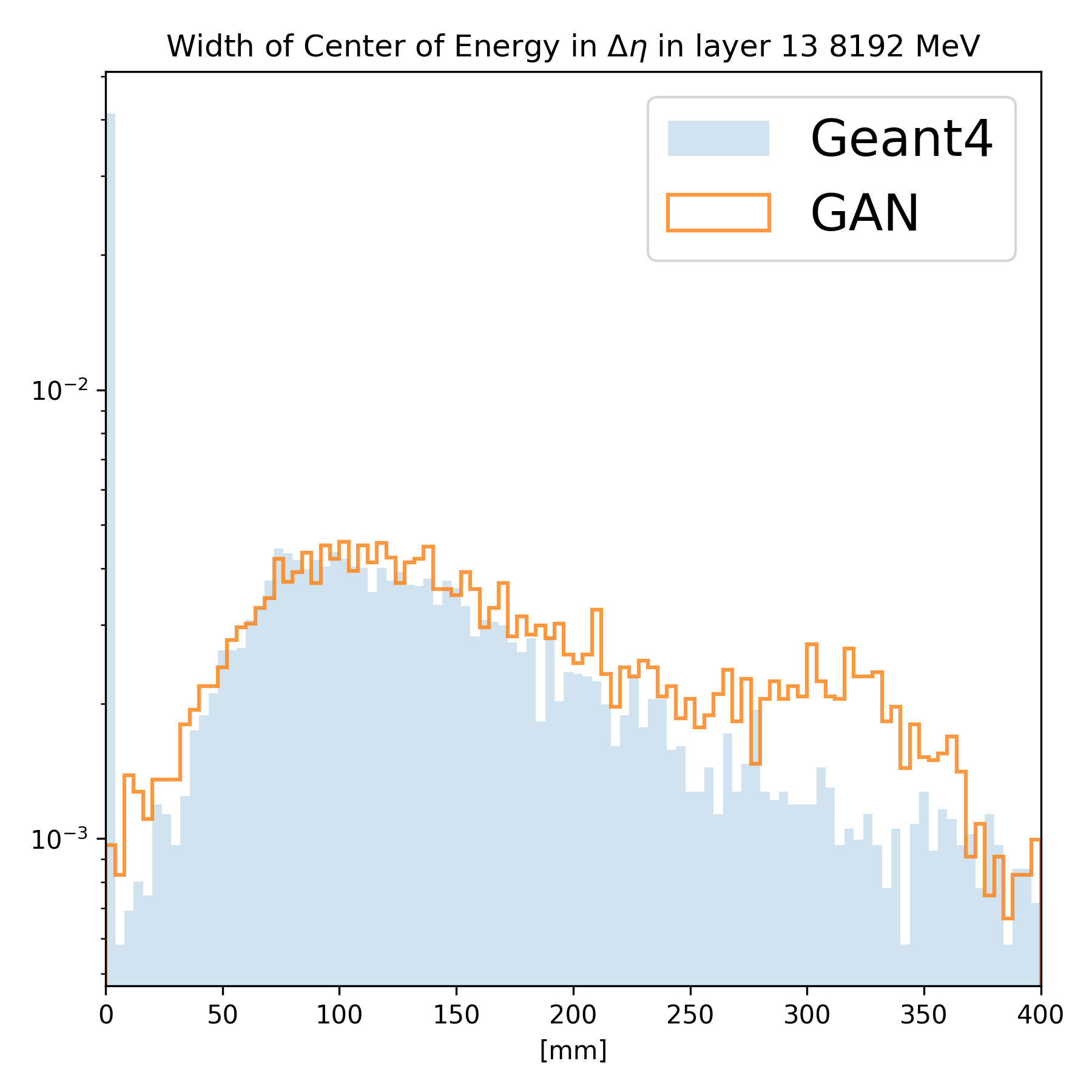} \\
    \includegraphics[width=0.3\textwidth]{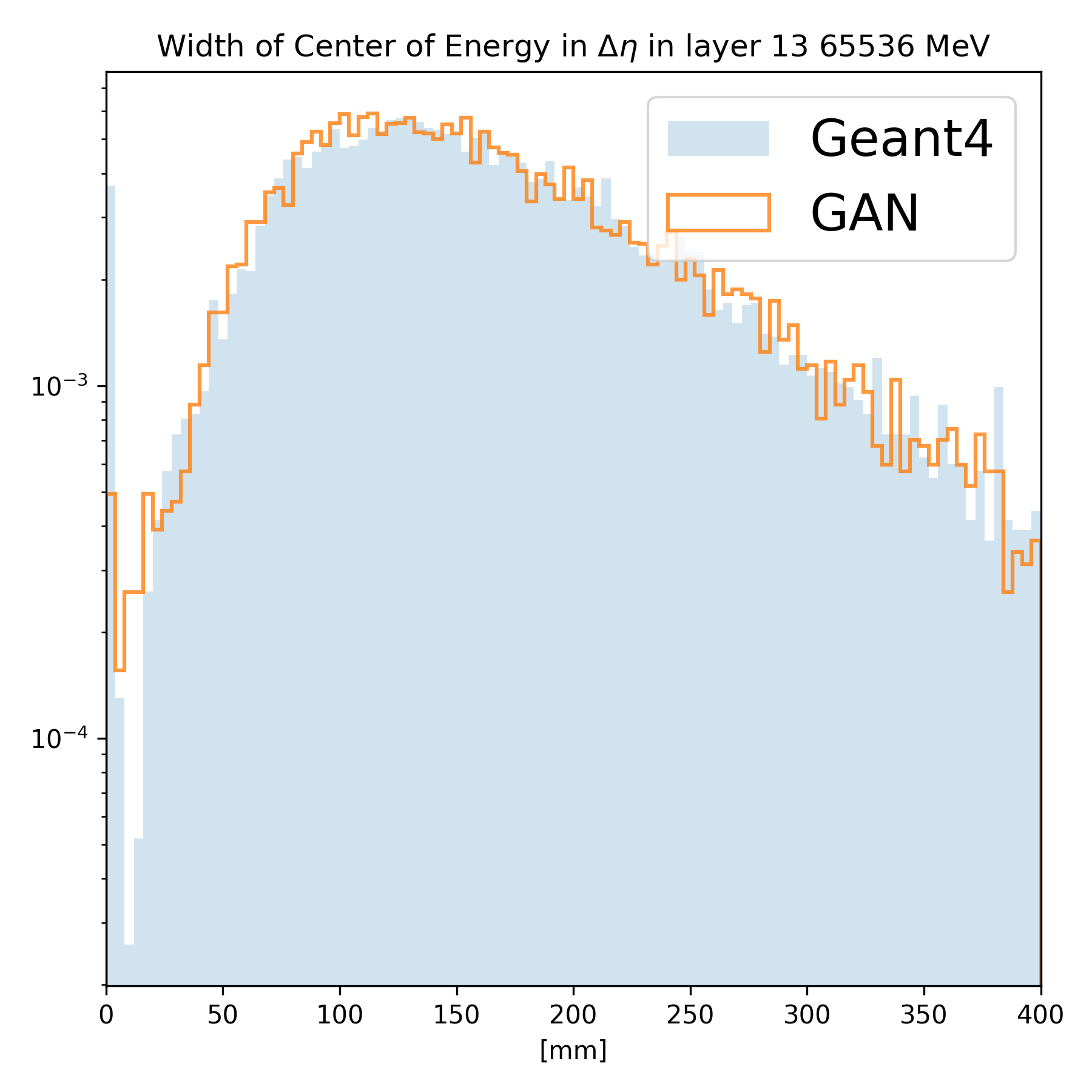}
    \includegraphics[width=0.3\textwidth]{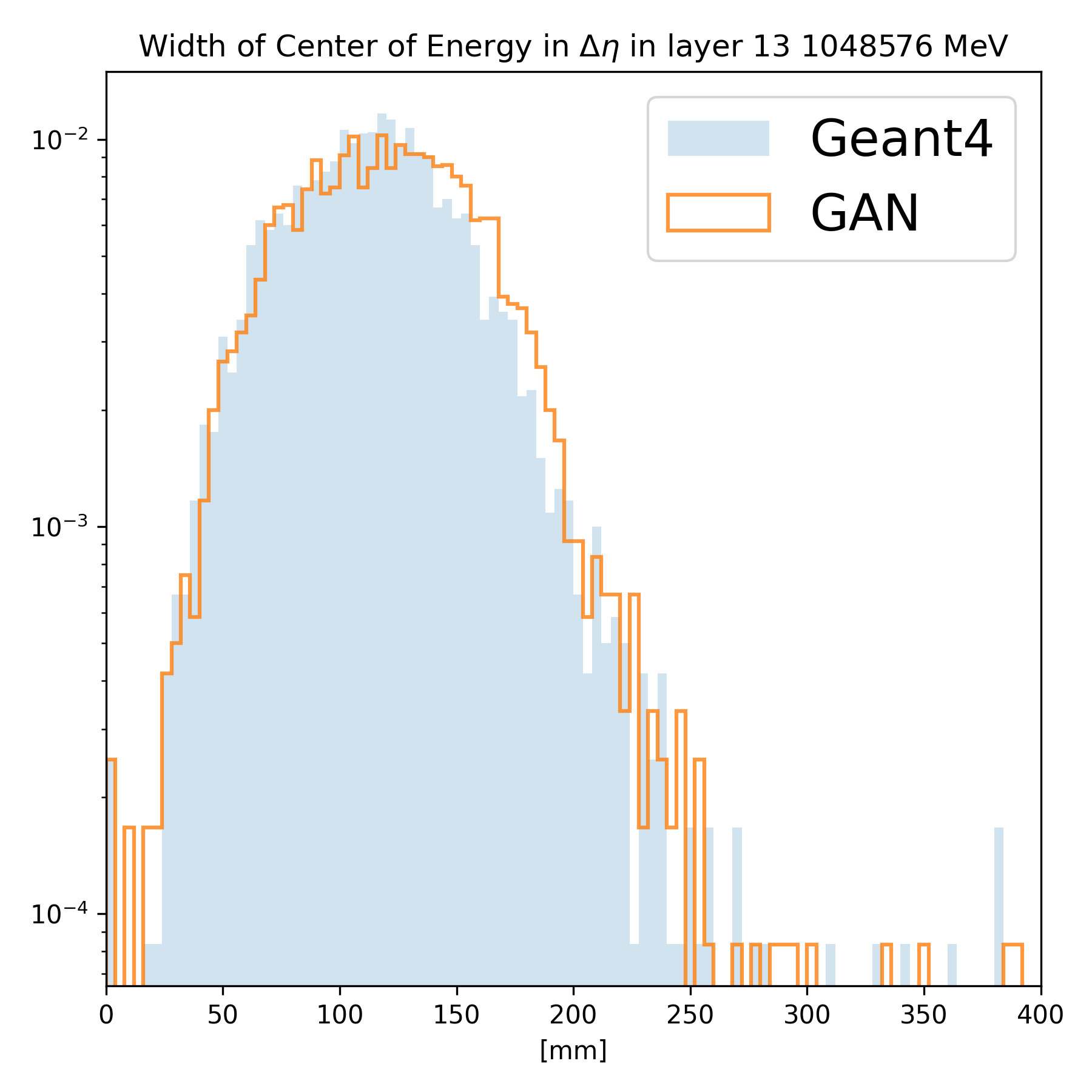}
    \caption{
        \textbf{Pion \CSG with layer-energy normalisation}: shower width compared to \GEANT simulation (solid area) in different incident energies of 512~\MeV, 4~\GeV, 8~\GeV, 65~\GeV and 1~\TeV in layers 12 (up) and 13 (down).
    }
    \label{fig:widthpionperenergy}
\end{figure}

Regarding the shower width, in the case of photons, good agreement is observed, with only minor deviations.
The width of pion showers, however, displays a more notable discrepancy between \CSG and \GEANT.
In particular, \CSG struggles to accurately reproduce extremely narrow showers.
These comes from the events with a deposit of a MIP deposit observed in \Fig{\ref{fig:layerpion}}.
A spurious peak within the distribution of width for layer 12 and 13 is observed in the pion GAN.
Further investigation indicates that this phenomenon arises from events within low-momentum samples, as shown in \Fig{\ref{fig:widthpionperenergy}}, which shows the width along $\eta$ in layer 12 and 13 for various momentum points.
This peak is visible exclusively in the 512~\MeV sample and vanishes entirely at 8~\GeV.
While this represents a limitation in \CSG, the low energy deposits in the affected layers suggest that any potential influence on physics performance would be nearly negligible.
On the other hand, \CSG excels at modelling showers with high momenta.
This capability holds promise for good jet substructure reconstructions by effectively mitigating challenges related to clustering and the undesired merging of adjacent clusters that should remain separate.
However, to fully assess these potential advantages, comprehensive simulations and reconstructions within an authentic detector framework are required, which is beyond the scope of this paper.

\FloatBarrier

\subsection{Energy in all voxels}
\label{sec:voxelenergy}

\begin{figure}[htp]
\centering
    \includegraphics[width=0.3\textwidth]{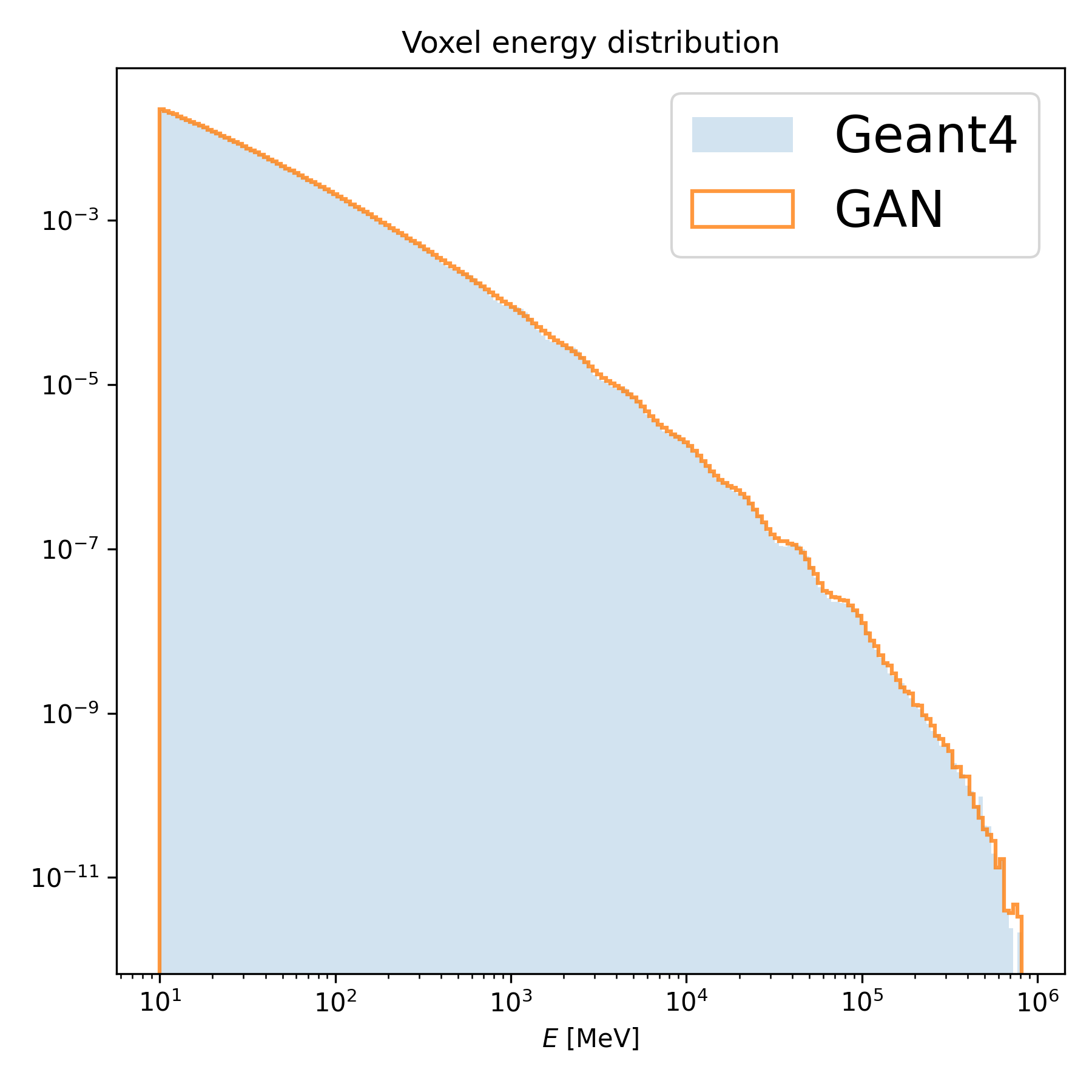} 
    \includegraphics[width=0.3\textwidth]{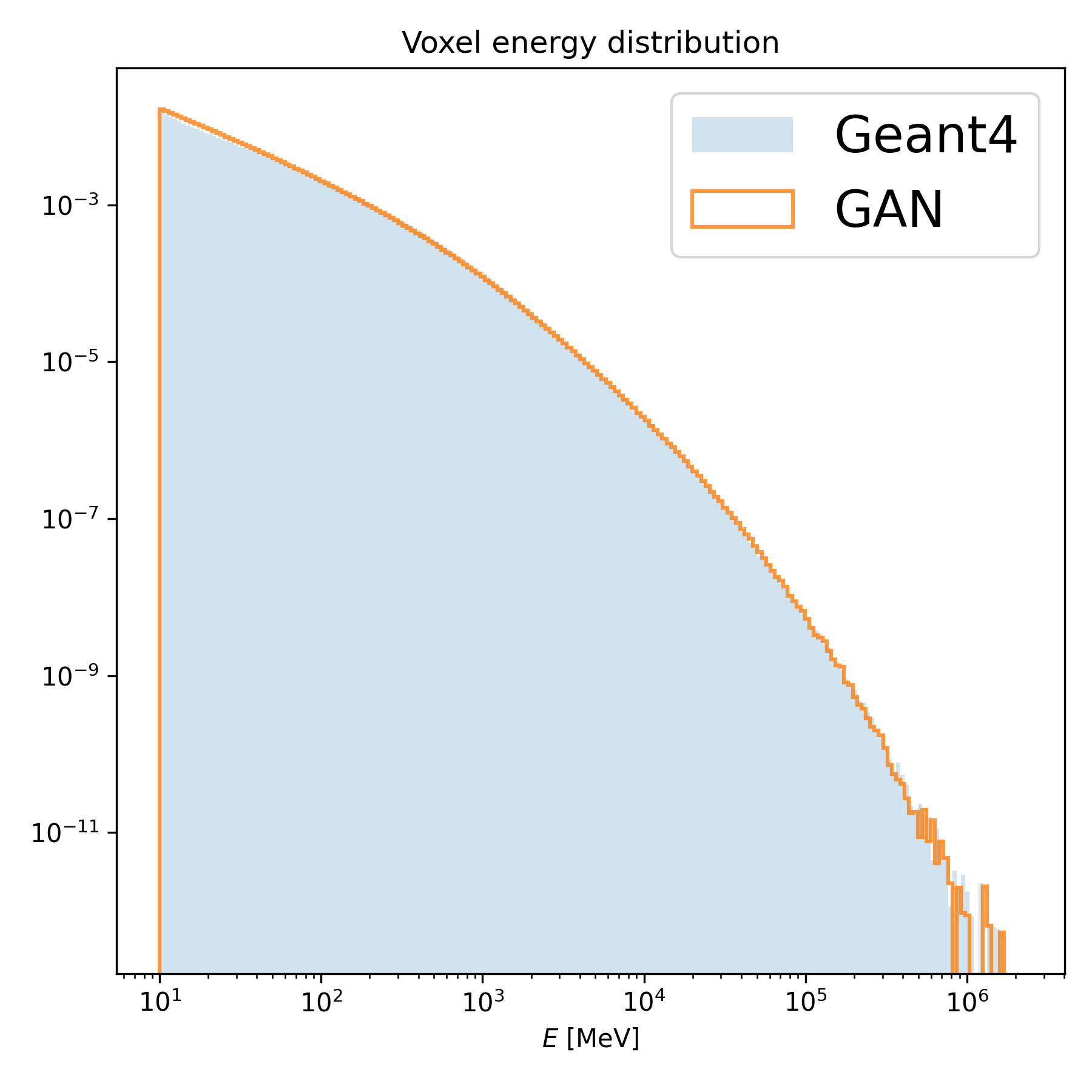}
    \caption{
        \textbf{Photons (left) and pion (right) \CSG with layer-energy normalisation}: distribution of the voxel energy.
        These distributions include contributions from all energies and all layers.
    }
    \label{fig:voxelEnergy}
\end{figure}

\begin{figure}[htp]
\centering
    \includegraphics[width=0.3\textwidth]{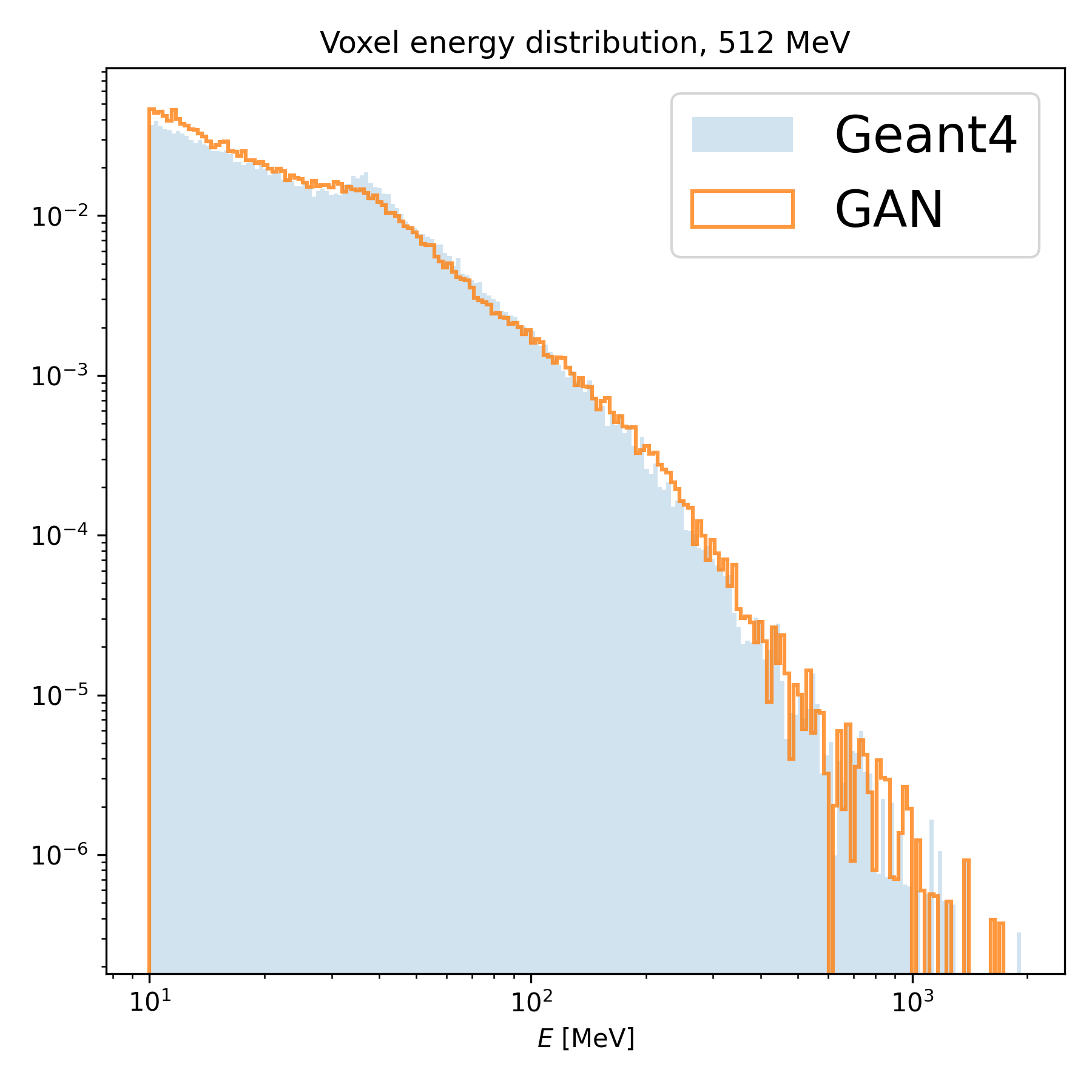} 
    \includegraphics[width=0.3\textwidth]{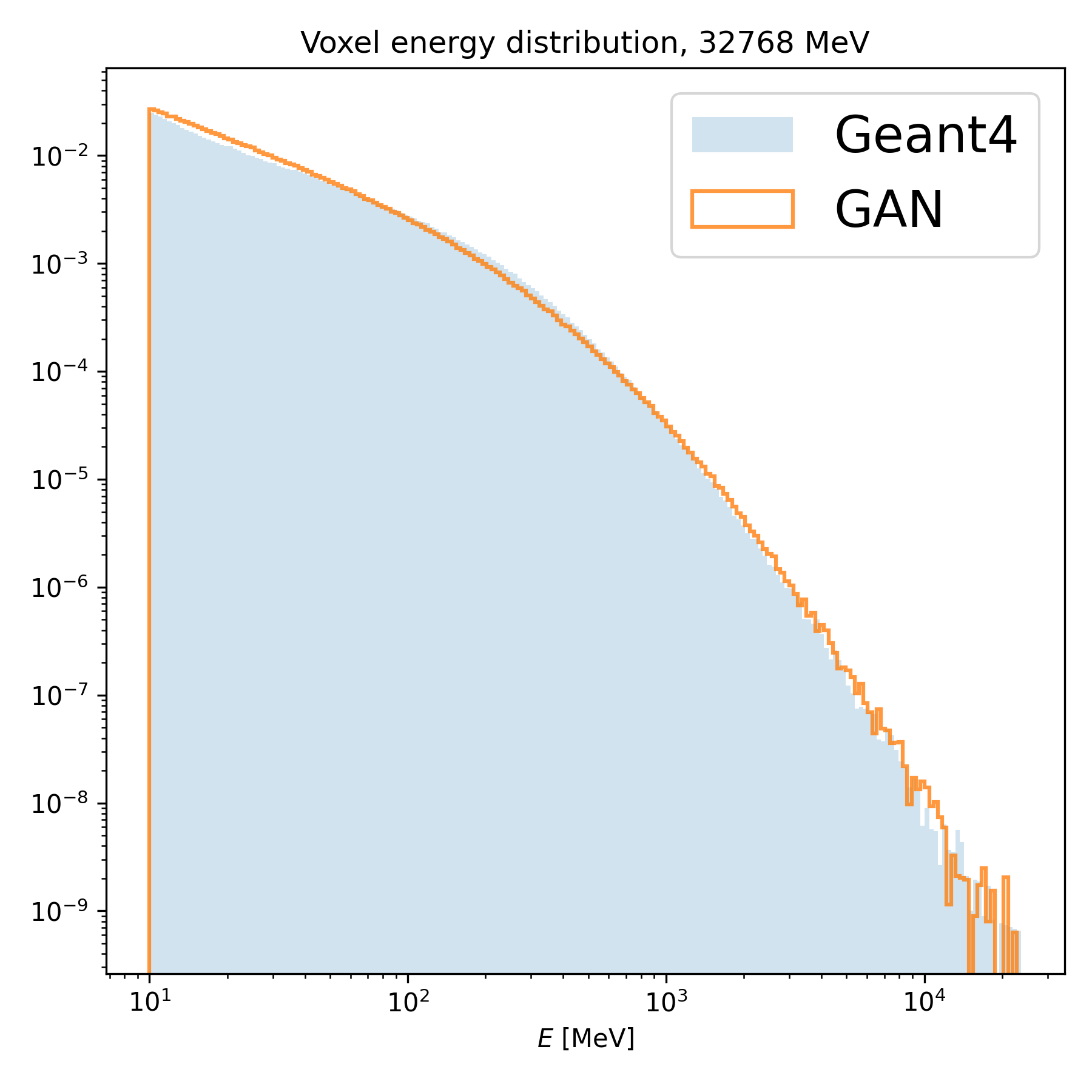}
    \includegraphics[width=0.3\textwidth]{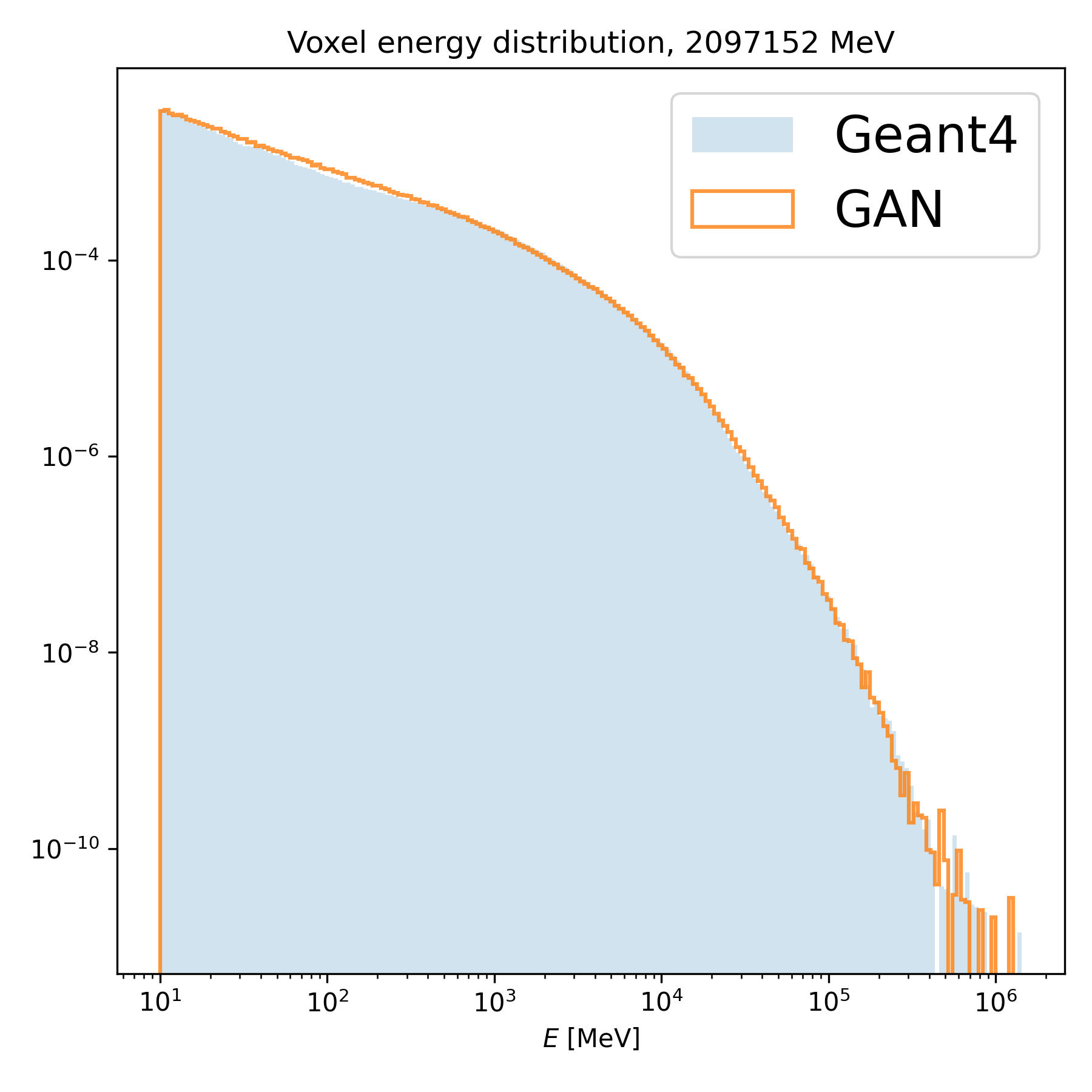} 
    \caption{
        \textbf{Pion \CSG with layer-energy normalisation}: distribution of the voxel energy of pions of 512~\MeV (top left), 32~\GeV (top right) and 2~\TeV (bottom).
        These distributions include contributions from all layers.
    }
    \label{fig:voxelEnergyperSample}
\end{figure}

The efficacy of a model can also be evaluated by considering low-level variables such as the energy in each voxel, in contrast to the high-level observables examined in previous sections.
The distribution of energy across all voxels from all samples depicted in \Fig{\ref{fig:voxelEnergy}} for photons and pions, reveals an impressive accord in these distributions by \CSG.
Furthermore, this agreement between \CSG and \GEANT is observed in individual incident momentum samples as illustrated in \Fig{\ref{fig:voxelEnergyperSample}}.

\FloatBarrier

\subsection{Generation time}
\label{sec:time}
The generation time for the GANs is assessed on CPUs and is summarised in \Tab{\ref{tab:time}} for different particles and momenta.
A single event is generated, mirroring the common scenario at LHC where events typically include only a few particles within the small $|\eta|$ range.
As expected, the generation time does not depend on the energy of the particle to be simulated but depends on the particle type due to the different network's complexity.
Therefore, the generation of pions demands relatively more time due to the larger network.

The per-particle generation time can be reduced by increasing the batchsize which refers to the number of particles generated simultaneously, as is indicated in \Tab{\ref{tab:timeBatch}}.
Although this strategy is presently not applicable within the ATLAS fast simulation framework, it can bring advantages if the experiment changes its strategy to parametrising the detector with larger $\eta$ slices.
For instance, wider regions could be defined to include the entire Barrel region ($|\eta| < 1.2$) or the entire EndCap region ($1.5 < |\eta| < 3$).
Such an approach would enable the simultaneous simulation of many particles within these regions.
This approach would facilitate the concurrent simulation of a bunch of particles within these specified regions, particularly when simulating a jet containing a spray of hadrons, which are currently all simulated using the pion GAN in AtlFast3.

\begin{table}[htb]
    \centering
    \caption{
        Generation time for a single particle.
        Results are averaged over 100 trials and the standard deviations are taken as errors, measured on i9-Intel(R) Core(TM) 9900K CPU @ 3.60GHz.
    }
    \begin{tabular}{cc
        S[round-mode=places, round-precision=2]
    }
        \toprule
        Particle    & {Energy}  & {Time [ms]} \\
        \midrule
        Photons     & 1~\GeV  & 4.5 \pm 0.2\\ 
                    & 65~\GeV & 4.9 \pm 0.2\\ 
                    & 1~\TeV  & 4.1 \pm 0.2\\ 
        \midrule
        Pions       & 1~\GeV  & 6.2 \pm 0.2\\ 
                    & 65~\GeV & 6.3 \pm 0.3\\
                    & 1~\TeV  & 6.3 \pm 0.3\\ 
        \bottomrule
    \end{tabular}
\label{tab:time}
\end{table}

\begin{table}[htbp]
    \centering
    \caption{
        Generation time for pions of 65~\GeV with various batch sizes.
        Results are averaged over 100 trials and the standard deviations are taken as errors, measured on i9-Intel(R) Core(TM) 9900K CPU @ 3.60GHz.
    }
    \begin{tabular}{c
        S[round-mode=places, round-precision=2]
    }
        \toprule
        {Batch size}  & {Time per batch [ms]} \\
        \midrule
        1           & 6.3 \pm 0.3\\ 
        10          & 6.8 \pm 1.5\\ 
        100         & 8.2 \pm 1.5\\ 
        1000        & 14.4 \pm 2.1\\ 
        10000       & 70.8 \pm 4.8\\ 
        \bottomrule
    \end{tabular}
\label{tab:timeBatch}
\end{table}

\subsection{Memory requirement}
\label{sec:memory}
It is crucial to make sure \CSG model is small enough to be usable.
Assessing memory consumption is complex and depends on the chosen inference tool within the production system.
Here the memory usage is estimated through the LWTNN library~\cite{LWTNN}, which is the inference tool used in AtlFast3.
LWTNN operates by storing weights in a JSON file, subsequently loaded into memory.
Although the size of this file does not directly translate to the exact memory demand, as optimisations can be made and there exist associated overheads, the ratio between these two files can offer a sense of the additional memory required by \CSG due to the enlarged networks and the additional GANs employed for photons (and also for electrons, given the identical network structure shared, as employed in FastCaloGAN).

The sizes of the JSON file for FastCaloGAN are approximately 3~MB for photons and 3.9~MB for pions, whereas in the case of \CSG, the networks necessitate 6.2~MB for photons and 5.6~MB for pions.
The value for photons is multiplied by 3 for optimal performance, though it could be scaled down by a factor of 2 with a minor reduction in quality when employing only two GANs.
In total, approximately 43~MB would be needed to parametrise a detector slice with \CSG, including a pion GAN alongside 3 GANs each for electrons and photons.
This constitutes roughly four times the memory demand of FastCaloGAN.

Scaling the estimation to the entire detection range, a preliminary approximation of the actual memory needed by \CSG can be derived based on the numbers provided in Ref.~\cite{ATL-SOFT-PUB-2018-001}.
Here the assumption is made that \CSG would be integrated into the ATLAS Athena framework as a replacement for FastCaloGAN.
And additional assumption is that the previously mentioned scaling in the size of the networks is the same for all detector regions.
The publication states that a pure parameterisation with FastCaloGAN requires 2.5~GB of memory.
Given that the ATLAS Athena framework~\cite{athena} consumes approximately 2~GB on its own, it can be inferred that the memory footprint of the pure FastCaloGAN parameterisation for the entire detector range is approximately 0.5~GB.
Consequently, the adoption of \CSG to exchange FastCaloGAN would necessitate roughly 2~GB for parameterisation, resulting in a total memory requirement of 4~GB for a fast simulation task, comfortably fitting within the memory capacity of the computing system used for ATLAS simulation jobs.
Therefore, \CSG could be seamlessly deployed without major modifications within the ATLAS production system.
Additionally, the memory footprint could be further reduced by utilising the ONNX library~\cite{onnxruntime} or other optimised inference libraries.

\FloatBarrier

\section{Future directions}
\label{sec:future}
The performance of \CSG holds potential for improvement through addressing the visible discrepancies presented in this paper.
For example, the response of low-momentum pions could be improved to address the width of the shower which is not well described in some calorimeter layers.
The \CSG also struggle to accurately replicate pions interactions as minimum ionising particles.
This could be addressed by categorising events depending on the starting position of hadronic shower.
However, implementing this solution is not straightforward, as it requires large training statistics for each category and multiple GANs to be used in the inference stage.

Likewise, although significant progress has been achieved, the precision of photon GANs still falls behind of the levels achieved by other models, in particularly in the medium momentum range where the modelling is comparatively less accurate.
Therefore, any future research efforts that prioritise this specific region, may lead to a refinement of the electromagnetic shower simulation and further enhance the capabilities of the photon GANs.

Considering that some of the GANs in \CSG have achieved a low \chisq close to 1, future research could involve the development of a more sophisticated figure of merit to select the optimal iteration.
This necessity arises due to the existence of multiple iterations that give close \chisq value in the total energy distributions but differ in their ability to describe the shapes of the simulated events.
One potential solution would be expanding the histograms considered in the \chisq calculation, including not only the total energy but also the shapes and/or the energy distributions in each layer.
This solution introduces computational complexity though, in particular that assessing the shapes requires expensive computations.
Moreover, this assessment must be repeated for all generated events at every energy point for each iteration.
Hence it was not employed in the present work.

Adopting a different figure of merit may also need a re-evaluation of the optimal training iteration count for GANs.
While an extended training is already adopted for pions, this adjustment could also serve as a straightforward approach to further enhance the performance of all GANs.

A natural expansion of \CSG, as highlighted in \Sect{\ref{sec:time}}, involves the potential to parametrise a wider detector range using a single GAN.
This advancement could be realised by incorporating an additional conditional parameter into the GAN input layer, specifically the $\eta$ value of the incident particle. 
Although it is not feasible to test on the CaloChallenge dataset due to the absence of this parameter, we maintain confidence that this strategy would prove successful given the condition of comparable responses across different $\eta$ values.
Although a single GAN might not adequately parametrise the entire detector range for a complex system like ATLAS, we anticipate that a modest estimate of around 10 GANs could potentially replace the current deployment of 100 GANs in AtlFast3.
Despite challenges introduced by the increased data volume and problem complexity, it would simplify the overall system and reduce the resources demanded during simulation.

The current approach, employing fixed momentum points with substantial statistics per point, has proven highly effective for GANs.
FastCaloGAN in ATLAS already demonstrated the efficacy of interpolating between momentum points, thereby satisfying the experiment's physics requisites.
An alternative strategy involves training on a continuous momentum distribution, similar to other datasets in the CaloChallenge.
While this approach assures improved momentum interpolation, it introduces a challenge in selecting the optimal training iteration.
A new metric would be required to replace the conventional \chisq.
Additional explorations and studies will be necessary when training with this type of dataset.

While refining \CSG, various data manipulations and training strategies employed by other models in the CaloChallenge are explored.
However, it was found that these approaches did not yield enhancements in performance:

Both diffusion models~\cite{amram2023calodiffusion} and normalising flow models~\cite{Krause:2021ilc} introduce noise into the dataset, which is then subtracted from the generated events.
Diverse levels of noise, ranging from 1~\keV to 1~\MeV, are tested, but no observable improvement in performance is detected; in fact, degradation is noted in most cases.
Similarly, ideas involving masking voxels through a range of thresholds are tested, yielding no discernible benefits.
While the best GANs do not incorporate either of these options, considering the advantageous outcomes witnessed with other approaches, further investigations in this direction remain a possibility.
Such studies could potentially enhance the performance of GANs or offer evidence that GANs exhibit robustness in handling low-momentum voxels compared to other models.

Another interesting area for future exploration involves the implementation of variable learning rates.
At present, a static learning rate is employed; thanks to the incremental nature of training, adapting the learning rate value as the training progresses could offer improved training stability and potentially yield enhanced results.
However, this concept is not included in the present results due to its unsuccessful initial trials and the already commendable performance achieved by \CSG in comparison to state-of-the-art benchmarks.

\section{Conclusions}
\label{sec:conclude}

In particle physics research, the need for fast and precise simulations is ever-growing.
Attention has shifted from traditional methods to machine learning-based methods.
The development of \CSG exploits Generative Adversarial Networks (GANs) and achieves a significant improvement with respect to the GAN-based method of FastCaloGAN used in the ATLAS experiment.
While many new types of generative models have been proposed in the last few years, \CSG underscores the enduring competitiveness of GANs and achieves a similar performance to the state-of-the-art generative models.
This accomplishment is achieved through the optimisation of the GAN architecture, the hyper-parameter optimisation, and, crucially, the pre-processing of the data.
While certain improvements are rooted in machine learning techniques, the majority stem from a profound knowledge of the calorimeter showering processes in the ATLAS detector.
These insights demonstrate how domain knowledge is still a crucial factor in maximising the efficacy of machine learning tools.
While the work presented can easily be applied to any calorimeter that uses a similar voxelisation strategy to the one implemented by ATLAS, it is crucial to emphasise that given the similarities to FastCaloGAN, this work can seamlessly be integrated into the ATLAS software framework.
This could potentially yield a substantial performance enhancement for the forthcoming generation of ATLAS simulations.

\section*{Acknowledgements}
We would like to thank the organisers of the CaloChallenge for creating the competition and for the useful discussions carried out while preparing this paper. In particular, we would like to thank Dalila Salimani for the fruitful exchange of ideas concerning the layer-energy normalisation approach. 
We wish to express our gratitude for the substantial efforts of the ATLAS collaboration in releasing the codebase and dataset to the public.
Our appreciation extends to the ATLAS fast calorimeter community for their valuable insights and information shared regarding the dataset.
This project has received funding from the European Union’s Horizon 2020 research and innovation programme under the Marie Skłodowska-Curie grant agreement No 754496.
RZ is supported by US High-Luminosity Upgrade of the Large Hadron Collider (HL-LHC) under Work Authorization No.\ KA2102021.
In this work, we used the NumPy 1.19.5~\cite{harris2020array}, Matplotlib 3.5.1~\cite{Hunter:2007}, sklearn 1.0.2~\cite{scikit-learn}, h5py 3.1.0~\cite{collette_python_hdf5_2014}, TensorFlow 2.6.0~\cite{tensorflow2015-whitepaper}, Pandas 1.4.1~\cite{reback2020pandas} software packages.
We are grateful to the developers of these packages.

\bibliographystyle{elsarticle-num}
\def\bibname{\Large\bf References}
\bibliography{main}

\end{document}